\documentclass[fleqn,usenatbib]{mnras}

\usepackage{newtxtext,newtxmath}
\usepackage[T1]{fontenc}

\DeclareRobustCommand{\VAN}[3]{#2}
\let\VANthebibliography\thebibliography
\def\thebibliography{\DeclareRobustCommand{\VAN}[3]{##3}\VANthebibliography}

\usepackage{graphicx}	% Including figure files
\usepackage{amsmath}	% Advanced maths commands
\usepackage{bm} % For bold maths symbols
\usepackage{aecompl}    % To allow rendering of figure axis titles

\title[High priority targets from glitching pulsars]{High priority targets for transient gravitational waves from glitching pulsars}

\author[Yim et al.]{
Garvin Yim,$^{1}$\thanks{E-mail: g.yim@pku.edu.cn (GY)}
Lijing Shao$^{1,2}$
and Renxin Xu$^{1,3}$
\\
$^{1}$Kavli Institute for Astronomy and Astrophysics, Peking University, Beijing 100871, China\\
$^{2}$National Astronomical Observatories, Chinese Academy of Sciences, Beijing 100012, China\\
$^{3}$Department of Astronomy, School of Physics, Peking University, Beijing 100871, China
}

\date{Accepted XXX. Received YYY; in original form ZZZ}

\pubyear{\the\year{}}

\begin{document}
\label{firstpage}
\pagerange{\pageref{firstpage}--\pageref{lastpage}}
\maketitle

\begin{abstract}
%This is a simple template for authors to write new MNRAS papers.
%The abstract should briefly describe the aims, methods, and main results of the paper.
%It should be a single paragraph not more than 250 words (200 words for Letters).
%No references should appear in the abstract.
Glitching pulsars are expected to be important sources of gravitational waves. In this paper, we explore six different models that propose the emission of transient continuous waves, lasting days to months, coincident with glitches. The maximal gravitational wave energy is calculated for each model, which is then used to determine whether associated gravitational waves could be detectable with LIGO-Virgo-KAGRA's O4 detectors. We provide an analytical approximation to calculate the signal-to-noise ratio which includes information about the source's sky position, improving on previous estimates that assume isotropic or sky and orientation averaged sensitivities. By analysing the entire glitching population, we find that certain models predict detectable signals in O4, whereas others do not. We also rank glitching pulsars by signal-to-noise ratio, based on archival data, and we find that for all models, the Vela pulsar (PSR~J0835$-$4510) would provide the strongest signal. Moreover, PSR~J0537$-$6910 is not expected to yield a detectable signal in O4, but will start becoming relevant for next generation detectors. Our analysis also extends to the entire pulsar population, regardless of whether they have glitched, and we provide a list of pulsars that would present a significant signal, if they were to glitch. Finally, we apply our analysis to the April 2024 Vela glitch and find that a signal should be detectable under certain models. The non-detection of a supposedly detectable signal would provide an efficiency factor that quantifies a model's contribution to gravitational wave emission, eventually leading to a differentiation of models and independent constraints on physical parameters.
\end{abstract}


%%%%%%%%%%%%%%%%%%%%%%%%%%%%%%%%%%%%%%%%%%%%%%%%%%

%%%%%%%%%%%%%%%%% BODY OF PAPER %%%%%%%%%%%%%%%%%%

\section{Introduction}

With LIGO-Virgo-KAGRA's (LVK's) Observing Run 4 (O4) currently underway, each day lies a new opportunity to detect a gravitational wave (GW) signal from the Universe's most energetic events. So far, we have detected at least 90 GW signals using ground-based detectors which have all come from the collision of binary compact objects such as black holes and neutron stars \citep[NSs;][]{LVK2023GWTC3}. Moreover, several pulsar timing array collaborations have recently reported evidence of a stochastic GW background \citep{nanograv2023, epta2023, ppta2023, cpta2023}. It is evident that the era of GW astronomy has well and truly begun. 

Although we have successfully made several detections, there are still some types of signals that we have not yet detected, namely, continuous GWs (CWs) and burst signals. Conventionally, CWs are quasi-monochromatic and have durations lasting much longer than a given observational period, such that they can be treated as quasi-infinite. These types of signals are therefore modelled as sinusoids so are relatively straightforward to understand. Burst signals on the other hand do not necessarily have such a simple form. They can be modelled, e.g.~a rapidly decaying sinusoid, or they can be unmodelled, which is what is typically done in practice.

In recent years, more interest has developed in \textit{transient} CWs, which are sometimes called ``long transient signals'' \citep{prixGiampanisMessenger2011, LVK2022magnetarbursts, LVK2022transientCWsearch}. Like conventional CWs, transient CWs are modelled, having the same quasi-monochromatic behaviour but having a finite duration which is longer than a burst signal. Typically, this means that they have a duration of minutes to months. Having a finite duration means that there must also be a start time. Often, transient CWs are treated like conventional CWs but with a window function that encompasses the signal, causing a time evolution to the amplitude that otherwise would not have been there.

Hypothetically, any transient event could trigger transient CWs (or a burst signal), but only certain astrophysical cases have been looked at in detail. One case comes from magnetar bursts and giant flares, which are radiative events categorised by having peak luminosities of $\lesssim 10^{43}~\text{erg~s}^{-1}$ and $> 10^{44}~\text{erg~s}^{-1}$, respectively \citep{kashiyamaIoka2011, kaspiBeloborodov2017}. 

The other well-known case, which will be the case that we will focus on here, is from NS glitches. A glitch refers to the sudden increase in the spin of a NS, thought to be due to the unpinning of superfluid vortices \citep{andersonItoh1975}. This unpinning causes a rapid transfer of angular momentum from the fast superfluid interior to the crust, which the magnetosphere and electromagnetic radiation are tied to. There is also another popular model, though disfavoured for NSs with large and frequent glitches, which is called the starquake model \citep{ruderman1969, baymPines1971}. This model attributes glitches to a rapid decrease in the moment of inertia of the NS when the crust's elastic limit is exceeded during the spin-down of a NS. A review of these glitch models and others can be found in \cite{haskellMelatos2015}. For some glitches, there is also a glitch recovery where the spin recovers back to, but not necessarily reaching, the pre-glitch spin frequency. This process normally takes days to months \citep{antonopoulouHaskellEspinoza2022, zhouetal2022}, similar to the timescales of interest that define transient CWs.

Several groups have extended these glitch models by considering the GWs that are emitted either during or immediately after a glitch. A recent review can be found in \cite{haskellJones2024}. Several of these models predict bursts of GWs \citep[e.g.][]{keerJones2015, hoetal2020, yimJones2023, lopezetal2022, pradhanPathakChatterjee2023, wilsonHo2024}, but we will only focus on models that predict transient CWs from now on. 

Transient CWs from glitches can come from: the liberation of pinned superfluid energy \citep{prixGiampanisMessenger2011}, Ekman pumping \citep{vanEysdenMelatos2008, bennettvanEysdenMelatos2010, singh2017}, transient mountains \citep{yimJones2020, moraguesetal2023} or trapped ejecta \citep{yimetal2024}. In this work, we will explore these models and what will be most important to us will be the amount of emitted GW energy, $E_\text{GW}$. We will evaluate model-independent estimates of $E_\text{GW}$ as well as model-specific values. 

The ultimate goal of this work is to provide a list of high priority targets for GW searches. This would allow for efficient use of limited computing power as well as provide electromagnetic observers more reason to obtain more accurate and frequent timing data. Furthermore, having a list of high priority targets will help guide GW searches, especially those using open data where archival data can (and should) be searched over.

In fact, there have already been two GW searches conducted for the signals that we are focusing on here. The first transient CW search from glitching pulsars was done by \cite{keiteletal2019} and they looked specifically at the Crab and Vela pulsars which glitched once each during Observing Run~2 (O2). No GW signal was detected so upper limits were set on the GW strain as a function of signal duration. A similar search was done by LVK and other astronomers \citep{LVK2022transientCWsearch, modafferietal2021} for Observing Run 3 (O3), where there were 9 glitches from 6 pulsars, but no GW signals were detected in that effort either.

Alongside trying to detect these signals, we can also make statements about the prospects of detecting  them with current and future GW detectors. This is what \cite{yimJones2020} did for the Crab and Vela pulsars using the transient mountain model. %They found that for current detectors, it is unlikely that glitches from the Crab pulsar will be detectable, but for Vela's glitches, they found that it could be detectable. Promisingly, they found that both the Crab and Vela's glitches will be detectable with next generation detectors. 
Their analysis was based on the signal-to-noise ratio (SNR), a proxy for the coherent $\mathcal{F}$-statistic \citep{jaranowskiKrolakSchutz1998, cutlerSchutz2005, dressigackerPrixWette2018}, with a fixed threshold for detection. \cite{moraguesetal2023} extended their work by considering the entire known glitching pulsar population using a model-agnostic approach, as well as specialising to the transient mountain model. In their analysis, they used the semi-coherent $\mathcal{F}$-statistic \citep{prixGiampanisMessenger2011} as their detection statistic and the detection threshold was carefully calculated according to \cite{tenorioetal2022}. 

In this work, we return to the relatively simple method of using the SNR to quantify detectability, but we will make some improvements that allow us to include previously neglected features, such as the source's sky position and orientation as well as that of the GW detector. Doing this allows the detectability to be approximated more accurately than before, c.f.\,\cite{yimJones2020}, but also allows for a relatively easy analysis of theoretical models without having to go into the complexities related to the GW search itself, like what was done in \cite{moraguesetal2023}. We will follow \cite{moraguesetal2023} by generalising our analysis to the entire glitching pulsar population as well as choosing to be agnostic to the specific transient CW model, in addition to specialising to specific models. We go one step further and also look at pulsars that have not glitched, assuming that one day they might do. For the first time, we assess the detectability of transient CWs from Ekman pumping. We also apply our analysis to the recent April 2024 glitch observed in the Vela pulsar \citep{zubietaetal2024, campbell-wilsonFlynnBateman2024, groveretal2024, palfreyman2024, wangetal2024}.

In Section~\ref{section_gravitational_wave_energy_budget}, we provide an overview of the different models used in this analysis. Then, in Section~\ref{section_signal_to_noise_ratio}, we present an analytic approximation for the SNR of transient CWs, which accounts for the position and orientation of the source and GW detector. In Section~\ref{section_population_study}, we use this approximation to calculate the detectability of all known glitching pulsars, but we also provide an analysis that is independent of whether the pulsar has been observed to glitch. Moreover, in Section~\ref{section_glitch_in_vela}, we use our analytic approximation to assess the detectability of the recent April 2024 glitch observed in the Vela pulsar. Finally, Section~\ref{section_discussion} contains some points of discussion and Section~\ref{section_conclusion} concludes this paper.

\section{Gravitational wave energy budget}
\label{section_gravitational_wave_energy_budget}

Broadly speaking, the energy budget for GW emission can come from two distinct parts of a glitch: the glitch rise and/or the glitch recovery. One common aspect of glitch rise models is that they are able to explain the short duration spin-up part of the glitch by enforcing the conservation of angular momentum. Post-glitch models are typically agnostic to what causes the spin-up; they are only concerned about the spin-down part which is seen as the recovery. The post-glitch models outlined here also consider angular momentum conservation, but only for the post-glitch regime. There has also been discussion about potential \textit{pre-glitch} GW emission \citep{yimJones2020} but we will not consider that possibility here. A recent review of GWs from glitching pulsars can be found in \cite{haskellJones2024}. 

In addition to these two types of models, there are also ``agnostic'' models which only consider energy conservation but not angular momentum conservation. These often give energy budgets that are larger than ones found by self-consistent calculations, but they are still useful in that they give firm upper limits on how much energy can be emitted as GWs.

In this section, we outline six models, the first two being ``glitch rise'' models, the next two being ``post-glitch'' models, and the last two are ``agnostic'' models:
\begin{itemize}
\item Starquake model (one component),
\item Vortex unpinning model (two components),
\item Transient mountain model,
\item Ekman pumping model,
\item Na\"ive model (one component),
\item Excess superfluid energy model (two components).
\end{itemize}
Of most interest to us will be the amount of energy that can be released in these models for GW emission, so for each model, we will approximate the (maximal) energy available for GW emission. The energy for each model will be presented in a table in Section~\ref{subsection_table_of_energy_budgets}. The energies will then be used to calculate the SNR later in Section~\ref{section_signal_to_noise_ratio}.

\subsection{Glitch rise models}
\subsubsection{Starquake model (one component)}

The starquake model \citep{ruderman1969, baymPines1971} begins with a newly-born rotating NS that is assumed to be initially entirely fluid. The centrifugal force causes an equatorial bulge and after some time, when the NS cools, the body starts to solidify eventually forming a crust that is oblate in shape. The shape at which this crust solidifies is known as the ``reference oblateness'' and it is defined as when there is no strain in the crust.

As the NS spins-down (primarily through magnetic dipole radiation), the centrifugal force becomes weaker, leading to the NS wanting to become more spherical. However, due to the rigidity of the crust, this is prevented from happening at the expense of elastic strain being built up. 

When the strain becomes too large and the elastic limit is reached (at the breaking strain), there is a catastrophic event, called a starquake, which causes the crust to break and the NS rapidly changes shape into one that is less oblate. The moment of inertia decreases when this happens and because angular momentum needs to be conserved, we observe that the NS must spin-up, i.e.~the NS has a glitch.

In the simplest starquake model, the NS can be treated as just being one component behaving as an elastic body. The moment of inertia about the rotation axis is $I$ and the angular frequency is given by $\Omega = 2\pi\nu$, where $\nu$ is the spin frequency of the NS. If we assume that all the excess rotational kinetic energy after the glitch goes into GW production, we find
\begin{gather}
E_\text{GW} = \Delta E_\text{rot} \equiv \frac{1}{2}(I_0+\Delta I)(\Omega_0+\Delta\Omega)^2 - \frac{1}{2} I_0 \Omega_0^2~, \\
\label{E_GW_starquake}
E_\text{GW} = \frac{1}{2} I_0 \Omega_0 \Delta\Omega~,
\end{gather}
where we have ensured that angular momentum has been conserved
\begin{align}
\Delta J \equiv (I_0+\Delta I)(\Omega_0+\Delta\Omega) - I_0 \Omega_0 = 0~,
\end{align}
where subscript `0' represents quantities before the glitch and `$\Delta$' quantities show the perturbed value immediately after the glitch, which are assumed to be much smaller than their respective pre-glitch values, i.e.~\mbox{$\Delta X/X_0 \ll 1$} for a generic variable $X$. To first order in small quantities, we find that the observed glitch size is given by a reduction in the moment of inertia, shown as
\begin{align}
\frac{\Delta\Omega}{\Omega_0} \approx -\frac{\Delta I}{I_0}~.
\end{align}
Note that eq.\,(\ref{E_GW_starquake}) is an exact result and does not require one to consider only first order terms. It is also worth comparing this energy budget with the ``na\"{i}ve'' estimate where only energy is conserved, which will be covered later. One finds that $E_\text{GW}$ from the starquake model is half the value expected from just energy conservation alone, c.f.~eq.\,(\ref{E_GW_naive}). The results calculated here agree with earlier calculations by \cite{sideryPassamontiAndersson2010} and \cite{LSC2011velaglitch}.

\subsubsection{Vortex unpinning model (two components)}

We will now move onto a more realistic two component set-up for describing NSs. There is great reason to consider NSs with (at least) two components with (at least) one of them being superfluid; without superfluidity, there are several observed phenomena that are difficult to explain such as: large and frequent glitches \citep{baymPines1971, pinesAlpar1985}, long post-glitch recovery timescales \citep{baymPethickPines1969, baymetal1969, alparetal1984, pinesAlpar1985}, and long thermal cooling timescales \citep{shibazakiLamb1989, pageApplegate1992, larsonLink1999}. More recently, observations of glitches with a ``delayed spin-up'' \citep{shawetal2018, ashtonetal2019, shawetal2021} give observational evidence for there being three components \citep{graberCummingAndersson2018},\footnote{See also \cite{haskelletal2018} who were able to use a two-component model to explain delayed spin-ups. The key ingredient that allowed them to do this was to have vortex accumulation and differential rotation, which affected the coupling of different parts of the star \citep{khomenkoHaskell2018}.} but we will not cover this here and leave it for future work (Yim \& Jones, in prep.). For our purposes, two components are enough as this is all that is needed to explain glitch rises and post-glitch recoveries \citep{andersonItoh1975, alparetal1984}. 

The way in which superfluidity describes glitches is as follows \citep{andersonItoh1975}. In the two component model, one component is a ``pinned'' superfluid inside the NS, and the other component represents everything else, including any ``unpinned'' superfluid in the crust or core, as well as the normal baryonic crust \citep{baymetal1969}. 

A rotating NS means that the superfluid component also rotates, but it can only carry angular momentum by forming an array of vortices where the vortices are assumed to be parallel to the rotation axis. The areal density of these vortices then determines the local rotational velocity of the superfluid, which is valid for regions encasing a sufficiently large number of vortices \citep{anderssonComer2021}.

As the NS spins-down on secular timescales, the areal density of vortices must also decrease, so it does this by having vortices migrate radially outwards (on average). However, during this migration, the vortices can become pinned in the core or in the inner crust, preventing further spin-down of the superfluid. This leads to the other (unpinned) component being the sole component that is spinning down. As a result, a lag develops between the two components.

As the NS spins-down more, the lag grows and it becomes harder for the vortices to remain pinned. At some critical lag, the vortices collectively unpin and rapidly transfer their angular momentum to the unpinned component which is what we observe as a glitch. Consequently, the pinned component spins-down and the unpinned component spins-up, resulting in the lag decreasing. Usually, the lag is assumed to be zero after the glitch so that both components co-rotate temporarily. However, this assumption is still unclear and model-dependent. For simplicity, we will assume co-rotation immediately after the glitch.

With this information, we can now calculate how much energy remains after accounting for the changes in rotational kinetic energy of the two components. To do this, we first write down the total angular momentum of the NS system
\begin{align}
\label{angular_momentum_total_two_component_model}
    J = I_\text{p} \Omega_\text{p} + I_\text{u} \Omega_\text{u}~,
\end{align}
where the subscripts `p' and `u' represent the pinned and unpinned components, respectively. The magnetosphere which is responsible for the observed rotational frequency is assumed to be frozen to the unpinned component so we have $\Omega_\text{u} = \Omega$.

We then perturb the above equation and set \mbox{$\Delta J = 0$} to find
\begin{align}
\Delta\Omega = \Delta\Omega_\text{u} = - \frac{I_\text{p}}{I_\text{u}} \Delta \Omega_\text{p}~,
\end{align}
where we have assumed that the moment of inertias remain constant throughout the glitch. This equation verifies our previous qualitative description, where it was said that the pinned component spins-down as the unpinned component spins-up.

Next, we look at the change in rotation kinetic energy of each component, labelled by $i = \text{p}, \text{u}$. If both components co-rotate after the glitch at angular frequency $\Omega_\text{co}$, then the change in rotational kinetic energy of each component is
\begin{align}
\Delta E_{\text{rot},i} = \frac{1}{2} I_i \left[\Omega_\text{co}^2 - (\Omega_\text{co} - \Delta \Omega_i)^2 \right]~,
\end{align}
and the total change in energy is the sum of these, giving 
\begin{align}
\label{change_in_rotational_kinetic_energy_two_components}
\Delta E_\text{rot} =  - \left[ \frac{1}{2} I_\text{u} (\Delta \Omega)^2 + \frac{1}{2} I_\text{p} (\Delta \Omega_\text{p})^2\right]~,
\end{align}
which is a result also obtained by \cite{prixGiampanisMessenger2011}. Note that the change in energy is negative showing that the final state has less energy than the initial state. We assume this released energy goes directly into GW production meaning 
\begin{align}
\label{E_GW_vortex_unpinning}
E_\text{GW} = \left|\Delta E_\text{rot}\right| = \frac{1}{2} I (\Delta \Omega)^2 \left(\frac{I}{I_\text{p}} - 1\right)~,
\end{align}
where we have used $I = I_\text{p} + I_\text{u}$. Typically, the fractional moment of inertia associated with the pinned superfluid $I_\text{p}/I$, is comparable to the observed fractional change in spin-down rate $\Delta \dot{\Omega}/\dot\Omega_0$ \citep{linkEpsteinLattimer1999, anderssonetal2012}, and from statistical analyses of glitch data, one finds $I_\text{p}/I \sim 0.01 - 0.1$ \citep{anderssonetal2012, hoetal2015, fuentesetal2017, hoetal2022}. It is also possible to calculate $I_\text{p}/I$ from the perspective of microphysics, where an equation of state and superfluid pairing gap model is needed \citep{anderssonComerGlampedakis2005, anderssonetal2012}. Depending on model assumptions, it is possible to get reasonable agreement between microphysical theory and glitch observations. However, there is ongoing debate on whether the superfluid in the inner crust is enough to provide the angular momentum reservoir required to power glitch activity when the effect of entrainment is included \citep{chamel2012, anderssonetal2012, piekarewiczFattoyevHorowitz2014}. For our purposes here, we neglect entrainment.

One notable point about eq.\,(\ref{E_GW_vortex_unpinning}) is that the energy obtainable in the vortex unpinning model is quadratic in the glitch size, $\Delta \Omega$, whereas it is linear for the starquake model. This means calculated energies will be a few orders of magnitude smaller. Aside from \cite{prixGiampanisMessenger2011}, this quadratic relation was also found by \cite{sideryPassamontiAndersson2010} and \cite{LSC2011velaglitch}, however, their coefficients were slightly different from what was derived here. We attribute this to a difference in the interpretation of the two components in the case for \cite{sideryPassamontiAndersson2010}, or simply from rounding for \cite{LSC2011velaglitch}.

Finally, one can also write eq.\,(\ref{change_in_rotational_kinetic_energy_two_components}) in terms of the lag between the two components, $\omega = \Omega_\text{p} - \Omega_\text{u}$, giving
\begin{align}
\label{change_in_rotational_kinetic_energy_two_components_lag}
\Delta E_\text{rot} =  - \frac{1}{2} I_\text{u} \omega_0 \Delta \Omega~,
\end{align}
where we have assumed co-rotation after the glitch, so that $\omega_0 \equiv \Omega_{\text{p}, 0} - \Omega_{\text{u}, 0} = \Delta \Omega_\text{u} - \Delta \Omega_\text{p}$, where $\omega_0$ is the lag just before the glitch. We choose to use eq.\,(\ref{E_GW_vortex_unpinning}) for our expression for $E_\text{GW}$ as there is only one uncertain parameter, $I_\text{p}$, rather than two as in the case for eq.\,(\ref{change_in_rotational_kinetic_energy_two_components_lag}) where $I_\text{u}$ and $\omega_0$ are uncertain.

\subsection{Post-glitch models}

\subsubsection{Transient mountain model}
\label{subsubsection_transient_mountain_model}

We now outline the first of two post-glitch models. For glitches that show a glitch recovery, the rate in change of angular frequency $\dot{\Omega} < 0$ (also called the spin-down rate $|\dot{\Omega}|$), usually shows a decrease immediately after the glitch, i.e.~$\Delta \dot{\Omega} < 0$. Or, in other words, the spin-down rate increases after the glitch. 

In the transient mountain model \citep{yimJones2020}, the increase in spin-down rate is ascribed to the formation of a NS mountain at the moment of the glitch, which radiates away GWs and causes a braking torque on the system. Then, as the glitch exponentially recovers, the mountain dissipates away, reducing the braking torque and causes the spin-down rate to recover back towards the pre-glitch spin-down rate. It is exactly analogous to the case when calculating the ``upper spin-down limit'' for secularly spinning-down pulsars, but here, it is specific for glitches.

During this recovery, transient CWs are emitted at twice the spin frequency of the NS, i.e.~$f = 2\nu$. If the recovery of the observed spin frequency is exponential with a recovery timescale of $\tau_\text{EM}$, then the amplitude of the transient CWs will also recover exponentially, but with a timescale of $\tau_\text{GW} = 2\tau_\text{EM}$. Explicitly, the GW amplitude evolves as
\begin{align}
h_0(t) = \sqrt{- \frac{5}{2}\frac{G}{c^3}\frac{I}{d^2}\frac{\dot{\Omega}_0}{\Omega_0}\left(\frac{\Delta \dot\Omega}{\dot\Omega_0}\right)} e^{-\frac{\Delta t}{2\tau_\text{EM}}}~,
\end{align}
where $d$ is the distance to the source and $\Delta t = t - t_\text{g}$ is the time that has elapsed after the glitch. The above equation only holds when there is little or no permanent change in the spin-down rate caused by the glitch. In the case that there is, then $\Delta\dot\Omega$ should be replaced by the ``transient'' part of $\Delta \dot\Omega$ that fully recovers on long timescales, i.e.~when $\Delta t\gg \tau_\text{EM}$.

The timing model of a glitch is often modelled as an instantaneous increase in the spin frequency at the time of the glitch $t_\text{g}$ followed by the sum of an exponential recovery, a linear recovery as well as a permanent offset \citep{edwardsHobbsManchester2006, yuetal2013}. Mathematically, this is shown as 
\begin{align}
\Delta\nu(t) = \left.
\begin{cases}
0~, & \text{if } t < t_\text{g} \\
\Delta\nu_\text{p} + \Delta\dot{\nu}_\text{p} \cdot \Delta t + \Delta\nu_\text{t}e^{-\frac{\Delta t}{\tau_\text{EM}}}~, & \text{if } t \ge t_\text{g} 
\end{cases}
\right.
\end{align}
where $\Delta \nu_\text{p}$ is the permanent offset in spin due to the glitch, $\Delta \dot\nu_\text{p}$ is the permanent offset in the spin-down rate due to the glitch, and $\Delta \nu_\text{t}$ is the transient change in spin that will fully recover long after the glitch.

One can see that the total spin-up at $t = t_\text{g}$ is $\Delta\nu(t_\text{g}) = \Delta\nu_\text{p} + \Delta\nu_\text{t}$ and it is standard to define the healing parameter as
\begin{align}
    Q \equiv \frac{\Delta\nu_\text{t}}{\Delta\nu(t_\text{g})}~,
\end{align}
which tells us the fraction of the total spin-up that recovers on timescales much greater than $\tau_\text{EM}$. For $Q = 0$, there is no recovery so the glitch is simply seen as a step function, whereas $Q = 1$ corresponds to a full recovery with no permanent offset. With this definition, we can write down the expected GW energy from the transient mountain model as \citep{yimJones2020}
\begin{align}
\label{E_GW_transient_mountain}
    E_\text{GW} = Q I_0 \Omega_0 \Delta \Omega~.
\end{align}
If one uses $Q = 1/2$, we get the same energy as the starquake model\footnote{This is only to recover the same energy and not a prediction of the starquake model in general.} and for $Q = 1$, we get the same as the ``na\"ive'' model that will be covered later in Section~\ref{subsubsection_naive_model}, where only energy conservation is considered. The value of $Q$ varies for different pulsars and even for different glitches from the same pulsar, but to get an idea of their values, the average $Q$ for the Crab and Vela pulsars are around $\langle Q\rangle \sim 0.8$ and $\langle Q\rangle\sim 0.2$, respectively \citep{crawfordDemianski2003, yimJones2020}.

\subsubsection{Ekman pumping model}

Ekman pumping is a concept taken from fluid mechanics that concerns what happens to a viscous fluid when acted upon by a tangential force at its boundary with a container. For a review, see \cite{bentonClark1974}. The analogy for NSs is the fluid interior reacting to the rapid spin-up of the solid crust during a glitch. At the fluid-container boundary, the fluid has a no-slip condition, meaning that if the fluid were to have a relative velocity with respect to the container, then there must be a tangential fluid flow gradient that develops. The region in which this gradient exists is called the Ekman layer and the gradients eventually cause meridional flows in the fluid, ``pumping'' fluid in cells around the container.

There are three phases of Ekman pumping that occur when the container spins-up. Firstly, there is the formation of the Ekman layer which occurs on the spin-up timescale, then there is the spin-up of the fluid in the Ekman layer on the order of the Ekman timescale $\tau_\text{E}$, and lastly, there is the viscous decay of the Ekman layer which occurs on some viscous timescale. For post-glitch recoveries, we are concerned about the second phase where the fluid is increasing its velocity on the Ekman timescale.

\cite{greenspanHoward1963} were pioneers in this field for the modern era, developing and writing down the governing equations for an incompressible and uniform fluid contained in a cylinder. \cite{walin1969} improved on this by exploring stratified fluids, and \cite{abneyEpstein1996} further extended to compressible fluids. The main conclusions were that stratification and compressibility decreased the size of the Ekman layer, effectively decreasing the amount of fluid that needs to spin-up, resulting in smaller Ekman timescales.

Over a decade later, \cite{vanEysdenMelatos2008} realised that it was possible to get GWs from Ekman pumping if non-axisymmetric meriodional flows were excited. They calculated the GW strain from the mass quadrupole and found that it peaked resonantly around frequencies of $f = 2\nu$ and $f = \nu$ (the $\nu$ radiation is only observable for a non-polar observer), with characteristic dimensionless GW strain 
\begin{align}
    h_0^\text{(mass)} =\frac{\pi \rho_0 G L^6 \Omega^4}{c^4 g d} \left(\frac{\Delta\Omega}{\Omega}\right)~,
\end{align}
where $\rho_0$ is the equilibrium mass density of the fluid ($\rho_0 \sim 10^{17}~\text{kg m}^{-3}$), $L$ is some typical length scale of the system ($L \sim 10^4~\text{m}$), $g$ is the gravitational acceleration at the boundary ($g \sim 10^{12}~\text{m s}^{-2}$) and $d$ is the distance to the source. The typical timescale for the GW emission is given by the rescaled Ekman timescale
\begin{align}
\label{ekman_timescale}
    \tau_{\text{E}, \alpha\gamma} = E^{-\frac{1}{2}} \Omega^{-1} \omega_{\alpha\gamma}^{-1}~,
\end{align}
where $E = \nu_\text{vis}/(L^2\Omega)$ is the dimensionless Ekman number with $\nu_\text{vis}$ being the viscosity of the fluid interior, and $\omega_{\alpha\gamma}$ is some dimensionless parameter that determines how quickly the Ekman flow decays away which depends on stratification, compressibility and how fast the system is spinning \citep{vanEysdenMelatos2008, singh2017}. $\alpha$ and $\gamma$ label what mode is excited and the leading quadrupole term is represented by $\alpha = 2$.

Soon after, \cite{bennettvanEysdenMelatos2010} found that the current quadrupole contributes more to GW radiation than the mass quadrupole does. The characteristic dimensionless GW strain of the current quadrupole is \citep{singh2017}
\begin{align}
    h_0^\text{(curr)} =\frac{2 \pi \rho_0 G L^6 \Omega^3}{3 c^5 d} \left(\frac{\Delta\Omega}{\Omega}\right)~,
\end{align}
and the GW timescale is the same as the rescaled Ekman timescale provided above in eq.\,(\ref{ekman_timescale}). Like the mass quadrupole case, emission is at $f = 2\nu$ and $f = \nu$, the latter being true only for non-polar observers. The only way to distinguish whether the GW radiation comes from a mass or current quadrupole is by analysing the difference in GW polarisation for a non-polar observer \citep{singh2017}.

Following these calculations, \cite{singh2017} was able to relax some assumptions and explored a wider parameter space. Crucially, he provided an estimate of the amount of GW energy expected to be radiated from Ekman pumping. First, he estimated the maximum energy that is available from the glitch, which is on the order of how much rotational kinetic energy the crust gains
\begin{align}
E_\text{glitch} \sim I_\text{crust} \Omega \Delta \Omega \sim \Gamma M L^2 \Omega \Delta \Omega~,
\end{align}
where $\Gamma$ is the fraction of the total mass contained within the crust which is around $\Gamma \sim 0.01$ \citep{lorenzRavenhallPethick1993, ravenhallPethick1994}. For simplicity, the system evaluated by \cite{singh2017} and others was assumed to be cylindrical in shape (of radius $L$ and height $2L$) so $M = 2\pi\rho_0L^3$. As a result, the glitch energy can now be written as
\begin{align}
E_\text{glitch} = 2 \pi \rho_0 \Gamma L^5 \Omega^2 \left(\frac{\Delta\Omega}{\Omega}\right)~.
\end{align}
Finally, after simulating the system, \cite{singh2017} found that only a small fraction $\eta_\text{Ek} \sim 10^{-7}$ - $10^{-5}$ (for \mbox{$d = 1$ - $10~\text{kpc}$}) of the glitch energy can be radiated away as GWs 
\begin{align}
\label{E_GW_ekman_pumping}
E_\text{GW} = 2 \pi \rho_0 \Gamma L^5 \eta_\text{Ek} \Omega^2 \left(\frac{\Delta\Omega}{\Omega}\right)~,
\end{align}
making the chance of detection slim. Instead, most of the glitch energy goes into the kinetic and potential energy of the fluid. Nevertheless, we can use eq.\,(\ref{E_GW_ekman_pumping}) later in our analysis, and we will show quantitatively that it will indeed be difficult to detect GWs from such a mechanism.

\subsection{Agnostic models}
\label{subsection_agnostic_models}
\subsubsection{Na\"ive model (one component)}
\label{subsubsection_naive_model}
% This same energy was used by Ho et al. (2020) but for f-modes
We will now move onto the energy budgets that are agnostic to the detailed mechanisms at play. They only consider energy conservation, as opposed to the aforementioned models, and they result in optimistic energy budgets. Nevertheless, they are interesting to calculate as they provide an upper limit on how much energy will be available for GW radiation during and immediately after a glitch.

We begin by looking at the simplest case of a one component system with constant moment of inertia $I$. If it were to spin up, then the amount of rotational kinetic energy that it has gained is simply 
\begin{align}
\label{E_GW_naive}
E_\text{GW} \approx I \Omega \Delta \Omega~,
\end{align}
to leading order in $\Delta \Omega$. One might believe that this is the maximal energy that could then go into GWs if all the gained rotational kinetic energy is converted into GWs. In such a case, the increase in rotational kinetic energy must come from somewhere but we remain agnostic to where this energy comes from. It is conceivable that this rotational kinetic energy is gained from mechanisms that occur in the NS's interior, but could equally be due to external factors too. The above equation is identical to the glitch energy used by \cite{hoetal2020} who were concerned about the excitation of $f$-modes by glitches. The transient CW sources considered here have durations that last much longer than the burst sources considered there but they are similar in the fact that both utilise glitches as their energy source.

\subsubsection{Excess superfluid energy model (two components)}
% Also include superfluid excess energy (by virtue of the superfluid spinning faster than the normal component - the extra energy that it has compared to if it were co-rotating)
% Doesn't necessarily depend on a glitch, it just says how much energy is available in the superfluid reservoir
Next, we look at the two component model where there is a relatively faster rotating pinned superfluid component contained within an unpinned component. The fact that the pinned superfluid is not co-rotating with the unpinned component means that there is potentially a reservoir of rotational kinetic energy that can be tapped into (or ``excess'' energy), given by \citep{prixGiampanisMessenger2011}
\begin{align}
E_\text{excess} = \frac{1}{2} I_\text{p} (\Omega_\text{p}^2 - \Omega^2) \approx I_\text{p} \Omega \omega~,
\end{align}
to leading order in the lag $\omega = \Omega_\text{p} - \Omega$, and where $I_\text{p}$ and $\Omega_\text{p} (>\Omega)$ are the moment of inertia and angular frequency of the pinned component. Note that the above energy estimate does not require a glitch to occur. One could imagine that this reservoir of energy is depleted slowly by some dissipative mechanism where energy is lost from the system but angular momentum is not. Clearly, if GW radiation was the underlying mechanism for depletion, then one must account for the angular momentum carried away by the GWs for a self-consistent calculation, but the point of the above calculation is that it provides a firm upper limit for $E_\text{GW}$ whilst disregarding angular momentum conservation. In this case, the only variable that changes is the lag (angular momentum conservation would require $\delta \Omega \ne 0$), and so the amount of energy obtainable in this agnostic model is
\begin{align}
E_\text{GW} = -\delta E_\text{excess} = - I_\text{p} \Omega \delta \omega~,
\end{align}
where `$\delta$' represents the change in a variable over timescales much longer than the glitch rise timescale (which is indicated by `$\Delta$'). We also know $\delta \omega < 0$ and the minus signs have been added so that $E_\text{GW}$ is positive. 

It is interesting to note that if one uses the conservation of angular momentum at this point and suppose that the change in the lag is due to a glitch, $\delta \omega = \Delta \omega$, then by conserving angular momentum in eq.\,(\ref{angular_momentum_total_two_component_model}), we find 
\begin{align}
\label{conservation_of_angular_momentum_two_component}
\delta \omega = \Delta \omega = - \frac{I}{I_\text{p}} \Delta\Omega~,
\end{align}
which leads to
\begin{align}
\label{E_GW_excess_superfluid}
E_\text{GW} = I \Omega \Delta \Omega~,
\end{align}
which is the same as what was calculated using the na\"ive model in eq.\,(\ref{E_GW_naive}), making the two component model no different from the one component model. This is precisely what was done in eq.\,(13) of \cite{prixGiampanisMessenger2011} and eq.\,(6) of \cite{moraguesetal2023}.

However, if one wanted to self-consistently account for the conservation of angular momentum, then it should have been done from the beginning, in which case, the GW energy should also include the effect of $\delta \Omega$ leading to
\begin{align}
E_\text{GW} = -\delta E_\text{excess} = - I_\text{p} \Omega_0 \delta \omega - I_\text{p} \omega_0 \delta \Omega~,
\end{align}
and when we assume $\delta \omega = \Delta \omega$ and $\delta \Omega = \Delta \Omega$, and using eq.\,(\ref{conservation_of_angular_momentum_two_component}), we get
\begin{align}
\label{self_consistent_E_GW_two_component}
E_\text{GW} \approx I \Omega_0 \left(1 - \frac{I_\text{p}}{I} \frac{\omega_0}{\Omega_0}\right) \Delta \Omega~.
\end{align}
If the pinned superfluid fraction $I_\text{p}/I$ and/or the normalised pre-glitch lag $\omega_0/\Omega_0$ is much smaller than one, then we recover the earlier estimate given by eq.\,(\ref{E_GW_excess_superfluid}).

In reality, any lag over a sufficiently long time will tend towards the steady-state lag $\omega_\infty$ which is non-zero so long as there is a spin-down torque acting on the NS. It is usually assumed that the lag is in steady-state just before the glitch, where some instability or event triggers the glitch \citep{alparetal1994, anderssonComerPrix2003, warszawskiMelatosBerloff2012}. Using the two component model \citep{baymetal1969}, one finds that the normalised steady-state lag is given by
\begin{align}
\frac{\omega_\infty}{\Omega_0} = \frac{1}{2} \frac{I_\text{p}}{I} \left(1 - \frac{I_\text{p}}{I}\right)^{-1} \frac{\tau_\text{coup}}{\tau_\text{age}}~,
\end{align}
where $\tau_\text{coup}$ is a coupling timescale between the pinned and unpinned component which is on the order of the glitch recovery timescale (days to months), and $\tau_\text{age} = - \Omega_0/2\dot\Omega_0$ is the characteristic age of the NS, see also Yim \& Jones (in prep.). Inputting a range of possible parameters, $I_\text{p}/I = [0.01, 0.1]$ and $\tau_\text{coup}/\tau_\text{age} = [10^{-3}, 10^{-4}]$, we find that the second term in the parenthesis in eq.\,(\ref{self_consistent_E_GW_two_component}) is smaller by a factor of $10^{-8} - 10^{-5}$ compared to the first term, and so can be safely ignored. This results in the GW energy being given by eq.\,(\ref{E_GW_excess_superfluid}), regardless of whether angular momentum conservation is self-consistently accounted for or not.

\subsection{Table of energy budgets}
\label{subsection_table_of_energy_budgets}
% Summarise all energies with coefficient beta being different for all models. Pull out factor of I Omega Delta Omega (such that naive model has value of 1).
% Provide estimates of all the different betas using typical inputs.
We have now looked at all the energy budget models in detail and we summarise the results in Table~\ref{table_of_E_GW}. In the table, we also provide expressions for $\kappa$ for the different models, which are the coefficients when compared to the agnostic models, defined as
\begin{align}
\label{E_GW_kappa}
E_\text{GW} \equiv \kappa I \Omega \Delta \Omega~,
\end{align}
where $\kappa = 1$ for the agnostic models of Section~\ref{subsection_agnostic_models}.

\begin{table*}
 \centering
 \caption{\label{table_of_E_GW}A summary of the GW energies calculated using different models. A comparison to agnostic estimates is given by $\kappa$, where agnostic models have $\kappa = 1$ and is defined in eq.\,(\ref{E_GW_kappa}). A detailed description of the model parameters can be found under the respective sections in the main text.}
 \begin{tabular}{ccccccc}
 \hline \hline
 & \multicolumn{2}{c}{Glitch rise} & \multicolumn{2}{c}{Post-glitch} & \multicolumn{2}{c}{Agnostic}\\
 & Starquake & Vortex unpinning & Transient mountain & Ekman pumping & One component & Two components \\ \hline \\ \vspace{-20pt} \\
 $E_\text{GW}$ & $\frac{1}{2} I \Omega \Delta\Omega$ & $\frac{1}{2} I (\Delta \Omega)^2 \left(\frac{I}{I_\text{p}} - 1\right)$ & $Q I \Omega \Delta \Omega$ & $2 \pi \rho_0 \Gamma L^5 \eta_\text{Ek} \Omega \Delta\Omega$ & $I \Omega \Delta \Omega$ & $I \Omega \Delta \Omega$\\
 $\kappa$ & $\frac{1}{2}$ & $\frac{1}{2} \left(\frac{\Delta\Omega}{\Omega}\right) \left(\frac{I}{I_\text{p}} - 1\right)$ & $Q$ & $\eta_\text{Ek} \frac{I_\text{crust}}{I}$ & $1$ & $1$ \\
 Typical value of $\kappa$ & $\frac{1}{2}$ & $10^{-5} - 10^{-4}$ & $10^{-3} - 1$ & $10^{-9} - 10^{-7}$ & $1$ & $1$ \\
 \hline \hline
 \end{tabular}
\end{table*}

One general pattern is that the glitch rise and post-glitch models all produce energies lower than the agnostic models. In this sense, we call the associated agnostic energies the ``upper energy limits'' that can be obtained from glitches. The ``upper spin-down limit'' for glitches is given by the transient mountain model. 

We can also rank the models by how much GW energy can be obtained. In descending order, we have the agnostic models (both one and two components) being the most optimistic, followed by the transient mountain and starquake models, followed by vortex unpinning and finally the most pessimistic model comes from Ekman pumping, which would typically give energies $10^{-9} - 10^{-7}$ smaller than the agnostic case.

In the following sections, we follow the energies obtained from three models: the agnostic case, the transient mountain case and vortex unpinning case. The starquake and Ekman pumping $E_\text{GW}$ can be simply calculated by multiplying the agnostic case by a factor of $1/2$ and $10^{-9} - 10^{-7}$, respectively, whereas the other cases have dependencies on observables which we have data for.

\section{Signal-to-noise ratio}
\label{section_signal_to_noise_ratio}

% Can write in terms of energy, first done by Cutler & Thorne (2002), later verified by Prix, Giampanis & Messenger (2011)
To connect the GW energy to the SNR, we follow a calculation first done by \cite{prixGiampanisMessenger2011}, as well as \cite{yimJones2020}, but we modify it to include geometric effects from the source's sky position and orientation as well as the GW detector's position and orientation on Earth. 

\subsection{Exact description}

As usual, the optimal SNR $\rho$ is defined as \citep{jaranowskiKrolakSchutz1998}
\begin{align}
\label{SNR_definition}
\rho \equiv \sqrt{(h|h)}~,
\end{align}
where the $(\cdot|\cdot)$ notation represents an inner product defined as
\begin{align}
(a|b) \equiv 4\mathrm{Re}\left(\int_0^{\infty}\frac{\tilde{a}(f)\tilde{b}^*(f)}{S_n(f)}\,df\right)~,
\end{align}
where $S_n(f)$ is the one-sided power spectral density at GW frequency $f$, the tilde ($\tilde{~}$) represents a Fourier transform, and the asterisk ($^*$) represents the complex conjugate.

The GW strain $h(t)$ (in general relativity) can always be decomposed into two independent polarisations $h_{+}$ and $h_{\times}$ as
\begin{align}
\label{strain_decomposed_into_polarisations}
h(t) = F_{+}(t)h_{+}(t) + F_{\times}(t)h_{\times}(t)~,
\end{align}
where $F_{+}$ and $F_{\times}$ are the detector's antenna pattern functions for the plus and cross polarisations, respectively. $h_{+}$ and $h_{\times}$ can be further decomposed into a GW amplitude $h_0$, a GW phase $\Psi$, and some functions that depend on the inclination angle $\iota$ (the angle between the source's angular momentum vector and our line of sight) and also the wobble angle $\theta$ (the angle between the source's angular momentum vector and its deformation axis), as illustrated on the right hand side of Fig.\,\ref{figure_geometric_parameters}. The relevant equations can be found in eqs.\,(21) and (22) of \cite{jaranowskiKrolakSchutz1998} but we will not explicitly need them here. Additionally, the GW amplitude can be written in terms of the ellipticity $\varepsilon$
\begin{align}
\label{h_0_epsilon}
h_0 = (2\pi)^2 \frac{G}{c^4} \frac{I f^2}{d} \varepsilon~,
\end{align}
where the ellipticity is defined as
\begin{align}
    \varepsilon \equiv \frac{|I_{xx} - I_{yy}|}{I_{zz}}~,
\end{align}
where $I_{ii}$ represents the source's moment of inertia about its $i$'th principal axis.

If there is a time dependence to the GW amplitude, then substituting eq.\,(\ref{strain_decomposed_into_polarisations}) into eq.\,(\ref{SNR_definition}) for a signal of duration $T_\text{GW}$ and using Parseval's theorem, we get 
\begin{align}
\label{SNR_exact_time_dependent_h_0}
    \rho^2 = \frac{\sin^4\theta}{S_n(f)}&\left[ \frac{1}{4}(1+\cos^2\iota)^2 \int_0^{T_\text{GW}} F_{+}^2(t) h_0^2(t)\,dt \right. \nonumber \\ 
    & ~~\left. + \cos^2\iota \int_0^{T_\text{GW}} F_{\times}^2(t)h_0^2(t)\,dt \right]~,
\end{align}
where we have specialised to GW radiation at twice the source's spin frequency, i.e.~$f = 2\nu$. If one wanted to calculate the SNR for $N$ identical detectors with detector arm opening angles of $\zeta$, then one should multiply the right hand side of this equation by $N \sin^2 \zeta$.

\subsection{Analytic approximations}

If the timescale associated with the evolution of the GW amplitude is long compared to the periodic sidereal variation of the antenna pattern functions, i.e.~\mbox{$\tau_\text{GW} \gg 1~\text{d}$}, then we can treat $h_0(t)$ as constant and pull it out from the integrals to get
\begin{align}
    \rho^2 \approx \frac{h_0^2\sin^4\theta}{S_n(f)}&\left[ \frac{1}{4}(1+\cos^2\iota)^2 \int_0^{T_\text{GW}} F_{+}^2(t)\,dt \right. \nonumber \\ 
    & ~~\left. + \cos^2\iota \int_0^{T_\text{GW}} F_{\times}^2(t)\,dt \right]~,
\end{align}
but we will keep in mind that there is still some time dependence in $h_0$ in reality. This will be important soon. \cite{jaranowskiKrolakSchutz1998} wrote down analytical expressions for the antenna pattern functions and after substituting them into the above equation, they found
\begin{align}
\label{SNR_A_2_B_2}
\rho^2 &= \left[A_2(\delta, \psi, \iota, \lambda, \gamma)T_\text{GW} \right. \nonumber \\
&\left. ~~~~~+ B_2(\alpha, \delta, \psi, \iota, \lambda, \gamma; T_\text{GW})\right]\frac{h_0^2\sin^4\theta}{S_n(f)}~,
\end{align}
where $A_2$ and $B_2$ are functions of geometric parameters like the right ascension of the source $\alpha$, the declination of the source $\delta$, the polarisation angle $\psi$, the source's inclination angle $\iota$, the detector's latitude $\lambda$ and the detector's orientation $\gamma$, measured as the angle between the bisector of the detector arms anticlockwise from East. These geometric parameters are shown visually in Fig.\,\ref{figure_geometric_parameters} and the full expressions for $A_2$ and $B_2$ can be found in Appendix~B of \cite{jaranowskiKrolakSchutz1998} but can also be found here in Appendix~\ref{appendix_A_2_B_2} for convenience.

\begin{figure*}
\centering
\includegraphics[width=\textwidth]{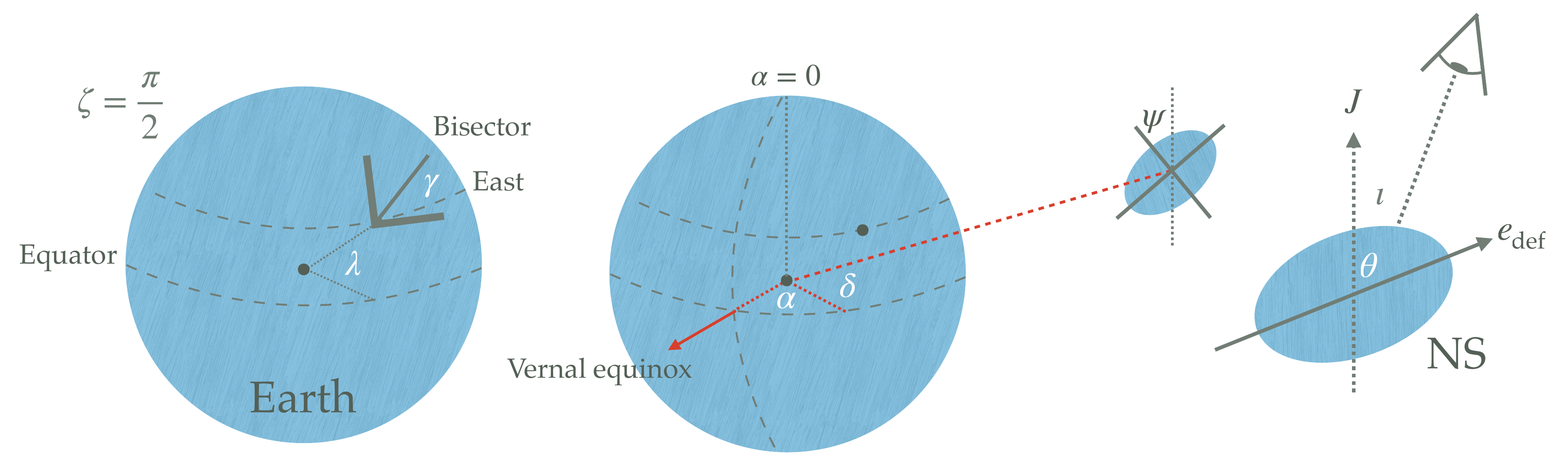}
\caption{\label{figure_geometric_parameters}A figure showing the different geometric parameters that concern the GW source and detector. Starting from the left, the diagram shows an L-shaped ($\zeta = \pi/2$) GW detector as bold solid lines. The angle between the bisector of the detector measured anticlockwise from East is $\gamma$. The central dot represents the centre of the Earth and $\lambda$ is the detector's latitute on Earth. Moving onto the central diagram, $\alpha$ and $\delta$ represent the source's right ascension and declination respectively, with the right ascension measured from the meridian that passes through the vernal equinox. $\psi$ is the polarisation angle and tells us how rotated the source is about our line of sight. Finally, on the right is a diagram of a deformed NS, which has a deformation axis $e_\text{def}$ and angular momentum vector $J$. The angle between the angular momentum vector and our line of sight is the inclination angle $\iota$ and the angle between $J$ and $e_\text{def}$ is the inclination $\theta$.}
\end{figure*}

There are several assumptions we can make to simplify eq.\,(\ref{SNR_A_2_B_2}) and they generally lead to the SNR being of the form
\begin{align}
\rho^2 \propto \frac{h_0^2 T_\text{GW} \sin^4\theta}{S_n(f)}~,
\end{align}
but now we reinstate the time dependence of $h_0$ by using $h_0^2 T_\text{GW} \rightarrow \int_0^{T_\text{GW}} h_0^2(t)\,dt$, which gives
\begin{align}
\label{SNR_beta}
\rho^2 = \beta \frac{\sin^4\theta}{S_n(f)} \int_0^{T_\text{GW}} h_0^2(t)\,dt~,
\end{align}
where the coefficient $\beta$ is determined by what assumptions are being made. For instance, $\beta = 1$ corresponds to the assumption that the GW detector is equally sensitive to all parts of the sky such that $F_{+} = F_{\times} = 1/\sqrt{2}$ (a constant) with the addition of $\iota = 0\text{\textdegree}$. This can be seen directly from eq.\,(\ref{SNR_exact_time_dependent_h_0}). In addition, there is the case where $\beta = 4/25$ which is the well-known coefficient derived by \cite{jaranowskiKrolakSchutz1998} when $\alpha$, $\delta$, $\psi$ and $\iota$, i.e.~the source's sky position and orientation, are averaged over. This latter option for $\beta$ was exactly what was used by \cite{prixGiampanisMessenger2011} and \cite{yimJones2020}.

Here, we derive yet another $\beta$ but specific for transient CW signals. It begins by realising that $B_2$ in eq.~(\ref{SNR_A_2_B_2}) is an oscillating term, with a period of two sidereal days. Therefore, it will always be bound, so for a sufficiently long $T_\text{GW}$, the $A_2 T_\text{GW}$ term will eventually dominate, leading to
\begin{align}
\label{beta_approximation}
\beta \approx A_2(\delta, \psi, \iota, \lambda, \gamma)~.
\end{align}
The explicit form of $A_2$ is provided in Appendix~\ref{appendix_A_2_B_2}. Eq.\,(\ref{beta_approximation}) is what we propose as a suitable choice for calculating the SNR of transient CWs, given $T_\text{GW}$ is sufficiently large. Moreover, this choice of $\beta$ allows for the inclusion of additional information, like the declination of the source, polarisation angle, inclination, detector latitude and detector orientation. Note that when using this approximation, the right ascension is not important. This makes sense as for a sufficiently long signal, the Earth would rotate about its axis at least once, meaning each value of the right ascension would carry equal weighting. Table~\ref{table_beta} summarises the different options for $\beta$, their respective assumptions as well as where each choice is applicable. 

\begin{table*}
\centering
\caption{A table containing the different options for $\beta$, their respective assumptions and the scenarios where their use may be appropriate.}
\label{table_beta}
\begin{tabular}{ccc}
\hline \hline
\centering$\beta$ & Assumptions & When applicable? \\ \hline \\ \vspace{-20pt} \\
\centering1 & Most optimistic & Order-of-magnitude calculations or short GW signals \\
\centering$\frac{4}{25}$ & Sky position and orientation averaged over & Same scenarios as $\beta = 1$ case \\
\centering$A_2$ & Sufficiently long transient signal & Have sky position and/or inclination angle of the source\\
\hline \hline
\end{tabular}
\end{table*}

\subsection{Accuracy of approximation}
\label{subsection_accuracy_of_approximation}

Next, one should ask how good of an approximation eq.\,(\ref{beta_approximation}) is. To do this, we explicitly evaluate the difference between the approximated SNR, where only the $A_2 T_\text{GW}$ term is considered in eq.~(\ref{SNR_A_2_B_2}), and the actual SNR, given precisely by eq.~(\ref{SNR_A_2_B_2}). The fractional difference is therefore given by
\begin{align}
\frac{\rho_\text{approx} - \rho_\text{exact}}{\rho_\text{exact}} = \frac{\sqrt{A_2 T_\text{GW}} - \sqrt{A_2 T_\text{GW} + B_2(T_\text{GW})}}{\sqrt{A_2 T_\text{GW} + B_2(T_\text{GW})}}~,
\end{align}
where we have suppressed some variable dependencies for clarity. In particular, recall that $A_2$ and $B_2$ depend on the detector's latitude and orientation, so we expect there to be different fractional differences for different detectors, even when the GW signal has the same properties. The values of $\lambda$ and $\gamma$ for LIGO Hanford, LIGO Livingston and Virgo are provided in Table~\ref{table_detector_parameters}.

\begin{table}
\centering
\caption{A table containing the latitude $\lambda$ and orientation $\gamma$ parameters of LIGO Hanford, LIGO Livingston and Virgo, taken from \citet{jaranowskiKrolakSchutz1998}.}
\label{table_detector_parameters}
\begin{tabular}{ccc}
\hline \hline
Detector & Latitude $\lambda$ (\textdegree) & Orientation $\gamma$ (\textdegree) \\ \hline \\ \vspace{-18pt} \\
LIGO Hanford & 46.45 & 171.8 \\
LIGO Livingston & 30.56 & 243.0 \\
Virgo & 43.63 & 116.5 \\
\hline \hline
\end{tabular}
\end{table}

\begin{figure}
\centering
\includegraphics[width=\columnwidth]{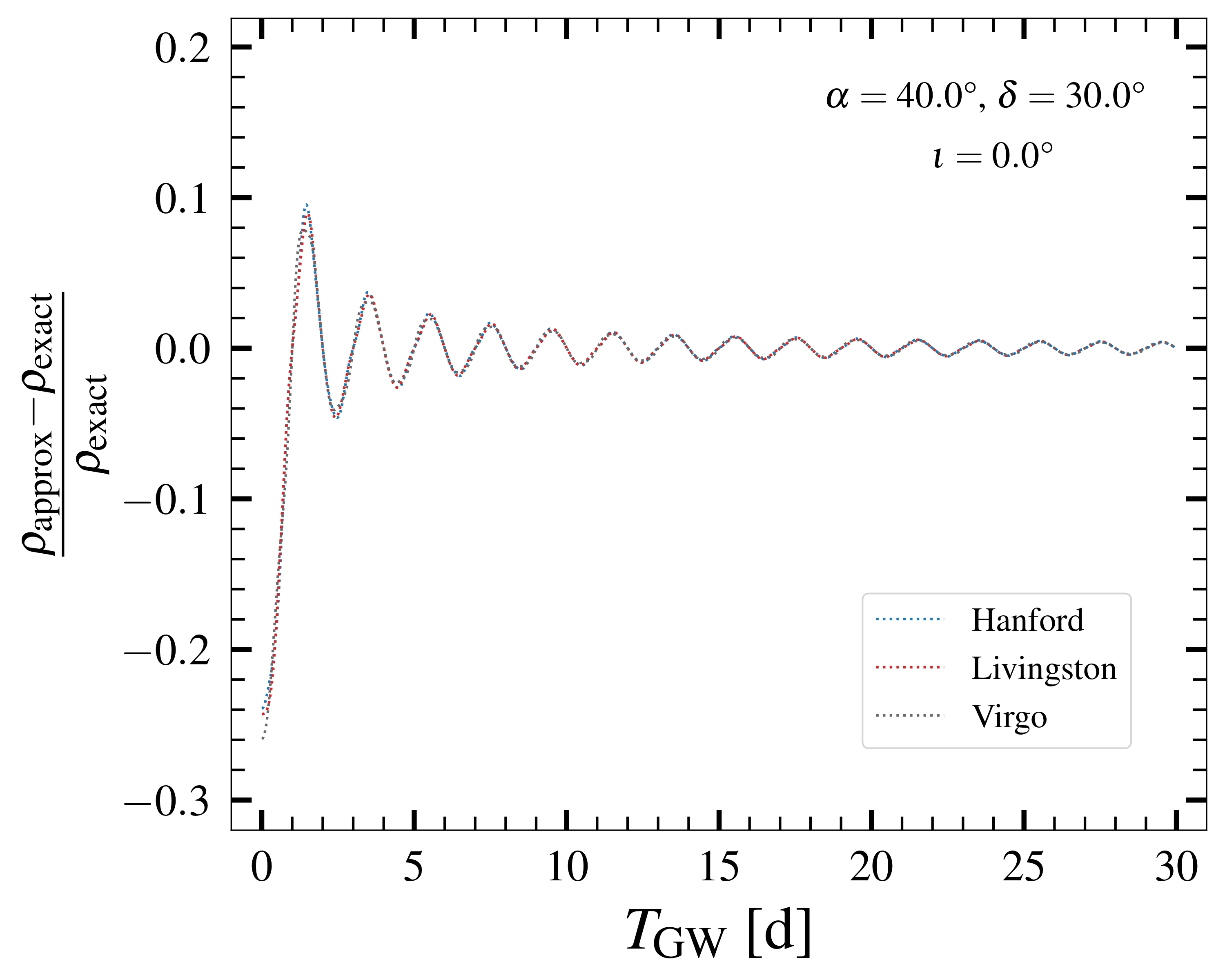}
\caption{\label{figure_alpha_40_delta_30}A figure showing an example of how the fractional difference between the approximated SNR and the exact SNR vary with GW signal duration. It can be clearly seen that the error induced by the approximation decreases for increasing $T_\text{GW}$. After around $T_\text{GW} \approx 1~\text{d}$, the error drops to less than 10\% for all three detectors. The parameters used to create this figure were $\alpha = 40\text{\textdegree}$, $\delta = 30\text{\textdegree}$ and $\iota = 0\text{\textdegree}$.}
\end{figure}

Using this, we plot the fractional difference in SNR as a function of GW duration $T_\text{GW}$ for different detectors. We find that for sufficiently long signals, the error of the approximation tends towards and oscillates around zero. See Fig.~\ref{figure_alpha_40_delta_30} for an example.

\begin{figure*}
\centering
\includegraphics[width=\textwidth]{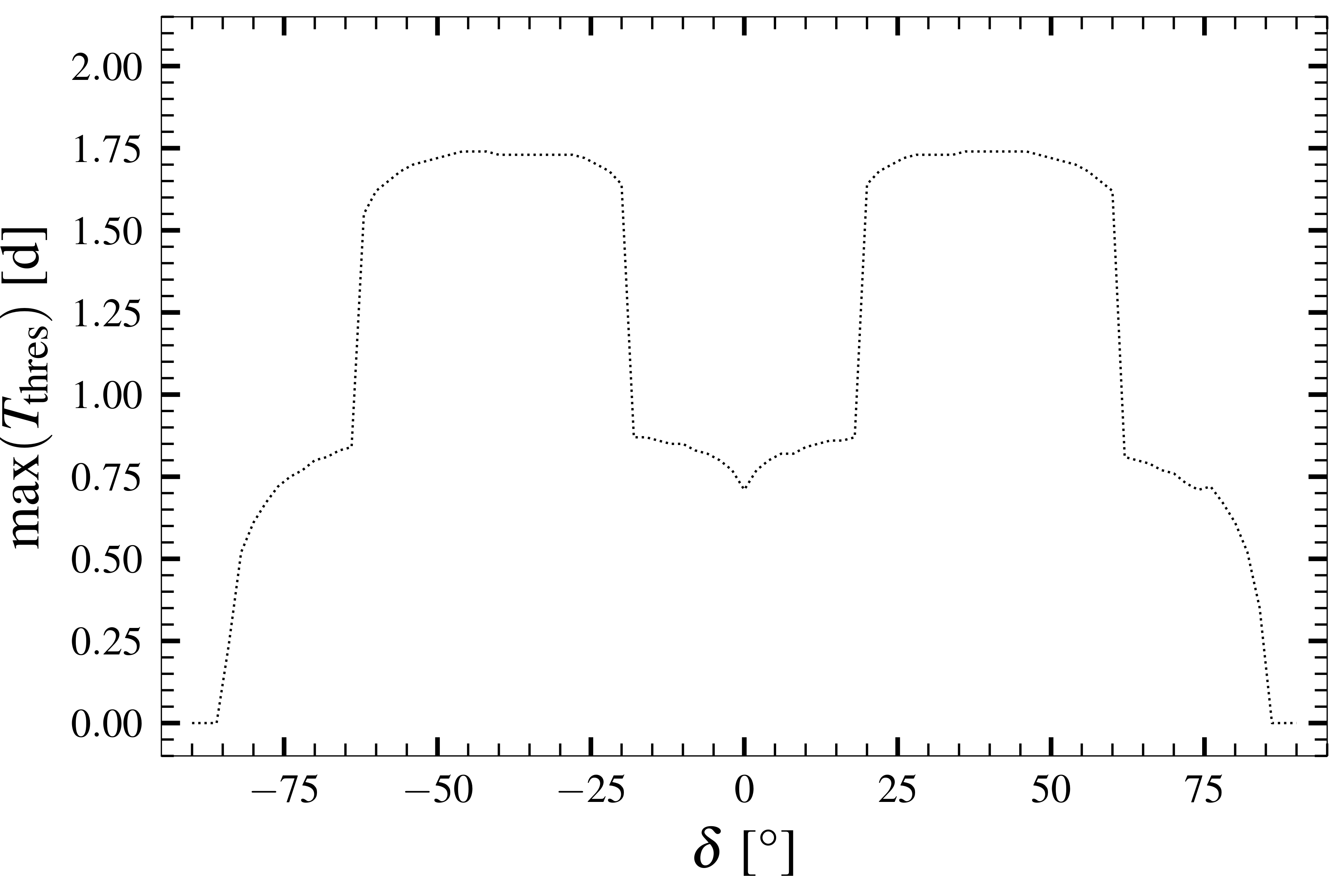}
\caption{\label{figure_max_T_thres}A figure showing the threshold GW duration a signal must have in order for the SNR approximation to have less than 10\% error, as a function of declination $\delta$. The maximisation on $T_\text{thres}$ was done across all values of right ascension $\alpha$ and for all three detectors found in Table~\ref{table_detector_parameters}. The inclination of the source was set to $\iota = 0\text{\textdegree}$ here.}
\end{figure*}

One can go further and explore, for a given $\iota$ and $\psi$, all \mbox{($\alpha$, $\delta$)} parameter space and find the minimum $T_\text{GW}$ required to have less than some percentage error, say 10\%. We will call this the threshold GW duration, $T_\text{thres}$. For instance, in Fig.\,\ref{figure_alpha_40_delta_30}, $T_\text{thres} \approx 1~\text{d}$ for all three detectors. For a given detector, each \mbox{($\alpha$, $\delta$)} maps onto a single value of $T_\text{thres}$ and in Fig.\,\ref{figure_max_T_thres}, we show the maximum $T_\text{thres}$ as a function of $\delta$, where the maximisation has been performed over all possible $\alpha$ and for all three detectors. In other words, for a given $\delta$, no $T_\text{thres}$ would be larger than $\text{max}(T_\text{thres})$ regardless of what value of $\alpha$ or detector one chooses. To create these figures, we selected $\iota = 0\text{\textdegree}$ resulting in there being no dependence on $\psi$. It also means that we only consider circularly polarised GW radiation, i.e.~\mbox{$h_{+} = h_{\times} = h_0$}. 

Working backwards, one can now use Fig.\,\ref{figure_max_T_thres} to find the minimum time the GW signal must last for the approximation to be accurate to more than 90\%. For example, a source with a declination of $\delta = 70\text{\textdegree}$ requires a GW signal to last for at least $T_\text{GW} \approx 0.75~\text{d}$ for the approximation error to be less than 10\%. One can also conclude that as long as the GW signal lasts for at least $T_\text{GW} \approx 1.74~\text{d}$, then the error would be less than 10\% regardless of sky position or detector used. Since glitch recoveries typically have timescales on the order of several days to months, then one can safely use the approximation proposed here to estimate the SNR.

\begin{figure*}
\centering
\includegraphics[width=\textwidth]{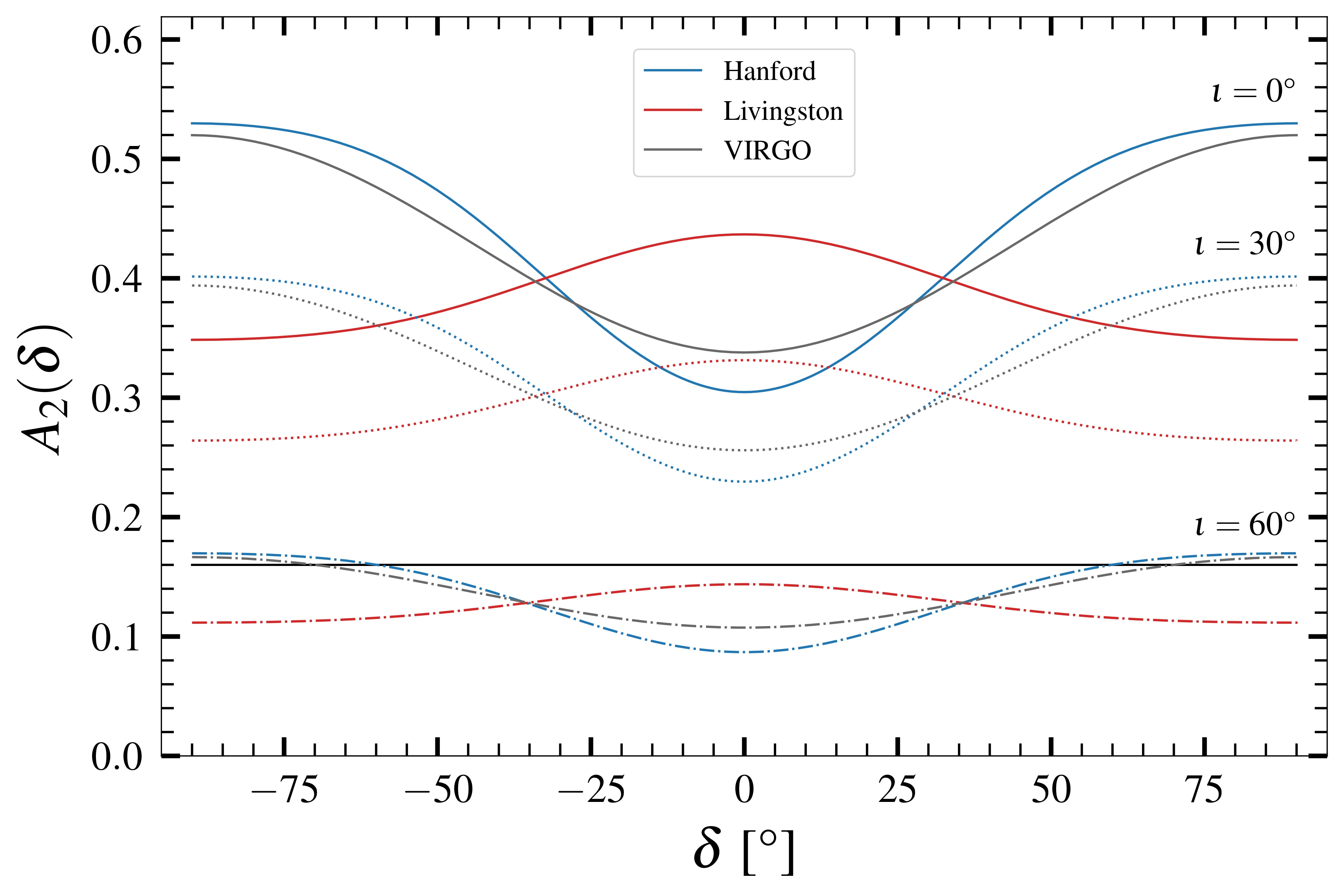}
\caption{\label{figure_A_2_delta}A figure showing $A_2$  as a function of declination $\delta$, for different inclinations $\iota$ and different GW detectors. Linestyles of solid, dotted and dash-dots represent values of $\iota = 0\text{\textdegree}, 30\text{\textdegree}, 60\text{\textdegree}$, respectively, and colours of blue, red and grey represent LIGO Hanford, LIGO Livingston and Virgo, respectively. The horizontal black line at $A_2 = 4/25$ represents the equivalent value one gets if one averages over all sky positions and orientations \citep{jaranowskiKrolakSchutz1998}.}
\end{figure*}

For ease of use, we plot $A_2~(\approx \beta)$ as a function of $\delta$ for different $\iota$ and for different detectors in Fig.\,\ref{figure_A_2_delta}, for $\psi = 0\text{\textdegree}$. One can directly read off the appropriate value of $A_2$ and use it in eq.\,(\ref{SNR_beta}) to estimate the SNR. 

\begin{figure}
\centering
\includegraphics[width=\columnwidth]{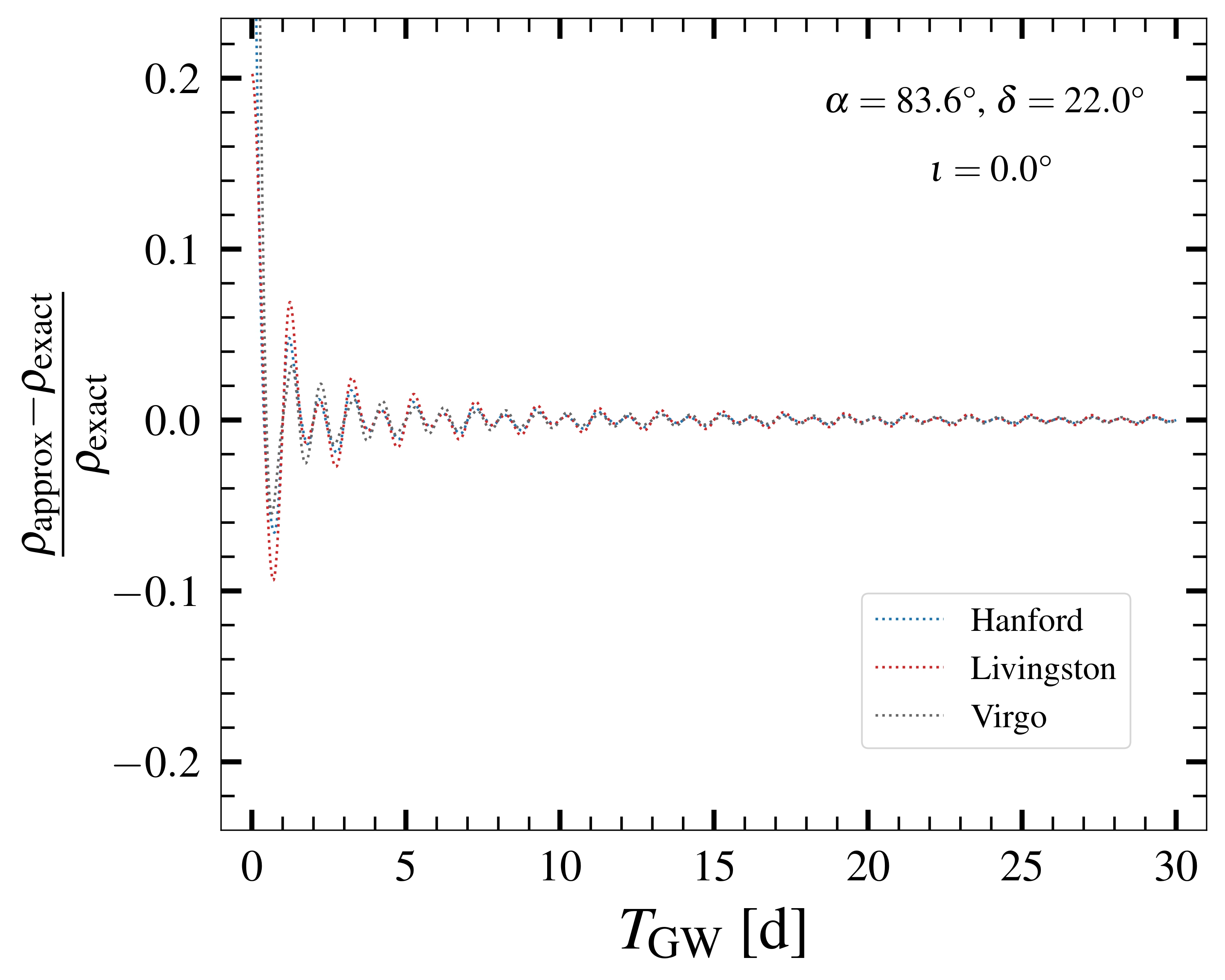}
\caption{\label{figure_crab}A figure showing the fractional difference between the approximate and exact SNR as a function of GW signal duration $T_\text{GW}$ for the Crab pulsar. Values of $\iota = 0\text{\textdegree}$, $\alpha = 83.6\text{\textdegree}$ and $\delta = 22.0\text{\textdegree}$ were used, where the latter two parameters were obtained from the ATNF Pulsar Catalogue \citep{manchesteretal2005}. One finds $T_\text{thres} = 0.36~\text{d}$ (for $<10$\% error) for the Crab pulsar.}
\end{figure}

\begin{figure}
\centering
\includegraphics[width=\columnwidth]{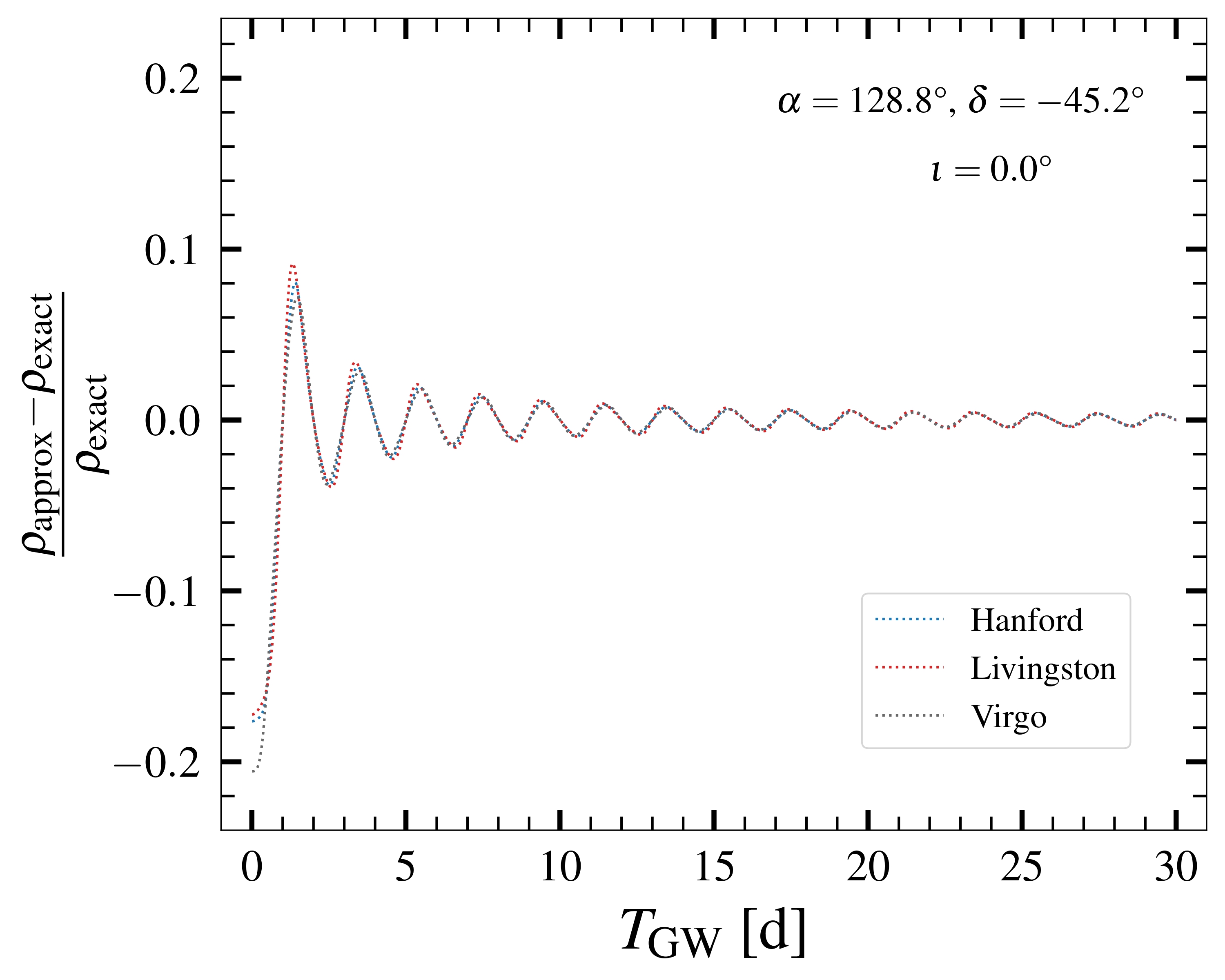}
\caption{\label{figure_vela}A figure showing the fractional difference between the approximate and exact SNR as a function of GW signal duration $T_\text{GW}$ for the Vela pulsar. Values of $\iota = 0\text{\textdegree}$, $\alpha = 128.8\text{\textdegree}$ and $\delta = -45.2\text{\textdegree}$ were used, where the latter two parameters were obtained from the ATNF Pulsar Catalogue \citep{manchesteretal2005}. One finds $T_\text{thres} = 0.77~\text{d}$ (for $<10$\% error) for the Vela pulsar.}
\end{figure}

\begin{figure}
\centering
\includegraphics[width=\columnwidth]{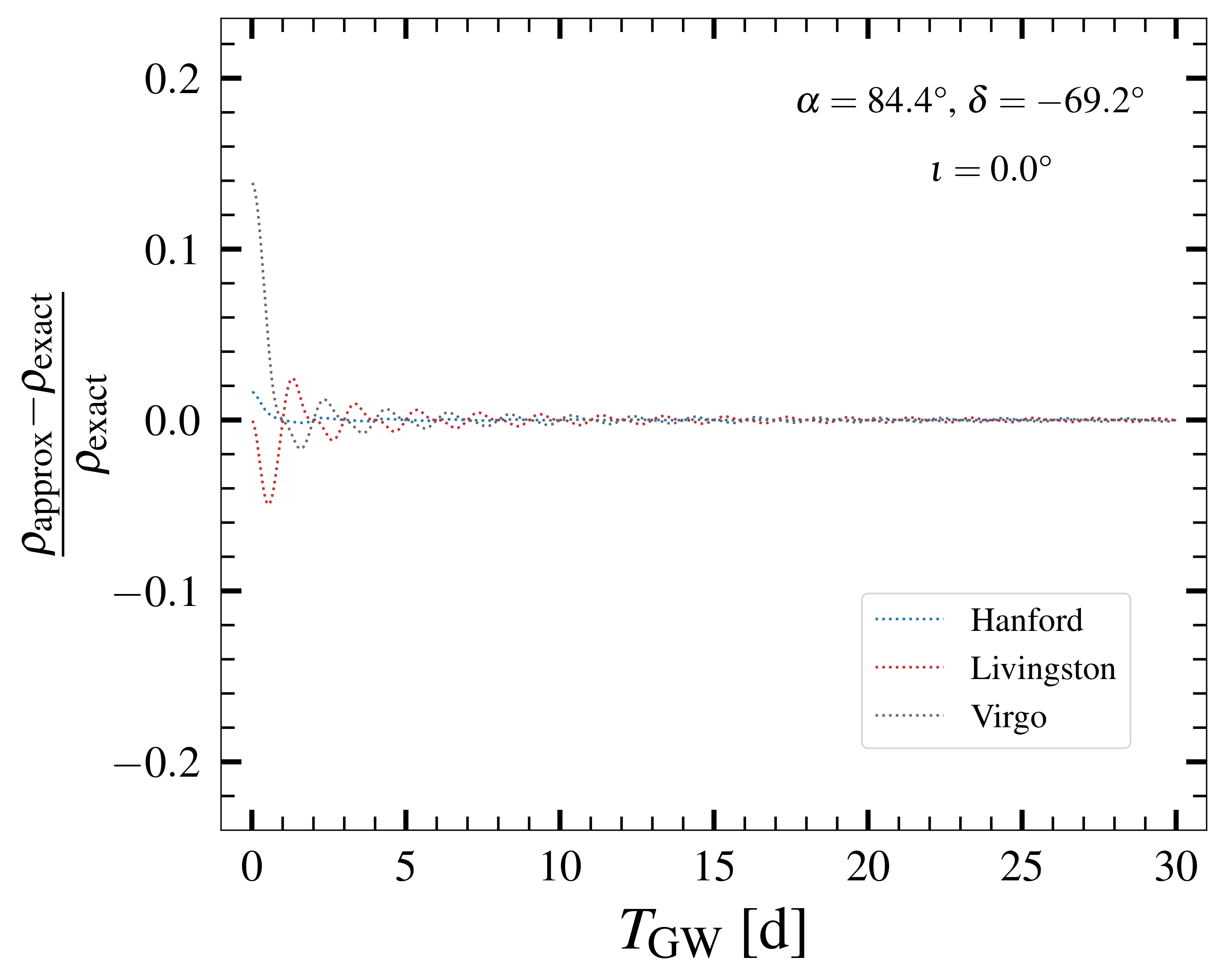}
\caption{\label{figure_J0537}A figure showing the fractional difference between the approximate and exact SNR as a function of GW signal duration $T_\text{GW}$ for PSR~J0537$-$6910. Values of $\iota = 0\text{\textdegree}$, $\alpha = 84.4\text{\textdegree}$ and $\delta = -69.2\text{\textdegree}$ were used, where the latter two parameters were obtained from the ATNF Pulsar Catalogue \citep{manchesteretal2005}. One finds $T_\text{thres} = 0.30~\text{d}$ (for $<10$\% error) for PSR~J0537$-$6910.}
\end{figure}

As it may be important for future observations, we also provide figures similar to Fig.\,\ref{figure_alpha_40_delta_30} but specifically for three pulsars of interest: the Crab pulsar, the Vela pulsar and PSR~J0537$-$6910 \citep{antonopoulouetal2018, ferdmanetal2018, hoetal2020J0537, hoetal2022, fesikPapa2020, LVK2021J0537spindown, LVK2021J0537rmodes}. These are shown in Figs.\,\ref{figure_crab} - \ref{figure_J0537}. One finds that the approximation, eq.\,(\ref{beta_approximation}), gives an error smaller than 10\% so long as $T_\text{GW} \geq 0.36~\text{d}, 0.77~\text{d}, 0.30~\text{d}$ for the Crab pulsar, Vela pulsar and PSR~J0537$-$6910, respectively. As for the values of $A_2$ for these three pulsars, for $\iota = 0\text{\textdegree}$, we have $A_2 = 0.355, 0.418, 0.365$ for the Crab pulsar, $A_2 = 0.456, 0.379, 0.432$ for the Vela pulsar and $A_2 = 0.518, 0.354, 0.498$ for PSR~J0537$-$6910, where each number in each list represents LIGO Hanford, LIGO Livingston and Virgo, respectively. 

\subsection{Connecting to E$_\text{GW}$}
%\subsection{Connecting to $\mathbf{E_\text{GW}}$}

Taking eq.\,(\ref{SNR_beta}) as our expression for the SNR, we can rewrite it in terms of the GW energy, $E_\text{GW}$. To do this, one recalls that the GW luminosity can be written as
\begin{align}
    L_\text{GW} = \frac{(2\pi)^6}{10} \frac{G}{c^5} I^2 f^6 \varepsilon^2~,
\end{align}
and if we impose $h_0 = h_0(t)$, i.e.~the GW amplitude has a time dependence, then $\varepsilon = \varepsilon(t)$ too, by eq.\,(\ref{h_0_epsilon}). One can then integrate the above equation with respect to time to give
\begin{align}
    E_\text{GW} = \frac{(2\pi)^6}{10} \frac{G}{c^5} I^2 f^6 \int_0^{T_\text{GW}}\varepsilon^2(t)\,dt~,
\end{align}
and the integral on the right hand side can be eliminated by using the integral of (the square of) eq.\,(\ref{h_0_epsilon}), resulting in
\begin{align}
    E_\text{GW} = \frac{2\pi^2}{5} \frac{c^3}{G} f^2 d^2 \int_0^{T_\text{GW}}h_0^2(t)\,dt~,
\end{align}
recovering eq.\,(3) of \cite{prixGiampanisMessenger2011}. Rearranging the above for the integral and substituting into our expression for the SNR, eq.\,(\ref{SNR_beta}), gives the final result
\begin{align}
\label{SNR_final_result}
\rho^2 = \frac{5}{2\pi^2} \frac{G}{c^3}\frac{\sin^4\theta}{S_n(f)}\frac{\beta E_\text{GW}}{f^2 d^2}~,
\end{align}
where $\beta = 1$ if assuming isotropic sensitivity, $\beta = 4/25$ if averaging over all sky positions and orientations, or $\beta = A_2$ for a sufficiently long transient CW signal. 

Putting in eq.\,(\ref{E_GW_kappa}) and setting $f = 2\nu$, we find
\begin{align}
\label{SNR_final_result_glitch_specific}
\rho^2 = \frac{5}{2} \frac{G}{c^3}\frac{\sin^4\theta}{S_n(f)}\frac{\beta \kappa I}{d^2} \left(\frac{\Delta \Omega}{\Omega}\right)~,
\end{align}
which describes the SNR for models based on glitches. There are a few interesting features to note. Firstly, there is no other dependency on the frequency $\Omega$ besides as a normalisation factor in the glitch size and in the power spectral density of the detector. Therefore, at fixed $\Delta \Omega/\Omega$, the detectability would be determined by the sensitivity of the detector, given all other parameters are fixed. Secondly, there is no explicit dependence on the spin derivative $\dot\Omega$. It is generally accepted that younger NSs, typically with larger spin-down rates, glitch more often and this should be accounted for if talking about the average chance of detecting a signal from a glitch, however, if we are only concerned about an individual glitch, then the detectability should not depend on $\dot\Omega$.

We will use eq.\,(\ref{SNR_final_result_glitch_specific}) in the subsequent section to calculate the approximate SNRs of all the models covered in Section~\ref{section_gravitational_wave_energy_budget} for all appropriate observed pulsars, using the assumption of $\beta \approx A_2$.

\section{Population study}
\label{section_population_study}

In this section, we combine all the previous information about different GW energy models and how to approximate the SNR and apply it to the observed pulsar population. The first subsection considers pulsars that have previously glitched, as recorded by the JBCA Glitch Catalogue\footnote{\url{http://www.jb.man.ac.uk/~pulsar/glitches/gTable.html}} \citep{espinozaetal2011, basuetal2022} and the ATNF Glitch Table\footnote{\url{https://www.atnf.csiro.au/research/pulsar/psrcat/glitchTbl.html}} \citep{manchesteretal2005}, and the second subsection drops this criterion, focusing on all 3000+ pulsars found in the ATNF Pulsar Catalogue\footnote{\url{https://www.atnf.csiro.au/research/pulsar/psrcat/}} \citep{manchesteretal2005}. 

In this analysis, we use the latest version of the ATNF Pulsar Catalogue, version 2.1.1, which contains 3534 pulsars. Since it is imperative to have the spin frequency and distance to calculate the SNR, we remove all pulsars that do not have these recorded, leaving 3402 pulsars remaining in our dataset.

\subsection{Including glitch data}

As for the glitch catalogues, we first analyse the JBCA Glitch Catalogue. Unprocessed, it contains 672 glitches from 207 pulsars.\footnote{When arranged by J-name, there are 208 pulsars, but this is due to a typo where the J-name of the MJD 55367 glitch of PSR~J1913+0446 has been accidentally entered as ``J1914+0446''. After correcting this, we count 207 unique J-names, not 225 as stated at the top of the catalogue.} We go through a process of cleaning which includes: removing glitches that do not provide $\Delta \nu/\nu$, removing antiglitches, removing pulsars with no distance measurement, and editing some J-names to be consistent with the other databases. The specific details of the cleaning procedure can be found in Appendix~\ref{appendix_cleaning}. After cleaning, we are left with 662 glitches from 202 pulsars.

We do a similar cleaning process for the ATNF Glitch Table, which originally contains 626 glitches from 211 pulsars,\footnote{We have not counted any extra recovery components as a glitch, as they are already attributed to an existing glitch in the dataset. Explicitly, the dataset contains 626 glitches, but 15 of these also have an extra recovery component (so are modelled by 2 exponential recoveries) and there is also a glitch that has 3 recovery components, and another with 4 recovery components. In total, the dataset has 626 glitches with 18 recovery components (making a total of 644 entries) from 211 pulsars.} but this is reduced to 616 glitches from 205 pulsars after cleaning. Again, the details of the cleaning procedure can be found in Appendix~\ref{appendix_cleaning}.

Clearly, there are some differences between the two glitch catalogues but one of the most obvious is the fact that the ATNF Glitch Table provides values of the healing parameter $Q$ alongside the glitches where $Q$ has been measured. This is particularly important for the transient mountain model (Section~\ref{subsubsection_transient_mountain_model}). As such, we create a separate dataset, which is a subset of the ATNF Glitch Table, that contains all the glitches that have $Q$. These glitches can sometimes have more than one recovery component, in which case, each component carries its own $Q$ and recovery timescale $\tau_\text{EM}$. In the end, there are 114 independent glitches, with some of these showing multiple glitch recovery components (18 extra components in total), making a total of 132 entries from 57 pulsars.

Of course, there will be a great deal of overlap between these glitch catalogues. We therefore need to choose which will be our ``base'' catalogue, where we obtain $\Delta \nu/\nu$ from by default, and which we will add data from. In this work, we choose the JBCA Glitch Catalogue as our base and add onto it any missing glitches that come from the ATNF Glitch Table. In Appendix~\ref{appendix_differences_between_glitch_catalogues}, we provide the list of pulsars that are missing from the JBCA Glitch Catalogue but are present in the ATNF Glitch Table (and vice versa). 

So, we will form two datasets that we will base our analysis on:
\begin{itemize}
    \item Dataset A ($\Delta \nu/\nu$): JBCA Glitch Catalogue (662 glitches from 202 pulsars) + Extras from ATNF Glitch Table (24 glitches from 17 pulsars),
    \item Dataset B ($\Delta \nu/\nu$, $Q$): JBCA Glitch Catalogue for $\Delta \nu/\nu$ and ATNF Glitch Table for $Q$ (114 glitches and 18 extra components from 57 pulsars).
\end{itemize}
Dataset A contains $\Delta \nu/\nu$ from 686 glitches from 219 pulsars and Dataset B contains both $\Delta \nu/\nu$ and $Q$ from 114 glitches (132 entries) from 57 pulsars. To both datasets, we append pulsar information such as $\nu$, $d$, $\alpha$ and $\delta$ from the ATNF Pulsar Catalogue. 

As mentioned at the end of Section~\ref{section_gravitational_wave_energy_budget}, we will explicitly consider only three models: the agnostic case ($\kappa = 1$), the vortex unpinning case ($\kappa = (1/2)(\Delta\nu/\nu)(I/I_\text{p} - 1)$) and the transient mountain case ($\kappa = Q$). The other models can be obtained by simple rescaling of the agnostic case. Dataset A will be used to calculate the first two cases and we will use $I_\text{p}/I = 0.01$ for the vortex unpinning model \citep{anderssonetal2012}. Dataset B will be used for the transient mountain model.

\begin{figure}
\centering
\includegraphics[width=\columnwidth]{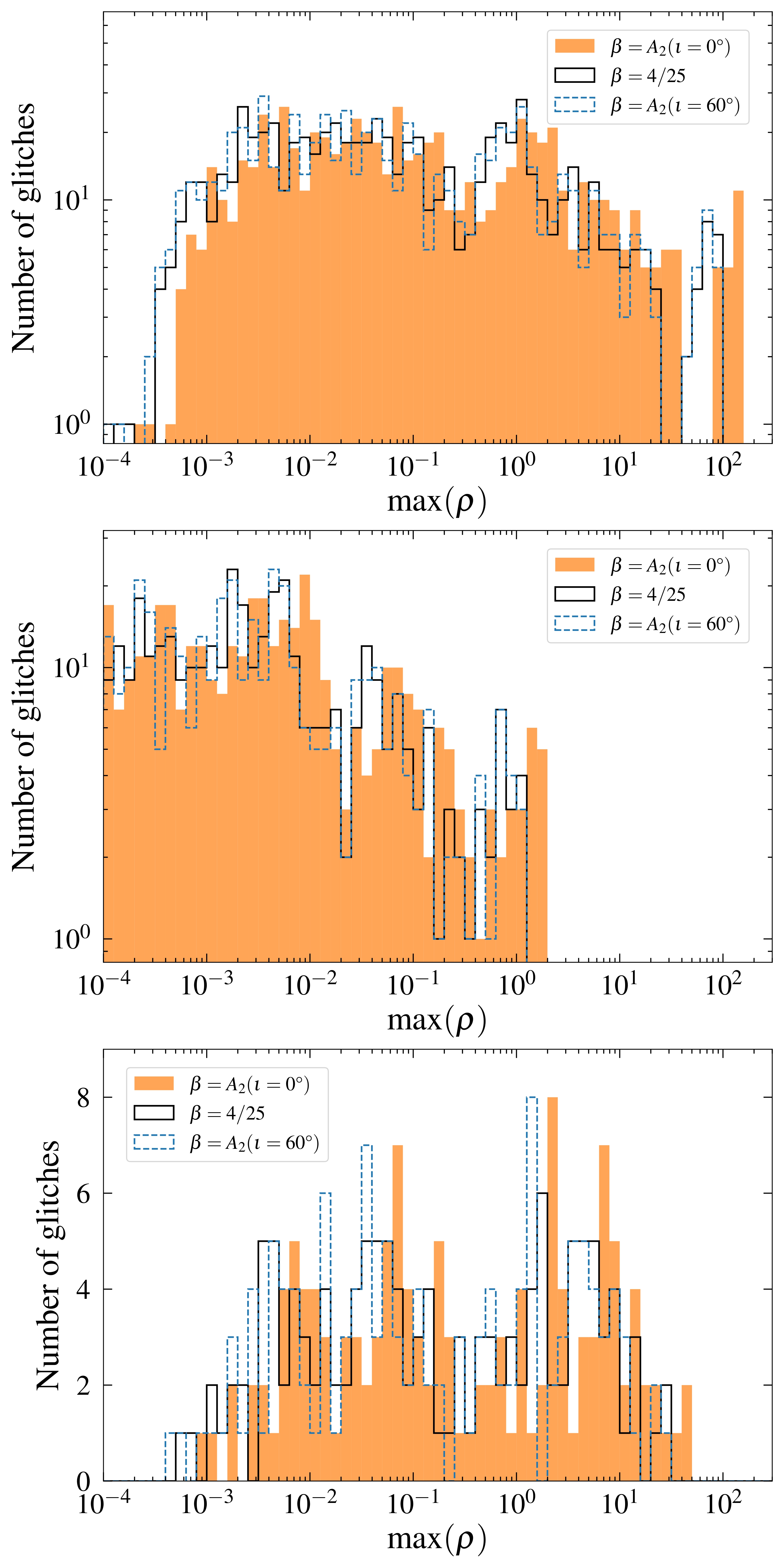}
\caption{\label{figure_SNR_histograms}Histograms of the maximum SNR, referring to the largest SNR from the three GW detectors. Top panel: agnostic models, middle panel: vortex unpinning model, bottom panel: transient mountain model. The orange-filled histograms represent the SNRs that are obtained when $\beta = A_2$ for $\iota = 0\text{\textdegree}$. The solid black line represents the case when $\beta = 4/25$, i.e.~when the source's sky position and orientation are averaged over. The dashed blue line represents the case when $\iota = 60\text{\textdegree}$ and $\psi = 0\text{\textdegree}$, which recovers a similar SNR distribution as the $\beta = 4/25$ case.}
\end{figure}

We present the results of the SNR calculation as histograms in Fig.\,\ref{figure_SNR_histograms}. When using $\beta = A_2$, each GW detector has a slightly different SNR due to the dependence of $A_2$ on $(\lambda, \gamma)$. Therefore, we choose to provide histograms for the largest SNR obtained from the three detectors. The sensitivities that we use are O4 sensitivities and we obtain the sensitivity curves from LIGO Document T2200043-v3.\footnote{\url{https://dcc.ligo.org/LIGO-T2200043/public}}

The orange-filled histograms represent the case where sky positions have been accounted for, by using $\beta = A_2$ in the most optimal orientation of $\iota = 0\text{\textdegree}$. In this case, one sees that the agnostic models (top panel) and the transient mountain model (bottom panel) predict a GW signal would have been detectable ($\rho \gtrsim 10$) if the glitch were to happen during O4. From eq.\,(\ref{SNR_final_result_glitch_specific}), one sees that $\rho \propto \sqrt{\kappa}$, and so to get to the starquake and Ekman pumping m odels, one rescales the agnostic SNR by $1/\sqrt{2}$ and $10^{-4.5} - 10^{-3.5}$, respectively. It is therefore apparent that the SNRs are still detectable for the starquake model, but Ekman pumping would give SNRs no larger than $\sim 0.1$, rendering the associated GWs completely undetectable with O4 detectors. Even with next generation detectors, where the sensitivity is an order of magnitude better, Ekman pumping will not give detectable GWs. Similarly, the vortex unpinning model (middle panel) will not give detectable GWs with O4 detectors, but with next generation detectors, there could potentially be a marginal detection. Not only will the improvement of the sensitivity matter, but the location of the detectors will also affect the detectability.

Also plotted (as solid black lines) are the histograms that would be obtained if one averaged over sky position and orientation, corresponding to $\beta = 4/25$. One finds that using this approximation shifts the SNR distribution towards lower values when compared to the $\beta = A_2(\iota = 0\text{\textdegree})$ case. This shifting to lower SNRs can also be achieved by having a larger value for $\iota$. Setting $\iota = 60\text{\textdegree}$ (and $\psi = 0\text{\textdegree}$) leads to a SNR distribution that is similar to the sky and orientation averaged case (dashed blue line). This is consistent with the $\iota = 60\text{\textdegree}$ curves in Fig.\,\ref{figure_A_2_delta} being close to the value of $4/25$. Therefore, one can conclude that unless $\iota \gtrsim 60\text{\textdegree}$, the $\beta = A_2$ approximation leads to larger SNRs than the sky and orientation averaged case. However, note that the SNR here only captures the $f = 2\nu$ radiation but when $\iota > 0\text{\textdegree}$, the $f = \nu$ radiation will contribute too. We do not worry about this here but it will have the effect of changing the value of $\iota$ that gives an equivalence to the $\beta = 4/25$ case.

We will now look at the most favourable pulsars that we expect to yield a detection. In Tables~\ref{table_naive_results} - \ref{table_mountain_results}, we present the top 15 pulsars that give that largest GW energies and SNRs assuming different energy models using $\beta = A_2(\iota = 0\text{\textdegree})$. For each pulsar, we also present its right ascension, declination, spin frequency, spin frequency derivative, distance, number of observed glitches, glitch size corresponding to the largest SNR, fractional change in the spin-down rate corresponding to the largest SNR, and for the transient mountain model, the healing parameter corresponding to the largest SNR. 

\begin{table*}
\centering
\caption{\label{table_naive_results}A table showing the top 15 pulsars that give the largest SNRs when using agnostic models. Starting with the left-most column, we have: pulsar J-name, right ascension, declination, spin frequency, spin frequency derivative, distance, number of glitches observed, glitch size that gives the largest SNR, fractional change in spin-down rate corresponding to the largest SNR, largest GW energy, largest SNR. Columns 2 - 6 were taken from the ATNF Pulsar Catalogue \citep{manchesteretal2005}, columns 7 - 9 were taken from the JBCA Glitch Catalogue \citep{espinozaetal2011, basuetal2022} and columns 10 - 11 were calculated using eqs.\,(\ref{E_GW_naive}) and (\ref{SNR_final_result}), respectively.}
\begin{tabular}{ccccccccccc}
\hline \hline
\multicolumn{11}{c}{Agnostic models} \\ \hline \\ \vspace{-20pt} \\
Pulsar J-name & $\alpha$ [\textdegree] & $\delta$ [\textdegree] & $\nu$ [Hz] & $\dot\nu$ [Hz s$^{-1}$] & $d$ [kpc] & $N_\text{g}$ & $\Delta \nu / \nu$ [10$^{-9}$] & $\Delta \dot\nu / \dot\nu$ [10$^{-3}$] & $E_\text{GW}$ [erg] & $\text{max}(\rho)$ \\ \hline \\ \vspace{-20pt} \\
J0835$-$4510 & $128.84$ & $-45.18$ & $11.195$ & $-1.57 \times 10^{-11}$ & $0.280$ & 24 & $3100$ & $148$ & $1.53 \times 10^{43}$ & $156.3$ \\
J0940$-$5428 & $145.24$ & $-54.48$ & $11.423$ & $-4.29 \times 10^{-12}$ & $0.377$ & 2 & $1573.9$ & $11$ & $8.11 \times 10^{42}$ & $95.9$ \\
J1952+3252 & $298.24$ & $32.88$ & $25.296$ & $-3.74 \times 10^{-12}$ & $3.000$ & 6 & $1489.9$ & $5.4$ & $3.76 \times 10^{43}$ & $38.5$ \\
J0205+6449 & $31.41$ & $64.83$ & $15.217$ & $-4.49 \times 10^{-11}$ & $3.200$ & 9 & $3800$ & $12$ & $3.47 \times 10^{43}$ & $36.6$ \\
J1813$-$1246 & $273.35$ & $-12.77$ & $20.802$ & $-7.60 \times 10^{-12}$ & $2.635$ & 1 & $1166$ & $6.4$ & $1.99 \times 10^{43}$ & $34.3$ \\
J2229+6114 & $337.27$ & $61.24$ & $19.362$ & $-2.90 \times 10^{-11}$ & $3.000$ & 9 & $1223.6$ & $13$ & $1.81 \times 10^{43}$ & $30.9$ \\
J1105$-$6107 & $166.36$ & $-61.13$ & $15.822$ & $-3.97 \times 10^{-12}$ & $2.360$ & 5 & $971.7$ & $0.1$ & $9.60 \times 10^{42}$ & $26.1$ \\
J0534+2200 & $83.63$ & $22.01$ & $29.947$ & $-3.78 \times 10^{-10}$ & $2.000$ & 30 & $516.37$ & $6.969$ & $1.83 \times 10^{43}$ & $24.0$ \\
J1028$-$5819 & $157.12$ & $-58.32$ & $10.941$ & $-1.93 \times 10^{-12}$ & $1.423$ & 1 & $2296.5$ & $35$ & $1.09 \times 10^{43}$ & $23.9$ \\
J1524$-$5625 & $231.21$ & $-56.42$ & $12.785$ & $-6.37 \times 10^{-12}$ & $3.378$ & 1 & $2977$ & $15.5$ & $1.92 \times 10^{43}$ & $22.5$ \\
J1531$-$5610 & $232.87$ & $-56.18$ & $11.876$ & $-1.95 \times 10^{-12}$ & $2.841$ & 1 & $2637$ & $25$ & $1.47 \times 10^{43}$ & $20.0$ \\
J1112$-$6103 & $168.06$ & $-61.06$ & $15.394$ & $-7.45 \times 10^{-12}$ & $4.464$ & 4 & $1825$ & $4.7$ & $1.71 \times 10^{43}$ & $18.3$ \\
J1617$-$5055 & $244.37$ & $-50.92$ & $14.418$ & $-2.81 \times 10^{-11}$ & $4.743$ & 6 & $2068$ & $13.2$ & $1.70 \times 10^{43}$ & $16.0$ \\
J1420$-$6048 & $215.03$ & $-60.80$ & $14.667$ & $-1.79 \times 10^{-11}$ & $5.632$ & 7 & $2019$ & $6.6$ & $1.71 \times 10^{43}$ & $13.9$ \\
J1809$-$1917 & $272.43$ & $-19.29$ & $12.084$ & $-3.73 \times 10^{-12}$ & $3.268$ & 1 & $1625.1$ & $7.8$ & $9.37 \times 10^{42}$ & $13.6$ \\
\hline \hline
\end{tabular}
\end{table*}

\begin{table*}
\centering
\caption{\label{table_superfluid_results}A table showing the top 15 pulsars that give the largest SNRs when using the vortex unpinning model. Starting with the left-most column, we have: pulsar J-name, right ascension, declination, spin frequency, spin frequency derivative, distance, number of glitches observed, glitch size that gives the largest SNR, fractional change in spin-down rate corresponding to the largest SNR, largest GW energy, largest SNR. Columns 2 - 6 were taken from the ATNF Pulsar Catalogue \citep{manchesteretal2005}, columns 7 - 9 were taken from the JBCA Glitch Catalogue \citep{espinozaetal2011, basuetal2022} and columns 10 - 11 were calculated using eqs.\,(\ref{E_GW_vortex_unpinning}) and (\ref{SNR_final_result}), respectively.}
\begin{tabular}{ccccccccccc}
\hline \hline
\multicolumn{11}{c}{Vortex unpinning model} \\ \hline \\ \vspace{-20pt} \\
Pulsar J-name & $\alpha$ [\textdegree] & $\delta$ [\textdegree] & $\nu$ [Hz] & $\dot\nu$ [Hz s$^{-1}$] & $d$ [kpc] & $N_\text{g}$ & $\Delta \nu / \nu$ [10$^{-9}$] & $\Delta \dot\nu / \dot\nu$ [10$^{-3}$] & $E_\text{GW}$ [erg] & $\text{max}(\rho)$ \\ \hline \\ \vspace{-20pt} \\
J0835$-$4510 & $128.84$ & $-45{.18}$ & $11.195$ & $-1.57 \times 10^{-11}$ & $0.280$ & 24 & $3100$ & $148$ & $2.35 \times 10^{39}$ & $1.94$ \\
J0940$-$5428 & $145.24$ & $-54{.48}$ & $11.423$ & $-4.29 \times 10^{-12}$ & $0.377$ & 2 & $1573.9$ & $11$ & $6.32 \times 10^{38}$ & $0.85$ \\
J0205+6449 & $31.41$ & $64.83$ & $15.217$ & $-4.49 \times 10^{-11}$ & $3.200$ & 9 & $3800$ & $12$ & $6.53 \times 10^{39}$ & $0.50$ \\
J1952+3252 & $298.24$ & $32.88$ & $25.296$ & $-3.74 \times 10^{-12}$ & $3.000$ & 6 & $1489.9$ & $5.4$ & $2.78 \times 10^{39}$ & $0.33$ \\
J1524$-$5625 & $231.21$ & $-56{.42}$ & $12.785$ & $-6.37 \times 10^{-12}$ & $3.378$ & 1 & $2977$ & $15.5$ & $2.83 \times 10^{39}$ & $0.27$ \\
J1813$-$1246 & $273.35$ & $-12{.77}$ & $20.802$ & $-7.60 \times 10^{-12}$ & $2.635$ & 1 & $1166$ & $6.4$ & $1.15 \times 10^{39}$ & $0.26$ \\
J1028$-$5819 & $157.12$ & $-58{.32}$ & $10.941$ & $-1.93 \times 10^{-12}$ & $1.423$ & 1 & $2296.5$ & $35$ & $1.23 \times 10^{39}$ & $0.25$ \\
J2229+6114 & $337.27$ & $61.24$ & $19.362$ & $-2.90 \times 10^{-11}$ & $3.000$ & 9 & $1223.6$ & $13$ & $1.10 \times 10^{39}$ & $0.24$ \\
J1531$-$5610 & $232.87$ & $-56{.18}$ & $11.876$ & $-1.95 \times 10^{-12}$ & $2.841$ & 1 & $2637$ & $25$ & $1.92 \times 10^{39}$ & $0.23$ \\
J1105$-$6107 & $166.36$ & $-61{.13}$ & $15.822$ & $-3.97 \times 10^{-12}$ & $2.360$ & 5 & $971.7$ & $0.1$ & $4.62 \times 10^{38}$ & $0.18$ \\
J1112$-$6103 & $168.06$ & $-61{.06}$ & $15.394$ & $-7.45 \times 10^{-12}$ & $4.464$ & 4 & $1825$ & $4.7$ & $1.54 \times 10^{39}$ & $0.17$ \\
J1617$-$5055 & $244.37$ & $-50{.92}$ & $14.418$ & $-2.81 \times 10^{-11}$ & $4.743$ & 6 & $2068$ & $13.2$ & $1.74 \times 10^{39}$ & $0.16$ \\
J1420$-$6048 & $215.03$ & $-60{.80}$ & $14.667$ & $-1.79 \times 10^{-11}$ & $5.632$ & 7 & $2019$ & $6.6$ & $1.71 \times 10^{39}$ & $0.14$ \\
J1809$-$1917 & $272.43$ & $-19{.29}$ & $12.084$ & $-3.73 \times 10^{-12}$ & $3.268$ & 1 & $1625.1$ & $7.8$ & $7.54 \times 10^{38}$ & $0.12$ \\
J0534+2200 & $83.63$ & $22.01$ & $29.947$ & $-3.78 \times 10^{-10}$ & $2.000$ & 30 & $516.37$ & $6.969$ & $4.67 \times 10^{38}$ & $0.12$ \\
\hline \hline
\end{tabular}
\end{table*}

\begin{table*}
\caption{\label{table_mountain_results}A table showing the top 15 pulsars that give the largest SNRs when using the transient mountain model. Starting with the left-most column, we have: pulsar J-name, right ascension, declination, spin frequency, spin frequency derivative, distance, number of glitches observed, glitch size that gives the largest SNR, fractional change in spin-down rate corresponding to the largest SNR, healing parameter that gives the largest SNR, largest GW energy, largest SNR. Columns 2 - 6 were taken from the ATNF Pulsar Catalogue \citep{manchesteretal2005}, columns 7 - 9 were taken from the JBCA Glitch Catalogue \citep{espinozaetal2011, basuetal2022}, column 10 was taken from the ATNF Glitch Table \citep{manchesteretal2005} and columns 11 - 12 were calculated using eqs.\,(\ref{E_GW_transient_mountain}) and (\ref{SNR_final_result}), respectively.}
\begin{tabular}{cccccccccccc}
\hline \hline
\multicolumn{12}{c}{Transient mountain model} \\ \hline \\ \vspace{-20pt} \\
Pulsar J-name & $\alpha$ [\textdegree] & $\delta$ [\textdegree] & $\nu$ [Hz] & $\dot\nu$ [Hz s$^{-1}$] & $d$ [kpc] & $N_\text{g}$ & $\Delta \nu / \nu$ [10$^{-9}$] & $\Delta \dot\nu / \dot\nu$ [10$^{-3}$] & $Q$ & $E_\text{GW}$ [erg] & $\text{max}(\rho)$ \\ \hline \\ \vspace{-20pt} \\
J0835$-$4510 & $128.84$ & $-45{.18}$ & $11.195$ & $-1.57 \times 10^{-11}$ & $0.280$ & $24$ & $1805.2$ & $77$ & $0.1684$ & $1.50 \times 10^{42}$ & $48.9$ \\
J0205+6449 & $31.41$ & $64.83$ & $15.217$ & $-4.49 \times 10^{-11}$ & $3.200$ & $9$ & $5400$ & $52$ & $0.77$ & $3.80 \times 10^{43}$ & $38.3$ \\
J0534+2200 & $83.63$ & $22.01$ & $29.947$ & $-3.78 \times 10^{-10}$ & $2.000$ & $30$ & $81$ & $3.4$ & $0.894$ & $2.56 \times 10^{42}$ & $9.0$ \\
J0940$-$5428 & $145.24$ & $-54{.48}$ & $11.423$ & $-4.29 \times 10^{-12}$ & $0.377$ & $2$ & $1573.9$ & $11$ & $0.0068$ & $5.51 \times 10^{40}$ & $7.9$ \\
J1617$-$5055 & $244.37$ & $-50{.92}$ & $14.418$ & $-2.81 \times 10^{-11}$ & $4.743$ & $6$ & $334$ & $13$ & $0.975$ & $2.67 \times 10^{42}$ & $6.4$ \\
J1028$-$5819 & $157.12$ & $-58{.32}$ & $10.941$ & $-1.93 \times 10^{-12}$ & $1.423$ & $1$ & $2296.5$ & $35$ & $0.0114$ & $1.24 \times 10^{41}$ & $2.6$ \\
J1112$-$6103 & $168.06$ & $-61{.06}$ & $15.394$ & $-7.45 \times 10^{-12}$ & $4.464$ & $4$ & $1202$ & $7$ & $0.022$ & $2.47 \times 10^{41}$ & $2.2$ \\
J1524$-$5625 & $231.21$ & $-56{.42}$ & $12.785$ & $-6.37 \times 10^{-12}$ & $3.378$ & $1$ & $2977.1$ & $15.6$ & $0.0058$ & $1.11 \times 10^{41}$ & $1.7$ \\
J1531$-$5610 & $232.87$ & $-56{.18}$ & $11.876$ & $-1.95 \times 10^{-12}$ & $2.841$ & $1$ & $2637$ & $25$ & $0.007$ & $1.03 \times 10^{41}$ & $1.7$ \\
J1420$-$6048 & $215.03$ & $-60{.80}$ & $14.667$ & $-1.79 \times 10^{-11}$ & $5.632$ & $7$ & $2019$ & $6.6$ & $0.008$ & $1.37 \times 10^{41}$ & $1.2$ \\
J1809$-$1917 & $272.43$ & $-19{.29}$ & $12.084$ & $-3.73 \times 10^{-12}$ & $3.268$ & $1$ & $1625.1$ & $7.8$ & $0.00602$ & $5.64 \times 10^{40}$ & $1.1$ \\
J1302$-$6350 & $195.70$ & $-63{.84}$ & $20.937$ & $-9.99 \times 10^{-13}$ & $2.632$ & $1$ & $2.3$ & $\dots$ & $0.36$ & $1.43 \times 10^{40}$ & $1.0$ \\
J1837$-$0604 & $279.43$ & $-6{.08}$ & $10.383$ & $-4.84 \times 10^{-12}$ & $4.779$ & $3$ & $1376$ & $8$ & $0.06$ & $3.51 \times 10^{41}$ & $0.9$ \\
J1709$-$4429 & $257.43$ & $-44{.49}$ & $9.760$ & $-8.86 \times 10^{-12}$ & $2.600$ & $5$ & $2872$ & $8$ & $0.0129$ & $1.39 \times 10^{41}$ & $0.8$ \\
J1826$-$1334 & $276.55$ & $-13{.58}$ & $9.853$ & $-7.31 \times 10^{-12}$ & $3.606$ & $7$ & $3581$ & $9.6$ & $0.0066$ & $9.06 \times 10^{40}$ & $0.5$ \\
\hline \hline
\end{tabular}
\end{table*}

One can immediately see results that are consistent with the conclusions drawn from Fig.\,\ref{figure_SNR_histograms}, namely, the agnostic and transient mountain models give detectable signals with O4 but the vortex unpinning model cannot. There are 15 pulsars that have $\text{max}(\rho) \geq 10$ using the agnostic model, none for the vortex unpinning model and two pulsars for the transient mountain model, PSR~J0835$-$4510 (Vela) and PSR~J0205+6449.

It is also worth looking at individual pulsars too, particularly the ones that have already been targeted by LVK. These are the Crab pulsar (PSR~J0534+2200), the Vela pulsar (PSR~J0835$-$4510) and the ``big glitcher'' (PSR~J0537$-$6910), of which the latter has been suggested to also emit GWs via $r$-modes \citep[e.g.][]{fesikPapa2020, LVK2021J0537rmodes}. Firstly, we track the Crab pulsar. It ranks 8th, 15th and 3rd in the agnostic, vortex unpinning and transient mountain models, respectively. The drop in ranking for the vortex unpinning model is due to the relatively small glitch sizes that the Crab exhibits, as the GW energy in this model is quadratic in the glitch size rather than linear. The Crab ranks relatively highly for the transient mountain model due to its large $Q$. Moving onto Vela, it remains the highest priority target for all models primarily due to its large glitches and close distance, holding also true for PSR~J0940$-$5428. 

Finally, we found that PSR~J0537$-$6910 is not included in the top 15 for any of our models due primarily to its large distance from us ($d = 49.7~\text{kpc}$). This is despite it having the largest $E_\text{GW} \sim 1\times10^{44}~\text{erg}$ in our dataset, which comes from it having the 3rd largest spin frequency, only to be beaten by millisecond pulsars PSR~J1824$-$2452A \citep[$\nu \approx 327.4~\text{Hz}$;][]{lyneetal1987, reardonetal2021} and PSR~J0613$-$0200 \citep[$\nu \approx 326.6~\text{Hz}$;][]{lorimeretal1995, reardonetal2021}. However, these millisecond pulsars exhibited the smallest glitches of our dataset, of sizes $\Delta \nu / \nu = 8\times10^{-12}$ \citep{cognardBacker2004, espinozaetal2011} and $\Delta \nu / \nu = 2.5\times10^{-12}$ \citep{mckeeetal2016}, respectively, meaning their GW energies are only $E_\text{GW}\sim10^{40}~\text{erg}$. One major reason that PSR~J0537$-$6910 is interesting is that it glitches on average every $\sim 100~\text{d}$ but with $\text{max}(\rho) = 2.7$ (using the agnostic models), any GWs are unlikely to be detected, even if we coherently stack multiple signals together. A detectable signal would require stacking more than 15 events which would take over 4 years. It is clear that glitches from PSR~J0537$-$6910 will only start being probed by GWs with next generation detectors, a conclusion in agreement with \cite{yim2022}.

\subsection{Ranking all pulsars}
% Fixed E_GW, evaluate SNR for differing E_GW
% Maybe similar to Lopez et al. (2022)?
In this section, we look at all pulsars from the ATNF Pulsar Catalogue (with known $\nu$ and $d$) and assume that at some point in the future, it will glitch, even if it has not glitched before. As a result, we will be able to rank which pulsars would give the most significant signal if a glitch were to occur.

To do this, we fix $E_\text{GW}$ (and $\theta$), and we calculate the parameter $\chi$ for each pulsar, defined as
\begin{align}
\chi \equiv \frac{1}{fd} \sqrt{\frac{5}{2\pi^2} \frac{G}{c^3} \frac{A_2}{S_n(f)}} ~,
\end{align}
such that the SNR can be calculated straightforwardly as
\begin{align}
\label{rho_chi}
\rho = \chi \sin^2\theta \sqrt{E_\text{GW}}~,
\end{align}
which comes from eq.\,(\ref{SNR_final_result}). For $\theta = \pi/2$, $\chi$ is simply the constant of proportionality between $\rho$ and $\sqrt{E_\text{GW}}$. Note that $\chi$ is a property that is unique for each pulsar and GW detector combination, as it depends on the source's sky location and orientation, as well as the detector's location, orientation and sensitivity.

For definiteness, we set $f = 2\nu$ and $\iota = 0\text{\textdegree}$, and we calculate $\chi$ for every pulsar in the ATNF catalogue. A given pulsar will have a different $\chi$ for each GW detector but we only report the largest value. The ranked results of the top 100 pulsars are presented in Table~\ref{table_chi}.

From here, we see that the Vela pulsar is the pulsar with the 8th highest $\chi$, with a value of $\chi \approx 3.99\times10^{-20}~\text{erg}^{-\frac{1}{2}}$. We also report the values for the Crab pulsar and PSR~J0537$-$6910 which are $\chi \approx 5.61\times10^{-21}~\text{erg}^{-\frac{1}{2}}$ and $\chi \approx 2.60\times10^{-22}~\text{erg}^{-\frac{1}{2}}$, respectively. We can verify our earlier calculations, for example, for Vela, by substituting $E_\text{GW} = 1.53\times10^{43}~\text{erg}$, $2.35\times10^{39}~\text{erg}$ and $1.50\times10^{42}~\text{erg}$ into eq.\,(\ref{rho_chi}) for the agnostic, vortex unpinning and transient mountain models and we recover the maximum SNRs in \mbox{Tables~\ref{table_naive_results} - \ref{table_mountain_results}} to within some rounding error. This makes $\chi$ a powerful tool to quickly estimate the SNR if $E_\text{GW}$ is provided. Additionally, the calculation no longer needs to be restricted to glitches but could also be used for other transient phenomena such as magnetar bursts or fast radio bursts, if one assumes $f = 2\nu$ radiation is emitted. %One could also use Table~\ref{table_chi} as a ranked list of targets for conventional CWs too if $h_0^2T_\text{GW}$ is assumed to be fixed between pulsars.

\begin{table*}
\caption{\label{table_chi}A table of the top 100 pulsars that give the largest $\chi$.}
\begin{tabular}{cccccp{10pt}ccccc}
\hline \hline
Rank & Pulsar J-name & $\nu$ [Hz] & $d$ [kpc] & $\chi$ [erg$^{-\frac{1}{2}}$] & & Rank & Pulsar J-name & $\nu$ [Hz] & $d$ [kpc] & $\chi$ [erg$^{-\frac{1}{2}}$] \\ \hline \\ \vspace{-20pt} \\
$1$ & J1658+3630 & $30.277$ & $0.224$ & $7.61 \times 10^{-20}$ & &$51$ & J2124$-$3358 & $202.794$ & $0.410$ & $1.02 \times 10^{-20}$ \\
$2$ & J1154$-$19 & $90.090$ & $0.121$ & $7.37 \times 10^{-20}$ & &$52$ & J1821+0155 & $29.602$ & $1.723$ & $1.01 \times 10^{-20}$ \\
$3$ & J2222$-$0137 & $30.471$ & $0.268$ & $6.59 \times 10^{-20}$ & &$53$ & J1405$-$4656 & $131.541$ & $0.669$ & $1.01 \times 10^{-20}$ \\
$4$ & J1745$-$0952 & $51.609$ & $0.226$ & $5.92 \times 10^{-20}$ & &$54$ & J1122$-$3546 & $127.582$ & $0.668$ & $9.91 \times 10^{-21}$ \\
$5$ & J1326+33 & $24.096$ & $0.344$ & $5.49 \times 10^{-20}$ & &$55$ & J1455$-$3330 & $125.200$ & $0.684$ & $9.77 \times 10^{-21}$ \\
$6$ & J0711$-$6830 & $182.117$ & $0.106$ & $4.93 \times 10^{-20}$ & &$56$ & J0900$-$3144 & $90.012$ & $0.900$ & $9.62 \times 10^{-21}$ \\
$7$ & J1045$-$0436 & $41.584$ & $0.329$ & $4.64 \times 10^{-20}$ & &$57$ & J1804$-$2717 & $107.032$ & $0.805$ & $9.52 \times 10^{-21}$ \\
$8$ & J0835$-$4510 & $11.195$ & $0.280$ & $3.99 \times 10^{-20}$ & &$58$ & J0348+0432 & $25.561$ & $2.100$ & $9.22 \times 10^{-21}$ \\
$9$ & J0453+1559 & $21.843$ & $0.522$ & $3.77 \times 10^{-20}$ & &$59$ & J1932+1756 & $23.906$ & $2.116$ & $9.18 \times 10^{-21}$ \\
$10$ & J0609+2130 & $17.954$ & $0.574$ & $3.38 \times 10^{-20}$ & &$60$ & J0605+3757 & $366.575$ & $0.215$ & $9.00 \times 10^{-21}$ \\
$11$ & J0940$-$5428 & $11.423$ & $0.377$ & $3.37 \times 10^{-20}$ & &$61$ & J0034+69 & $27.171$ & $2.283$ & $8.80 \times 10^{-21}$ \\
$12$ & J0437$-$4715 & $173.688$ & $0.157$ & $3.33 \times 10^{-20}$ & &$62$ & J1744$-$1134 & $245.426$ & $0.395$ & $8.50 \times 10^{-21}$ \\
$13$ & J1439$-$5501 & $34.922$ & $0.655$ & $2.73 \times 10^{-20}$ & &$63$ & J1757$-$5322 & $112.740$ & $0.945$ & $8.43 \times 10^{-21}$ \\
$14$ & J1518+4904 & $24.429$ & $0.806$ & $2.51 \times 10^{-20}$ & &$64$ & J1105$-$6107 & $15.822$ & $2.360$ & $8.43 \times 10^{-21}$ \\
$15$ & J0921$-$5202 & $103.308$ & $0.355$ & $2.41 \times 10^{-20}$ & &$65$ & J1727$-$2946 & $36.923$ & $1.881$ & $8.38 \times 10^{-21}$ \\
$16$ & J1528$-$3146 & $16.441$ & $0.774$ & $2.34 \times 10^{-20}$ & &$66$ & J1400$-$1431 & $324.230$ & $0.278$ & $8.36 \times 10^{-21}$ \\
$17$ & J1756$-$2251 & $35.135$ & $0.730$ & $2.25 \times 10^{-20}$ & &$67$ & J1302$-$6350 & $20.937$ & $2.632$ & $8.36 \times 10^{-21}$ \\
$18$ & J1434+7257 & $23.957$ & $0.974$ & $2.21 \times 10^{-20}$ & &$68$ & J2016+1948 & $15.399$ & $2.162$ & $8.34 \times 10^{-21}$ \\
$19$ & J1157$-$5112 & $22.941$ & $0.961$ & $2.18 \times 10^{-20}$ & &$69$ & J1806+2819 & $66.297$ & $1.330$ & $8.23 \times 10^{-21}$ \\
$20$ & J1829+2456 & $24.384$ & $0.909$ & $2.09 \times 10^{-20}$ & &$70$ & J1906+0757g & $17.486$ & $2.465$ & $7.92 \times 10^{-21}$ \\
$21$ & J1537+1155 & $26.382$ & $0.935$ & $2.04 \times 10^{-20}$ & &$71$ & J1300+1240 & $160.810$ & $0.709$ & $7.70 \times 10^{-21}$ \\
$22$ & J1045$-$4509 & $133.793$ & $0.340$ & $1.97 \times 10^{-20}$ & &$72$ & J1813$-$1246 & $20.802$ & $2.635$ & $7.69 \times 10^{-21}$ \\
$23$ & J2145$-$0750 & $62.296$ & $0.625$ & $1.88 \times 10^{-20}$ & &$73$ & J1911+1252g & $36.711$ & $2.167$ & $7.47 \times 10^{-21}$ \\
$24$ & J1753$-$28 & $11.647$ & $0.693$ & $1.85 \times 10^{-20}$ & &$74$ & J1402+1306 & $169.710$ & $0.696$ & $7.41 \times 10^{-21}$ \\
$25$ & J1517$-$32 & $15.527$ & $0.966$ & $1.83 \times 10^{-20}$ & &$75$ & J2229+6114 & $19.362$ & $3.000$ & $7.27 \times 10^{-21}$ \\
$26$ & J1729$-$2117 & $15.085$ & $0.969$ & $1.80 \times 10^{-20}$ & &$76$ & J1028$-$5819 & $10.941$ & $1.423$ & $7.25 \times 10^{-21}$ \\
$27$ & J1737$-$0811 & $239.520$ & $0.206$ & $1.72 \times 10^{-20}$ & &$77$ & J1454$-$5846 & $22.100$ & $2.988$ & $7.19 \times 10^{-21}$ \\
$28$ & J1022+1001 & $60.779$ & $0.725$ & $1.65 \times 10^{-20}$ & &$78$ & J1231$-$1411 & $271.453$ & $0.420$ & $7.10 \times 10^{-21}$ \\
$29$ & J1411+2551 & $16.012$ & $1.131$ & $1.64 \times 10^{-20}$ & &$79$ & J0827+53 & $74.074$ & $1.571$ & $7.02 \times 10^{-21}$ \\
$30$ & J1255$-$46 & $19.231$ & $1.359$ & $1.52 \times 10^{-20}$ & &$80$ & J1012+5307 & $190.268$ & $0.700$ & $6.95 \times 10^{-21}$ \\
$31$ & J1719$-$1438 & $172.707$ & $0.336$ & $1.50 \times 10^{-20}$ & &$81$ & J0307+7443 & $316.848$ & $0.386$ & $6.89 \times 10^{-21}$ \\
$32$ & J0709+0458 & $29.045$ & $1.203$ & $1.45 \times 10^{-20}$ & &$82$ & J0645+5158 & $112.950$ & $1.220$ & $6.49 \times 10^{-21}$ \\
$33$ & J0214+5222 & $40.691$ & $1.161$ & $1.40 \times 10^{-20}$ & &$83$ & J1012$-$4235 & $322.477$ & $0.372$ & $6.44 \times 10^{-21}$ \\
$34$ & J1018$-$1523 & $12.026$ & $1.068$ & $1.40 \times 10^{-20}$ & &$84$ & J1952+3252 & $25.296$ & $3.000$ & $6.27 \times 10^{-21}$ \\
$35$ & J1709$-$0333 & $283.800$ & $0.213$ & $1.32 \times 10^{-20}$ & &$85$ & J0205+6449 & $15.217$ & $3.200$ & $6.20 \times 10^{-21}$ \\
$36$ & J1908+1035g & $93.545$ & $0.671$ & $1.31 \times 10^{-20}$ & &$86$ & J1516$-$43 & $27.761$ & $3.037$ & $6.19 \times 10^{-21}$ \\
$37$ & J1420$-$5625 & $29.311$ & $1.328$ & $1.30 \times 10^{-20}$ & &$87$ & J1753$-$1914 & $15.884$ & $2.915$ & $6.14 \times 10^{-21}$ \\
$38$ & J1925+1636g & $20.117$ & $1.536$ & $1.29 \times 10^{-20}$ & &$88$ & J2042+0246 & $220.569$ & $0.640$ & $6.07 \times 10^{-21}$ \\
$39$ & J0030+0451 & $205.531$ & $0.329$ & $1.29 \times 10^{-20}$ & &$89$ & J1629$-$6902 & $166.650$ & $0.962$ & $6.01 \times 10^{-21}$ \\
$40$ & J0737$-$3039A & $44.054$ & $1.100$ & $1.28 \times 10^{-20}$ & &$90$ & J1918$-$0642 & $130.790$ & $1.111$ & $5.93 \times 10^{-21}$ \\
$41$ & J2235+1506 & $16.732$ & $1.540$ & $1.27 \times 10^{-20}$ & &$91$ & J1936+217 & $31.664$ & $2.968$ & $5.74 \times 10^{-21}$ \\
$42$ & J0621+1002 & $34.657$ & $1.351$ & $1.24 \times 10^{-20}$ & &$92$ & J2023+2853g & $88.261$ & $1.573$ & $5.65 \times 10^{-21}$ \\
$43$ & J1802$-$2124 & $79.066$ & $0.800$ & $1.23 \times 10^{-20}$ & &$93$ & J0534+2200 & $29.947$ & $2.000$ & $5.61 \times 10^{-21}$ \\
$44$ & J0407+1607 & $38.908$ & $1.337$ & $1.17 \times 10^{-20}$ & &$94$ & J0358+6627 & $10.928$ & $1.884$ & $5.57 \times 10^{-21}$ \\
$45$ & J2017$-$0414 & $24.631$ & $1.771$ & $1.11 \times 10^{-20}$ & &$95$ & J1750$-$2536 & $28.778$ & $3.227$ & $5.44 \times 10^{-21}$ \\
$46$ & J1730$-$2304 & $123.110$ & $0.620$ & $1.11 \times 10^{-20}$ & &$96$ & J1936+18 & $17.139$ & $3.596$ & $5.43 \times 10^{-21}$ \\
$47$ & J2018$-$0414 & $24.623$ & $1.806$ & $1.09 \times 10^{-20}$ & &$97$ & J2302+4442 & $192.592$ & $0.863$ & $5.39 \times 10^{-21}$ \\
$48$ & J0405+3347 & $15.636$ & $1.655$ & $1.08 \times 10^{-20}$ & &$98$ & J1643$-$1224 & $216.373$ & $0.740$ & $5.36 \times 10^{-21}$ \\
$49$ & J0509+3801 & $13.065$ & $1.562$ & $1.06 \times 10^{-20}$ & &$99$ & J1646$-$2142 & $170.849$ & $0.965$ & $5.23 \times 10^{-21}$ \\
$50$ & J1056$-$7117 & $38.009$ & $1.689$ & $1.02 \times 10^{-20}$ & &$100$ & J1531$-$5610 & $11.876$ & $2.841$ & $5.22 \times 10^{-21}$ \\
\hline \hline
\end{tabular}
\end{table*}

% Fixed glitch size
We can also do the same by fixing $\Delta\Omega/\Omega$ and in a similar way, we define
\begin{align}
\Upsilon \equiv \frac{1}{d} \sqrt{\frac{5}{2} \frac{G}{c^3} \frac{A_2 I}{S_n(f)}}~, 
\end{align}
such that the SNR can be calculated as
\begin{align}
\label{rho_upsilon}
\rho = \Upsilon \sin^2\theta \sqrt{\kappa} \sqrt{\frac{\Delta\Omega}{\Omega}}~.
\end{align}

In Table~\ref{table_upsilon}, we present the top 100 pulsars with the highest $\Upsilon$, assuming $f = 2\nu$ and $\iota = 0\text{\textdegree}$. Note that none of these top 100 pulsars are known glitchers and in fact, the first glitcher in the ranked list comes in 165th place, which is the Vela pulsar with $\Upsilon \approx 8.86\times10^{4}$. The Crab pulsar has $\Upsilon \approx 3.34\times10^{4}$ and PSR~J0537$-$6910 has $\Upsilon \approx 3.21\times10^{3}$. Better electromagnetic timing of these top 100 pulsars would unveil better constraints on the glitch mechanism if a glitch were to occur and it was coincident with observing GW detectors.

\begin{table*}
\caption{\label{table_upsilon}A table of the top 100 pulsars that give the largest $\Upsilon$. }
\begin{tabular}{cccccp{10pt}ccccc}
\hline \hline
Rank & Pulsar J-name & $\nu$ [Hz] & $d$ [kpc] & $\Upsilon$ & & Rank & Pulsar J-name & $\nu$ [Hz] & $d$ [kpc] & $\Upsilon$ \\ \hline \\ \vspace{-20pt} \\
$1$ & J0711$-$6830 & $182.117$ & $0.106$ & $1.79 \times 10^{6}$ & &$51$ & J0653+4706 & $210.167$ & $0.913$ & $1.95 \times 10^{5}$ \\
$2$ & J1154$-$19 & $90.090$ & $0.121$ & $1.32 \times 10^{6}$ & &$52$ & J0613$-$0200 & $326.601$ & $0.780$ & $1.94 \times 10^{5}$ \\
$3$ & J0437$-$4715 & $173.688$ & $0.157$ & $1.15 \times 10^{6}$ & &$53$ & J1802$-$2124 & $79.066$ & $0.800$ & $1.92 \times 10^{5}$ \\
$4$ & J1737$-$0811 & $239.520$ & $0.206$ & $8.16 \times 10^{5}$ & &$54$ & J1439$-$5501 & $34.922$ & $0.655$ & $1.89 \times 10^{5}$ \\
$5$ & J1709$-$0333 & $283.800$ & $0.213$ & $7.42 \times 10^{5}$ & &$55$ & J1757$-$5322 & $112.740$ & $0.945$ & $1.89 \times 10^{5}$ \\
$6$ & J0605+3757 & $366.575$ & $0.215$ & $6.55 \times 10^{5}$ & &$56$ & J1903$-$7051 & $277.940$ & $0.930$ & $1.86 \times 10^{5}$ \\
$7$ & J1745$-$0952 & $51.609$ & $0.226$ & $6.07 \times 10^{5}$ & &$57$ & J1036$-$8317 & $293.427$ & $0.934$ & $1.84 \times 10^{5}$ \\
$8$ & J1400$-$1431 & $324.230$ & $0.278$ & $5.39 \times 10^{5}$ & &$58$ & J1514$-$4946 & $278.603$ & $0.908$ & $1.81 \times 10^{5}$ \\
$9$ & J0030+0451 & $205.531$ & $0.329$ & $5.27 \times 10^{5}$ & &$59$ & J1120$-$3618 & $179.953$ & $0.954$ & $1.78 \times 10^{5}$ \\
$10$ & J1045$-$4509 & $133.793$ & $0.340$ & $5.23 \times 10^{5}$ & &$60$ & J1646$-$2142 & $170.849$ & $0.965$ & $1.78 \times 10^{5}$ \\
$11$ & J1719$-$1438 & $172.707$ & $0.336$ & $5.16 \times 10^{5}$ & &$61$ & J1847+01 & $288.761$ & $0.893$ & $1.74 \times 10^{5}$ \\
$12$ & J0921$-$5202 & $103.308$ & $0.355$ & $4.94 \times 10^{5}$ & &$62$ & J0900$-$3144 & $90.012$ & $0.900$ & $1.72 \times 10^{5}$ \\
$13$ & J1658+3630 & $30.277$ & $0.224$ & $4.58 \times 10^{5}$ & &$63$ & J1933$-$6211 & $282.212$ & $1.000$ & $1.69 \times 10^{5}$ \\
$14$ & J0307+7443 & $316.848$ & $0.386$ & $4.34 \times 10^{5}$ & &$64$ & J2355+0051 & $268.890$ & $0.958$ & $1.68 \times 10^{5}$ \\
$15$ & J1744$-$1134 & $245.426$ & $0.395$ & $4.14 \times 10^{5}$ & &$65$ & J2322+2057 & $207.968$ & $1.011$ & $1.68 \times 10^{5}$ \\
$16$ & J1012$-$4235 & $322.477$ & $0.372$ & $4.13 \times 10^{5}$ & &$66$ & J0453+1559 & $21.843$ & $0.522$ & $1.64 \times 10^{5}$ \\
$17$ & J2124$-$3358 & $202.794$ & $0.410$ & $4.09 \times 10^{5}$ & &$67$ & J2234+0611 & $279.597$ & $0.971$ & $1.62 \times 10^{5}$ \\
$18$ & J2222$-$0137 & $30.471$ & $0.268$ & $3.99 \times 10^{5}$ & &$68$ & J1536$-$4948 & $324.684$ & $0.978$ & $1.61 \times 10^{5}$ \\
$19$ & J0646$-$54 & $398.406$ & $0.367$ & $3.93 \times 10^{5}$ & &$69$ & J1658$-$5324 & $409.954$ & $0.880$ & $1.60 \times 10^{5}$ \\
$20$ & J1045$-$0436 & $41.584$ & $0.329$ & $3.84 \times 10^{5}$ & &$70$ & J1828+0625 & $275.667$ & $1.000$ & $1.60 \times 10^{5}$ \\
$21$ & J1231$-$1411 & $271.453$ & $0.420$ & $3.83 \times 10^{5}$ & &$71$ & J0447+2447 & $333.858$ & $0.925$ & $1.58 \times 10^{5}$ \\
$22$ & J1710+4923 & $310.537$ & $0.506$ & $3.16 \times 10^{5}$ & &$72$ & J1756$-$2251 & $35.135$ & $0.730$ & $1.57 \times 10^{5}$ \\
$23$ & J0742+4110 & $318.559$ & $0.523$ & $2.93 \times 10^{5}$ & &$73$ & J1216$-$6410 & $282.536$ & $1.098$ & $1.55 \times 10^{5}$ \\
$24$ & J1730$-$2304 & $123.110$ & $0.620$ & $2.71 \times 10^{5}$ & &$74$ & J1918$-$0642 & $130.790$ & $1.111$ & $1.54 \times 10^{5}$ \\
$25$ & J2042+0246 & $220.569$ & $0.640$ & $2.66 \times 10^{5}$ & &$75$ & J1905+0400 & $264.242$ & $1.064$ & $1.53 \times 10^{5}$ \\
$26$ & J1405$-$4656 & $131.541$ & $0.669$ & $2.63 \times 10^{5}$ & &$76$ & J1625$-$0021 & $352.906$ & $0.951$ & $1.53 \times 10^{5}$ \\
$27$ & J1012+5307 & $190.268$ & $0.700$ & $2.63 \times 10^{5}$ & &$77$ & J1125$-$6014 & $380.173$ & $0.988$ & $1.52 \times 10^{5}$ \\
$28$ & J1326+33 & $24.096$ & $0.344$ & $2.63 \times 10^{5}$ & &$78$ & J0101$-$6422 & $388.628$ & $1.001$ & $1.50 \times 10^{5}$ \\
$29$ & J1036$-$4353 & $595.200$ & $0.408$ & $2.56 \times 10^{5}$ & &$79$ & J1911$-$1114 & $275.805$ & $1.069$ & $1.49 \times 10^{5}$ \\
$30$ & J1122$-$3546 & $127.582$ & $0.668$ & $2.51 \times 10^{5}$ & &$80$ & J0645+5158 & $112.950$ & $1.220$ & $1.46 \times 10^{5}$ \\
$31$ & J1402+1306 & $169.710$ & $0.696$ & $2.50 \times 10^{5}$ & &$81$ & J1751$-$2857 & $255.436$ & $1.087$ & $1.46 \times 10^{5}$ \\
$32$ & J1300+1240 & $160.810$ & $0.709$ & $2.46 \times 10^{5}$ & &$82$ & J1857+0943 & $186.494$ & $1.200$ & $1.44 \times 10^{5}$ \\
$33$ & J2214+3000 & $320.592$ & $0.600$ & $2.45 \times 10^{5}$ & &$83$ & J1801$-$1417 & $275.855$ & $1.105$ & $1.43 \times 10^{5}$ \\
$34$ & J1455$-$3330 & $125.200$ & $0.684$ & $2.43 \times 10^{5}$ & &$84$ & J1207$-$5050 & $206.494$ & $1.267$ & $1.43 \times 10^{5}$ \\
$35$ & J1908+1035g & $93.545$ & $0.671$ & $2.43 \times 10^{5}$ & &$85$ & J1944+0907 & $192.857$ & $1.218$ & $1.43 \times 10^{5}$ \\
$36$ & J0614$-$3329 & $317.594$ & $0.630$ & $2.34 \times 10^{5}$ & &$86$ & J1622$-$0315 & $260.049$ & $1.142$ & $1.43 \times 10^{5}$ \\
$37$ & J2145$-$0750 & $62.296$ & $0.625$ & $2.33 \times 10^{5}$ & &$87$ & J1024$-$0719 & $193.716$ & $1.220$ & $1.43 \times 10^{5}$ \\
$38$ & J1643$-$1224 & $216.373$ & $0.740$ & $2.30 \times 10^{5}$ & &$88$ & J0751+1807 & $287.458$ & $1.100$ & $1.41 \times 10^{5}$ \\
$39$ & J0636$-$3044 & $253.437$ & $0.679$ & $2.29 \times 10^{5}$ & &$89$ & J1858$-$2216 & $419.461$ & $0.921$ & $1.40 \times 10^{5}$ \\
$40$ & J2234+0944 & $275.708$ & $0.714$ & $2.23 \times 10^{5}$ & &$90$ & J0732+2314 & $244.491$ & $1.151$ & $1.40 \times 10^{5}$ \\
$41$ & J0125$-$2327 & $272.081$ & $0.714$ & $2.21 \times 10^{5}$ & &$91$ & J0740+6620 & $346.532$ & $1.150$ & $1.39 \times 10^{5}$ \\
$42$ & J0636+5128 & $348.559$ & $0.714$ & $2.15 \times 10^{5}$ & &$92$ & J2019+2425 & $254.160$ & $1.163$ & $1.38 \times 10^{5}$ \\
$43$ & J1614$-$2230 & $317.379$ & $0.700$ & $2.13 \times 10^{5}$ & &$93$ & J1653$-$0158 & $508.212$ & $0.840$ & $1.37 \times 10^{5}$ \\
$44$ & J2302+4442 & $192.592$ & $0.863$ & $2.06 \times 10^{5}$ & &$94$ & J1453+1902 & $172.643$ & $1.269$ & $1.36 \times 10^{5}$ \\
$45$ & J1125+7819 & $238.004$ & $0.903$ & $2.05 \times 10^{5}$ & &$95$ & J2339$-$0533 & $346.713$ & $1.100$ & $1.33 \times 10^{5}$ \\
$46$ & J1804$-$2717 & $107.032$ & $0.805$ & $2.02 \times 10^{5}$ & &$96$ & J1852$-$1310 & $231.768$ & $1.266$ & $1.32 \times 10^{5}$ \\
$47$ & J1022+1001 & $60.779$ & $0.725$ & $2.00 \times 10^{5}$ & &$97$ & J1923+2515 & $263.981$ & $1.201$ & $1.32 \times 10^{5}$ \\
$48$ & J1629$-$6902 & $166.650$ & $0.962$ & $1.99 \times 10^{5}$ & &$98$ & J1124$-$3653 & $415.011$ & $0.987$ & $1.32 \times 10^{5}$ \\
$49$ & J0509+0856 & $246.558$ & $0.817$ & $1.99 \times 10^{5}$ & &$99$ & J1551$-$0658 & $141.044$ & $1.325$ & $1.32 \times 10^{5}$ \\
$50$ & J0312$-$0921 & $269.954$ & $0.817$ & $1.97 \times 10^{5}$ & &$100$ & J0318+0253 & $192.684$ & $1.328$ & $1.31 \times 10^{5}$ \\
\hline \hline
\end{tabular}
\end{table*}

\section{April 2024 glitch in Vela}
\label{section_glitch_in_vela}
In this section, we apply our analysis to the recent glitch that was observed from the Vela pulsar \citep{zubietaetal2024, campbell-wilsonFlynnBateman2024, groveretal2024, palfreyman2024, wangetal2024}. Although the glitch was first announced by \cite{zubietaetal2024}, we use the glitch parameters reported by \cite{palfreyman2024} as his data and analysis was able to narrow the glitch time to within a $\sim7$ second window. The glitch occurred on 29th April 2024 between 20:52:11.4 and 20:52:18.1 and the glitch size was $\Delta \Omega / \Omega \approx 2.4 \times 10^{-6}$. 

Fortunately, LIGO Hanford, LIGO Livingston and Virgo were all observing when the glitch happened, as part of O4. At that time, the sensitivities of the LIGO Hanford, LIGO Livingston and Virgo detectors, at $f = 2\nu \approx 22~\text{Hz}$, were $\sqrt{S_n(f)} \approx 8 \times 10^{-23}~\text{Hz}^{-\frac{1}{2}}$, $\sqrt{S_n(f)} \approx 4 \times 10^{-23}~\text{Hz}^{-\frac{1}{2}}$ and $\sqrt{S_n(f)} \approx 3 \times 10^{-22}~\text{Hz}^{-\frac{1}{2}}$, respectively.\footnote{\url{https://gwosc.org/detector_status/day/20240429/}}

Using the approximation of $\beta = A_2$, with $\iota = 0\text{\textdegree}$, we calculate the maximum SNR (maximised over the three detectors), $\text{max}(\rho)$, for this event and we find 
\begin{itemize}
    \item Agnostic model: $\text{max}(\rho) = 137.8$, 
    \item Vortex unpinning model: $\text{max}(\rho) = 1.5$,
    \item Transient mountain model: $\text{max}(\rho) = 61.6$,
\end{itemize}
where we used $I_\text{p}/I = 0.01$ for the vortex unpinning model and $Q = 0.2$ for the transient mountain model, which is the historical average for the Vela pulsar. Excitingly, some of these transient CW models can now be probed with the latest Vela glitch. In the event of a non-detection, upper limits can be placed on how much each of these models can contribute towards GW radiation which in turn puts constraints on the physical parameters that govern the models.

However, it should be noted that the SNRs calculated here are likely to be overestimates since there are other ways for the glitch energy to be used up. For instance, it is believed that $f$-modes of the NS could be excited during a glitch \citep{hoetal2020, yimJones2023, wilsonHo2024}. Interestingly, this could also radiate GWs. Following \cite{yimJones2023}, we calculate the associated SNR coming from the excitation and decay of $f$-modes and find that one could get a burst signal, with frequency $f\sim 2~\text{kHz}$ and duration $\tau \sim 0.1~\text{s}$, with a SNR of $\rho = 25, 50, 7$ for LIGO Hanford, LIGO Livingston and Virgo, respectively. Note that this $f$-mode calculation used $\beta = 1$ as the signal is not long enough for the $\beta = A_2$ approximation to be valid. This can only give a rough estimate of the SNR so the $f$-mode calculation should only be taken as indicative.

\section{Discussion}
\label{section_discussion}
In this work, we have often taken the most optimistic scenario for detecting transient CWs, such as the inclination angle being zero, i.e.~$\iota = 0\text{\textdegree}$. Since we observe pulsars pulse, we know that we cannot have $\iota = 0\text{\textdegree}$, meaning that we will never truly have just pure $f=2\nu$ radiation but rather a mix containing $f=\nu$ radiation as well. One should be able to replicate the analysis presented here but track the $f=\nu$ radiation too, but we will leave this for a future study. 

If $\iota$ is non-zero, then one might ask what a suitable value would be. We can obtain this from electromagnetic observations, specifically from radio polarisation data, which are often fit with the ``rotating vector model'' \citep{radhakrishnanCooke1969, everettWeisberg2001, jones2007}. Using this model, one obtains a value of the magnetic inclination angle ($\alpha_\text{RVM}$, the angle between the pulsar's rotation axis and magnetic axis) and the viewing angle ($\beta_\text{RVM}$, the angle between the magnetic axis and the observer's line of sight). The inclination $\iota$ is therefore just the sum of these two angles, $\iota = \alpha_\text{RVM} + \beta_\text{RVM}$ (or $\iota = 180\text{\textdegree} - \alpha_\text{RVM} - \beta_\text{RVM}$, see \cite{jones2015} and \cite{LVC2017}). Future work should incorporate this inclination dependence into the analysis to get a more faithful SNR calculation. 

In any case, one can still take the assumption of $\iota = 0\text{\textdegree}$ and define upper limits. The agnostic models ($\kappa = 1$) define an ``upper energy limit'' and the transient mountain model defines an ``upper spin-down limit''. In the case where one of these models predicts a detectable signal but is, in reality, not detected, then one can define an efficiency factor $\eta$ that tells us the largest fraction of, say, the upper energy limit that goes into GW emission. For instance, if the value of $E_\text{GW}$ corresponds to a detectable signal ($\rho > \rho_\text{thres}$), but a signal was not detected in reality, then we know the energy emitted must be smaller than $\eta E_\text{GW}$, where $\eta < 1$. Specifically, the efficiency factor can be calculated as
\begin{align}
\eta_\kappa \leq \frac{\rho_\text{thres}^2}{\rho_\kappa^2}~,
\end{align}
where the subscript $\kappa$ shows variables that are model-dependent and $\rho_\text{thres}$ is the threshold SNR used to claim a detection. 

For instance, for the recent Vela glitch covered in Section~\ref{section_glitch_in_vela}, the agnostic model predicted a maximum SNR of 137.8 (where the maximisation refers to maximising across all three detectors) and the transient mountain model predicted a maximum SNR of 61.6. Let us take $\rho_\text{thres} = 10$. If after searching, we do not find a signal, then we are able to place limits on the efficiencies of each model: $\eta_\text{en} \le 0.00526$ and $\eta_\text{sd} \le 0.02635$ for the agnostic (energy upper limit) and transient mountain (spin-down limit) models, respectively. Note that one can easily convert from upper energy efficiencies to upper spin-down efficiencies by simply dividing by $Q$
\begin{align}
\eta_\text{sd} = \frac{\eta_\text{en}}{Q}~.
\end{align}

In the event that we do get a detection, then we must be able to tackle in inverse problem, i.e.~being able to infer what the physical parameters of the signal are. The models covered here are analytical but are oversimplified, and so more work should be done on incorporating more realistic physics and exploring what can be learnt from a detection. One conclusion that can be drawn from the analysis here is that if a glitch-induced transient CW detection is made using current generation detectors, then we know it cannot be from the vortex unpinning or Ekman pumping models, unless they were modified. 

\section{Conclusion}
\label{section_conclusion}
The analysis presented here has shown how to go from simple glitch models that predict transient CWs to calculating whether those signals would be detectable or not. We began by looking at six different models, two that covered the glitch rise, two that covered the glitch recovery, and two that were agnostic to the exact mechanisms at play. We found that the agnostic models provided the most optimistic estimates of how much energy can be radiated as GWs, and we found that the one- and two-component agnostic models predict the same amount of GW energy. We call this the ``upper energy limit'' for glitches. Similarly, one can define the ``upper spin-down limit'' for glitches which comes from the transient mountain model. For the first time, we included Ekman pumping into detectability calculations.

Using these GW energies, we were able to derive eq.\,(\ref{SNR_final_result}) which estimated the SNR using an analytical approximation that allowed us to include the effects of the source's sky position and orientation and the detector's position and orientation. We showed that the SNR error when using this approximation is less than 10\% for any source, so long as the transient CW lasts longer than 1.74~d, regardless of whether the detector used is LIGO Hanford, LIGO Livingston or Virgo. Since most glitch recoveries have timescales much longer than this, we argue that the approximation is suitable, especially if only getting a sense of detectability.

Using this novel approximation, we applied the detectability calculation to the known population of glitching pulsars. We found that some models (agnostic, transient mountain, starquakes) predict a detectable signal in O4, whereas the vortex unpinning and Ekman pumping models do not predict detectable signals in O4. However, with next generation detectors, it is possible that the vortex unpinning model could provide a marginal detection, but the Ekman pumping model would not. 

We compared the SNRs obtained from the novel $\beta = A_2$ approximation with a commonly-used assumption of sky and orientation averaging ($\beta = 4/25$) and we found that the $\beta = A_2$ approximation leads to more optimistic SNRs, so long as the source's inclination is smaller than around $\sim$60\textdegree.

From these SNR calculations, we were also able to provide tables of the highest priority targets, dependent on the energy model assumed. In all models, the Vela pulsar was the top priority as it gave the largest detectability estimates. The Crab pulsar was also interesting, especially with the number of glitches that has been observed coming from it, but we also suggest PSR~J0940$-$5428, PSR~J1952+3252, PSR~J0205+6449 and PSR~J1813$-$1246 could potentially be even more interesting. Unfortunately, glitch-induced transient CWs from PSR~J0537$-$6910 do not appear to be detectable with current generation detectors, with it only becoming constraining for models during the next generation of detectors.

Next, we no longer restricted ourselves to the population of glitching pulsars and applied the analysis to the entire pulsar population, assuming that one day, some of these pulsars may glitch. We provided two lists of high priority sources (Tables~\ref{table_chi} and \ref{table_upsilon}), and in the latter, the top 164 pulsars have not been observed to glitch before. We advocate for better timing programmes for these potentially interesting pulsars.

Also, we formulated two new parameters, $\chi$ and $\Upsilon$, which can be used to quickly calculate the SNR for a given pulsar-detector combination. The SNR can be calculated by multiplying $\chi$ by the square root of the GW energy, or similarly, from multiplying $\Upsilon$ by the square root of the dimensionless glitch size. 

Finally, we applied our analysis to the recent April 2024 glitch from the Vela pulsar, which was coincident with O4. We found that the agnostic and transient mountain models give a detectable signal, whereas the vortex unpinning model does not. It is likely that LVK will conduct a search for this GW signal and in the event of a non-detection, we will be able to place upper limits on how much energy is radiated away as GWs, placing constraints on the models. In this case, this event would correspond to the first time the upper energy (and spin-down) limit for glitches would have been beaten.

From this analysis and Vela's most recent glitch, it is clear that the study of GWs from pulsar glitches is promising and timely. It will offer a new opportunity to independently probe NSs and with current and upcoming GW detectors, we should be able to place constraints on physical parameters that allow us to test our understanding of superfluidity, elasticity/plasticity, viscosity, magnetic fields, temperature gradients, and so on. More resources need to be dedicated towards this effort so that hopefully, when a signal is detected, we will be ready to maximise the science that can come out from it.

% % Example figure
% \begin{figure}
% 	% To include a figure from a file named example.*
% 	% Allowable file formats are eps or ps if compiling using latex
% 	% or pdf, png, jpg if compiling using pdflatex
% 	\includegraphics[width=\columnwidth]{example}
%     \caption{This is an example figure. Captions appear below each figure.
% 	Give enough detail for the reader to understand what they're looking at,
% 	but leave detailed discussion to the main body of the text.}
%     \label{fig:example_figure}
% \end{figure}

% % Example table
% \begin{table}
% 	\centering
% 	\caption{This is an example table. Captions appear above each table.
% 	Remember to define the quantities, symbols and units used.}
% 	\label{tab:example_table}
% 	\begin{tabular}{lccr} % four columns, alignment for each
% 		\hline
% 		A & B & C & D\\
% 		\hline
% 		1 & 2 & 3 & 4\\
% 		2 & 4 & 6 & 8\\
% 		3 & 5 & 7 & 9\\
% 		\hline
% 	\end{tabular}
% \end{table}

\section*{Acknowledgements}

This work was supported by the National Natural Science Foundation of China (12247180, 11991053), the National SKA Program of China (2020SKA0120300, 2020SKA0120100), the Beijing Natural Science Foundation (1242018), the Max Planck Partner Group Program funded by the Max Planck Society, and the High-Performance Computing Platform of Peking University.

%%%%%%%%%%%%%%%%%%%%%%%%%%%%%%%%%%%%%%%%%%%%%%%%%%
\section*{Data Availability}

The data used in this article are cited accordingly in the main text and are all publicly available. The main sources used were the ATNF Pulsar Catalogue \citep{manchesteretal2005}, JBCA Glitch Catalogue \citep{espinozaetal2011, basuetal2022} and the ATNF Glitch Table \citep{manchesteretal2005}. The new data generated in this work will be shared upon reasonable request by emailing the corresponding author. 



\bibliographystyle{mnras}
\bibliography{main} % if your bibtex file is called example.bib

\providecommand{\noopsort}[1]{}\providecommand{\singleletter}[1]{#1}%
\begin{thebibliography}{}
\makeatletter
\relax
\def\mn@urlcharsother{\let\do\@makeother \do\$\do\&\do\#\do\^\do\_\do\%\do\~}
\def\mn@doi{\begingroup\mn@urlcharsother \@ifnextchar [ {\mn@doi@}
  {\mn@doi@[]}}
\def\mn@doi@[#1]#2{\def\@tempa{#1}\ifx\@tempa\@empty \href
  {http://dx.doi.org/#2} {doi:#2}\else \href {http://dx.doi.org/#2} {#1}\fi
  \endgroup}
\def\mn@eprint#1#2{\mn@eprint@#1:#2::\@nil}
\def\mn@eprint@arXiv#1{\href {http://arxiv.org/abs/#1} {{\tt arXiv:#1}}}
\def\mn@eprint@dblp#1{\href {http://dblp.uni-trier.de/rec/bibtex/#1.xml}
  {dblp:#1}}
\def\mn@eprint@#1:#2:#3:#4\@nil{\def\@tempa {#1}\def\@tempb {#2}\def\@tempc
  {#3}\ifx \@tempc \@empty \let \@tempc \@tempb \let \@tempb \@tempa \fi \ifx
  \@tempb \@empty \def\@tempb {arXiv}\fi \@ifundefined
  {mn@eprint@\@tempb}{\@tempb:\@tempc}{\expandafter \expandafter \csname
  mn@eprint@\@tempb\endcsname \expandafter{\@tempc}}}

\bibitem[\protect\citeauthoryear{{Abadie} et~al.}{{Abadie}
  et~al.}{2011}]{LSC2011velaglitch}
{Abadie} J.,  et~al., 2011, \mn@doi [\prd] {10.1103/PhysRevD.83.042001}, \href
  {https://ui.adsabs.harvard.edu/abs/2011PhRvD..83d2001A} {83, 042001}

\bibitem[\protect\citeauthoryear{{Abbott} et~al.}{{Abbott}
  et~al.}{2017}]{LVC2017}
{Abbott} B.~P.,  et~al., 2017, \mn@doi [\apj] {10.3847/1538-4357/aa677f}, \href
  {https://ui.adsabs.harvard.edu/abs/2017ApJ...839...12A} {839, 12}

\bibitem[\protect\citeauthoryear{{Abbott} et~al.}{{Abbott}
  et~al.}{2021a}]{LVK2021J0537spindown}
{Abbott} R.,  et~al., 2021a, \mn@doi [\apjl] {10.3847/2041-8213/abffcd}, \href
  {https://ui.adsabs.harvard.edu/abs/2021ApJ...913L..27A} {913, L27}

\bibitem[\protect\citeauthoryear{{Abbott} et~al.}{{Abbott}
  et~al.}{2021b}]{LVK2021J0537rmodes}
{Abbott} R.,  et~al., 2021b, \mn@doi [\apj] {10.3847/1538-4357/ac0d52}, \href
  {https://ui.adsabs.harvard.edu/abs/2021ApJ...922...71A} {922, 71}

\bibitem[\protect\citeauthoryear{{Abbott} et~al.}{{Abbott}
  et~al.}{2022a}]{LVK2022magnetarbursts}
{Abbott} R.,  et~al., 2022a, \mn@doi [arXiv e-prints]
  {10.48550/arXiv.2210.10931}, \href
  {https://ui.adsabs.harvard.edu/abs/2022arXiv221010931T} {p. arXiv:2210.10931}

\bibitem[\protect\citeauthoryear{{Abbott} et~al.}{{Abbott}
  et~al.}{2022b}]{LVK2022transientCWsearch}
{Abbott} R.,  et~al., 2022b, \mn@doi [\apj] {10.3847/1538-4357/ac6ad0}, \href
  {https://ui.adsabs.harvard.edu/abs/2022ApJ...932..133A} {932, 133}

\bibitem[\protect\citeauthoryear{Abbott et~al.}{Abbott
  et~al.}{2023}]{LVK2023GWTC3}
Abbott R.,  et~al., 2023, \mn@doi [Phys. Rev. X] {10.1103/PhysRevX.13.041039},
  13, 041039

\bibitem[\protect\citeauthoryear{{Abney} \& {Epstein}}{{Abney} \&
  {Epstein}}{1996}]{abneyEpstein1996}
{Abney} M.,  {Epstein} R.~I.,  1996, \mn@doi [Journal of Fluid Mechanics]
  {10.1017/S0022112096002030}, \href
  {https://ui.adsabs.harvard.edu/abs/1996JFM...312..327A} {312, 327}

\bibitem[\protect\citeauthoryear{{Agazie} et~al.}{{Agazie}
  et~al.}{2023}]{nanograv2023}
{Agazie} G.,  et~al., 2023, \mn@doi [\apjl] {10.3847/2041-8213/acdac6}, \href
  {https://ui.adsabs.harvard.edu/abs/2023ApJ...951L...8A} {951, L8}

\bibitem[\protect\citeauthoryear{{Alpar}, {Pines}, {Anderson}  \&
  {Shaham}}{{Alpar} et~al.}{1984}]{alparetal1984}
{Alpar} M.~A.,  {Pines} D.,  {Anderson} P.~W.,   {Shaham} J.,  1984, \mn@doi
  [\apj] {10.1086/161616}, \href
  {https://ui.adsabs.harvard.edu/abs/1984ApJ...276..325A} {276, 325}

\bibitem[\protect\citeauthoryear{{Alpar}, {Chau}, {Cheng}  \& {Pines}}{{Alpar}
  et~al.}{1994}]{alparetal1994}
{Alpar} M.~A.,  {Chau} H.~F.,  {Cheng} K.~S.,   {Pines} D.,  1994, \mn@doi
  [\apjl] {10.1086/187357}, \href
  {https://ui.adsabs.harvard.edu/abs/1994ApJ...427L..29A} {427, L29}

\bibitem[\protect\citeauthoryear{{Anderson} \& {Itoh}}{{Anderson} \&
  {Itoh}}{1975}]{andersonItoh1975}
{Anderson} P.~W.,  {Itoh} N.,  1975, \mn@doi [\nat] {10.1038/256025a0}, \href
  {https://ui.adsabs.harvard.edu/abs/1975Natur.256...25A} {256, 25}

\bibitem[\protect\citeauthoryear{{Andersson} \& {Comer}}{{Andersson} \&
  {Comer}}{2021}]{anderssonComer2021}
{Andersson} N.,  {Comer} G.~L.,  2021, \mn@doi [Living Reviews in Relativity]
  {10.1007/s41114-021-00031-6}, \href
  {https://ui.adsabs.harvard.edu/abs/2021LRR....24....3A} {24, 3}

\bibitem[\protect\citeauthoryear{{Andersson}, {Comer}  \& {Prix}}{{Andersson}
  et~al.}{2003}]{anderssonComerPrix2003}
{Andersson} N.,  {Comer} G.~L.,   {Prix} R.,  2003, \mn@doi [\prl]
  {10.1103/PhysRevLett.90.091101}, \href
  {https://ui.adsabs.harvard.edu/abs/2003PhRvL..90i1101A} {90, 091101}

\bibitem[\protect\citeauthoryear{{Andersson}, {Comer}  \&
  {Glampedakis}}{{Andersson} et~al.}{2005}]{anderssonComerGlampedakis2005}
{Andersson} N.,  {Comer} G.~L.,   {Glampedakis} K.,  2005, \mn@doi [\nphysa]
  {10.1016/j.nuclphysa.2005.08.012}, \href
  {https://ui.adsabs.harvard.edu/abs/2005NuPhA.763..212A} {763, 212}

\bibitem[\protect\citeauthoryear{{Andersson}, {Glampedakis}, {Ho}  \&
  {Espinoza}}{{Andersson} et~al.}{2012}]{anderssonetal2012}
{Andersson} N.,  {Glampedakis} K.,  {Ho} W.~C.~G.,   {Espinoza} C.~M.,  2012,
  \mn@doi [\prl] {10.1103/PhysRevLett.109.241103}, \href
  {https://ui.adsabs.harvard.edu/abs/2012PhRvL.109x1103A} {109, 241103}

\bibitem[\protect\citeauthoryear{{Antoniadis} et~al.}{{Antoniadis}
  et~al.}{2023}]{epta2023}
{Antoniadis} J.,  et~al., 2023, \mn@doi [\aap] {10.1051/0004-6361/202346844},
  \href {https://ui.adsabs.harvard.edu/abs/2023A&A...678A..50E} {678, A50}

\bibitem[\protect\citeauthoryear{{Antonopoulou}, {Espinoza}, {Kuiper}  \&
  {Andersson}}{{Antonopoulou} et~al.}{2018}]{antonopoulouetal2018}
{Antonopoulou} D.,  {Espinoza} C.~M.,  {Kuiper} L.,   {Andersson} N.,  2018,
  \mn@doi [\mnras] {10.1093/mnras/stx2429}, \href
  {https://ui.adsabs.harvard.edu/abs/2018MNRAS.473.1644A} {473, 1644}

\bibitem[\protect\citeauthoryear{{Antonopoulou}, {Haskell}  \&
  {Espinoza}}{{Antonopoulou} et~al.}{2022}]{antonopoulouHaskellEspinoza2022}
{Antonopoulou} D.,  {Haskell} B.,   {Espinoza} C.~M.,  2022, \mn@doi [Reports
  on Progress in Physics] {10.1088/1361-6633/ac9ced}, \href
  {https://ui.adsabs.harvard.edu/abs/2022RPPh...85l6901A} {85, 126901}

\bibitem[\protect\citeauthoryear{{Archibald} et~al.,}{{Archibald}
  et~al.}{2013}]{archibaldetal2013}
{Archibald} R.~F.,  et~al., 2013, \mn@doi [\nat] {10.1038/nature12159}, \href
  {https://ui.adsabs.harvard.edu/abs/2013Natur.497..591A} {497, 591}

\bibitem[\protect\citeauthoryear{{Ashton}, {Lasky}, {Graber}  \&
  {Palfreyman}}{{Ashton} et~al.}{2019}]{ashtonetal2019}
{Ashton} G.,  {Lasky} P.~D.,  {Graber} V.,   {Palfreyman} J.,  2019, \mn@doi
  [Nature Astronomy] {10.1038/s41550-019-0844-6}, \href
  {https://ui.adsabs.harvard.edu/abs/2019NatAs...3.1143A} {3, 1143}

\bibitem[\protect\citeauthoryear{{Basu} et~al.,}{{Basu}
  et~al.}{2022}]{basuetal2022}
{Basu} A.,  et~al., 2022, \mn@doi [\mnras] {10.1093/mnras/stab3336}, \href
  {https://ui.adsabs.harvard.edu/abs/2022MNRAS.510.4049B} {510, 4049}

\bibitem[\protect\citeauthoryear{{Baym} \& {Pines}}{{Baym} \&
  {Pines}}{1971}]{baymPines1971}
{Baym} G.,  {Pines} D.,  1971, \mn@doi [Annals of Physics]
  {10.1016/0003-4916(71)90084-4}, \href
  {https://ui.adsabs.harvard.edu/abs/1971AnPhy..66..816B} {66, 816}

\bibitem[\protect\citeauthoryear{{Baym}, {Pethick}  \& {Pines}}{{Baym}
  et~al.}{1969a}]{baymPethickPines1969}
{Baym} G.,  {Pethick} C.,   {Pines} D.,  1969a, \mn@doi [\nat]
  {10.1038/224673a0}, \href
  {https://ui.adsabs.harvard.edu/abs/1969Natur.224..673B} {224, 673}

\bibitem[\protect\citeauthoryear{{Baym}, {Pethick}, {Pines}  \&
  {Ruderman}}{{Baym} et~al.}{1969b}]{baymetal1969}
{Baym} G.,  {Pethick} C.,  {Pines} D.,   {Ruderman} M.,  1969b, \mn@doi [\nat]
  {10.1038/224872a0}, \href
  {https://ui.adsabs.harvard.edu/abs/1969Natur.224..872B} {224, 872}

\bibitem[\protect\citeauthoryear{{Bennett}, {van Eysden}  \&
  {Melatos}}{{Bennett} et~al.}{2010}]{bennettvanEysdenMelatos2010}
{Bennett} M.~F.,  {van Eysden} C.~A.,   {Melatos} A.,  2010, \mn@doi [\mnras]
  {10.1111/j.1365-2966.2010.17416.x}, \href
  {https://ui.adsabs.harvard.edu/abs/2010MNRAS.409.1705B} {409, 1705}

\bibitem[\protect\citeauthoryear{{Benton} \& {Clark}}{{Benton} \&
  {Clark}}{1974}]{bentonClark1974}
{Benton} E.~R.,  {Clark} A.,  1974, \mn@doi [Annual Review of Fluid Mechanics]
  {10.1146/annurev.fl.06.010174.001353}, \href
  {https://ui.adsabs.harvard.edu/abs/1974AnRFM...6..257B} {6, 257}

\bibitem[\protect\citeauthoryear{{Burgay} et~al.}{{Burgay}
  et~al.}{2019}]{burgayetal2019}
{Burgay} M.,  et~al., 2019, \mn@doi [\mnras] {10.1093/mnras/stz401}, \href
  {https://ui.adsabs.harvard.edu/abs/2019MNRAS.484.5791B} {484, 5791}

\bibitem[\protect\citeauthoryear{{Campbell-Wilson}, {Flynn}  \&
  {Bateman}}{{Campbell-Wilson} et~al.}{2024}]{campbell-wilsonFlynnBateman2024}
{Campbell-Wilson} D.,  {Flynn} C.,   {Bateman} T.,  2024, The Astronomer's
  Telegram, \href {https://ui.adsabs.harvard.edu/abs/2024ATel16610....1C}
  {16610, 1}

\bibitem[\protect\citeauthoryear{{Chamel}}{{Chamel}}{2012}]{chamel2012}
{Chamel} N.,  2012, \mn@doi [\prc] {10.1103/PhysRevC.85.035801}, \href
  {https://ui.adsabs.harvard.edu/abs/2012PhRvC..85c5801C} {85, 035801}

\bibitem[\protect\citeauthoryear{{Cognard} \& {Backer}}{{Cognard} \&
  {Backer}}{2004}]{cognardBacker2004}
{Cognard} I.,  {Backer} D.~C.,  2004, \mn@doi [\apjl] {10.1086/424692}, \href
  {https://ui.adsabs.harvard.edu/abs/2004ApJ...612L.125C} {612, L125}

\bibitem[\protect\citeauthoryear{{Crawford} \& {Demia{\'n}ski}}{{Crawford} \&
  {Demia{\'n}ski}}{2003}]{crawfordDemianski2003}
{Crawford} F.,  {Demia{\'n}ski} M.,  2003, \mn@doi [\apj] {10.1086/377470},
  \href {https://ui.adsabs.harvard.edu/abs/2003ApJ...595.1052C} {595, 1052}

\bibitem[\protect\citeauthoryear{{Cutler} \& {Schutz}}{{Cutler} \&
  {Schutz}}{2005}]{cutlerSchutz2005}
{Cutler} C.,  {Schutz} B.~F.,  2005, \mn@doi [\prd]
  {10.1103/PhysRevD.72.063006}, \href
  {https://ui.adsabs.harvard.edu/abs/2005PhRvD..72f3006C} {72, 063006}

\bibitem[\protect\citeauthoryear{{Deller} et~al.,}{{Deller}
  et~al.}{2019}]{delleretal2019}
{Deller} A.~T.,  et~al., 2019, \mn@doi [\apj] {10.3847/1538-4357/ab11c7}, \href
  {https://ui.adsabs.harvard.edu/abs/2019ApJ...875..100D} {875, 100}

\bibitem[\protect\citeauthoryear{{Dreissigacker}, {Prix}  \&
  {Wette}}{{Dreissigacker} et~al.}{2018}]{dressigackerPrixWette2018}
{Dreissigacker} C.,  {Prix} R.,   {Wette} K.,  2018, \mn@doi [\prd]
  {10.1103/PhysRevD.98.084058}, \href
  {https://ui.adsabs.harvard.edu/abs/2018PhRvD..98h4058D} {98, 084058}

\bibitem[\protect\citeauthoryear{{Edwards}, {Hobbs}  \& {Manchester}}{{Edwards}
  et~al.}{2006}]{edwardsHobbsManchester2006}
{Edwards} R.~T.,  {Hobbs} G.~B.,   {Manchester} R.~N.,  2006, \mn@doi [\mnras]
  {10.1111/j.1365-2966.2006.10870.x}, \href
  {https://ui.adsabs.harvard.edu/abs/2006MNRAS.372.1549E} {372, 1549}

\bibitem[\protect\citeauthoryear{{Espinoza}, {Lyne}, {Stappers}  \&
  {Kramer}}{{Espinoza} et~al.}{2011}]{espinozaetal2011}
{Espinoza} C.~M.,  {Lyne} A.~G.,  {Stappers} B.~W.,   {Kramer} M.,  2011,
  \mn@doi [\mnras] {10.1111/j.1365-2966.2011.18503.x}, \href
  {https://ui.adsabs.harvard.edu/abs/2011MNRAS.414.1679E} {414, 1679}

\bibitem[\protect\citeauthoryear{{Everett} \& {Weisberg}}{{Everett} \&
  {Weisberg}}{2001}]{everettWeisberg2001}
{Everett} J.~E.,  {Weisberg} J.~M.,  2001, \mn@doi [\apj] {10.1086/320652},
  \href {https://ui.adsabs.harvard.edu/abs/2001ApJ...553..341E} {553, 341}

\bibitem[\protect\citeauthoryear{{Ferdman}, {Archibald}, {Gourgouliatos}  \&
  {Kaspi}}{{Ferdman} et~al.}{2018}]{ferdmanetal2018}
{Ferdman} R.~D.,  {Archibald} R.~F.,  {Gourgouliatos} K.~N.,   {Kaspi} V.~M.,
  2018, \mn@doi [\apj] {10.3847/1538-4357/aaa198}, \href
  {https://ui.adsabs.harvard.edu/abs/2018ApJ...852..123F} {852, 123}

\bibitem[\protect\citeauthoryear{{Fesik} \& {Papa}}{{Fesik} \&
  {Papa}}{2020}]{fesikPapa2020}
{Fesik} L.,  {Papa} M.~A.,  2020, \mn@doi [\apj] {10.3847/1538-4357/ab8193},
  \href {https://ui.adsabs.harvard.edu/abs/2020ApJ...895...11F} {895, 11}

\bibitem[\protect\citeauthoryear{{Fuentes}, {Espinoza}, {Reisenegger}, {Shaw},
  {Stappers}  \& {Lyne}}{{Fuentes} et~al.}{2017}]{fuentesetal2017}
{Fuentes} J.~R.,  {Espinoza} C.~M.,  {Reisenegger} A.,  {Shaw} B.,  {Stappers}
  B.~W.,   {Lyne} A.~G.,  2017, \mn@doi [\aap] {10.1051/0004-6361/201731519},
  \href {https://ui.adsabs.harvard.edu/abs/2017A&A...608A.131F} {608, A131}

\bibitem[\protect\citeauthoryear{{Graber}, {Cumming}  \& {Andersson}}{{Graber}
  et~al.}{2018}]{graberCummingAndersson2018}
{Graber} V.,  {Cumming} A.,   {Andersson} N.,  2018, \mn@doi [\apj]
  {10.3847/1538-4357/aad776}, \href
  {https://ui.adsabs.harvard.edu/abs/2018ApJ...865...23G} {865, 23}

\bibitem[\protect\citeauthoryear{{Greenspan} \& {Howard}}{{Greenspan} \&
  {Howard}}{1963}]{greenspanHoward1963}
{Greenspan} H.~P.,  {Howard} L.~N.,  1963, \mn@doi [Journal of Fluid Mechanics]
  {10.1017/S0022112063001415}, \href
  {https://ui.adsabs.harvard.edu/abs/1963JFM....17..385G} {17, 385}

\bibitem[\protect\citeauthoryear{{Grover}, {Krishnakumar}, {Joshi}  \&
  {Arumugam}}{{Grover} et~al.}{2024}]{groveretal2024}
{Grover} H.,  {Krishnakumar} M.~A.,  {Joshi} B.~C.,   {Arumugam} P.,  2024, The
  Astronomer's Telegram, 16611, 1

\bibitem[\protect\citeauthoryear{{Haskell} \& {Jones}}{{Haskell} \&
  {Jones}}{2024}]{haskellJones2024}
{Haskell} B.,  {Jones} D.~I.,  2024, \mn@doi [Astroparticle Physics]
  {10.1016/j.astropartphys.2023.102921}, \href
  {https://ui.adsabs.harvard.edu/abs/2024APh...15702921H} {157, 102921}

\bibitem[\protect\citeauthoryear{{Haskell} \& {Melatos}}{{Haskell} \&
  {Melatos}}{2015}]{haskellMelatos2015}
{Haskell} B.,  {Melatos} A.,  2015, \mn@doi [International Journal of Modern
  Physics D] {10.1142/S0218271815300086}, \href
  {https://ui.adsabs.harvard.edu/abs/2015IJMPD..2430008H} {24, 1530008}

\bibitem[\protect\citeauthoryear{{Haskell}, {Khomenko}, {Antonelli}  \&
  {Antonopoulou}}{{Haskell} et~al.}{2018}]{haskelletal2018}
{Haskell} B.,  {Khomenko} V.,  {Antonelli} M.,   {Antonopoulou} D.,  2018,
  \mn@doi [\mnras] {10.1093/mnrasl/sly175}, \href
  {https://ui.adsabs.harvard.edu/abs/2018MNRAS.481L.146H} {481, L146}

\bibitem[\protect\citeauthoryear{{Ho}, {Espinoza}, {Antonopoulou}  \&
  {Andersson}}{{Ho} et~al.}{2015}]{hoetal2015}
{Ho} W.~C.~G.,  {Espinoza} C.~M.,  {Antonopoulou} D.,   {Andersson} N.,  2015,
  \mn@doi [Science Advances] {10.1126/sciadv.1500578}, \href
  {https://ui.adsabs.harvard.edu/abs/2015SciA....1E0578H} {1, e1500578}

\bibitem[\protect\citeauthoryear{{Ho}, {Jones}, {Andersson}  \&
  {Espinoza}}{{Ho} et~al.}{2020a}]{hoetal2020}
{Ho} W. C.~G.,  {Jones} D.~I.,  {Andersson} N.,   {Espinoza} C.~M.,  2020a,
  \mn@doi [\prd] {10.1103/PhysRevD.101.103009}, \href
  {https://ui.adsabs.harvard.edu/abs/2020PhRvD.101j3009H} {101, 103009}

\bibitem[\protect\citeauthoryear{{Ho} et~al.,}{{Ho}
  et~al.}{2020b}]{hoetal2020J0537}
{Ho} W. C.~G.,  et~al., 2020b, \mn@doi [\mnras] {10.1093/mnras/staa2640}, \href
  {https://ui.adsabs.harvard.edu/abs/2020MNRAS.498.4605H} {498, 4605}

\bibitem[\protect\citeauthoryear{{Ho} et~al.}{{Ho} et~al.}{2022}]{hoetal2022}
{Ho} W. C.~G.,  et~al., 2022, \mn@doi [\apj] {10.3847/1538-4357/ac8743}, \href
  {https://ui.adsabs.harvard.edu/abs/2022ApJ...939....7H} {939, 7}

\bibitem[\protect\citeauthoryear{{I{\c{c}}dem}, {Baykal}  \&
  {Inam}}{{I{\c{c}}dem} et~al.}{2012}]{icdemBaykalInam2012}
{I{\c{c}}dem} B.,  {Baykal} A.,   {Inam} S.~{\c{C}}.,  2012, \mn@doi [\mnras]
  {10.1111/j.1365-2966.2011.19953.x}, \href
  {https://ui.adsabs.harvard.edu/abs/2012MNRAS.419.3109I} {419, 3109}

\bibitem[\protect\citeauthoryear{{Jaranowski}, {Kr{\'o}lak}  \&
  {Schutz}}{{Jaranowski} et~al.}{1998}]{jaranowskiKrolakSchutz1998}
{Jaranowski} P.,  {Kr{\'o}lak} A.,   {Schutz} B.~F.,  1998, \mn@doi [\prd]
  {10.1103/PhysRevD.58.063001}, \href
  {https://ui.adsabs.harvard.edu/abs/1998PhRvD..58f3001J} {58, 063001}

\bibitem[\protect\citeauthoryear{{Jones}}{{Jones}}{2007}]{jones2007}
{Jones} D.~I.,  2007, \mn@doi [\apss] {10.1007/s10509-007-9324-z}, \href
  {https://ui.adsabs.harvard.edu/abs/2007Ap&SS.308..125J} {308, 125}

\bibitem[\protect\citeauthoryear{{Jones}}{{Jones}}{2015}]{jones2015}
{Jones} D.~I.,  2015, \mn@doi [\mnras] {10.1093/mnras/stv1584}, \href
  {https://ui.adsabs.harvard.edu/abs/2015MNRAS.453...53J} {453, 53}

\bibitem[\protect\citeauthoryear{{Kashiyama} \& {Ioka}}{{Kashiyama} \&
  {Ioka}}{2011}]{kashiyamaIoka2011}
{Kashiyama} K.,  {Ioka} K.,  2011, \mn@doi [\prd] {10.1103/PhysRevD.83.081302},
  \href {https://ui.adsabs.harvard.edu/abs/2011PhRvD..83h1302K} {83, 081302}

\bibitem[\protect\citeauthoryear{{Kaspi} \& {Beloborodov}}{{Kaspi} \&
  {Beloborodov}}{2017}]{kaspiBeloborodov2017}
{Kaspi} V.~M.,  {Beloborodov} A.~M.,  2017, \mn@doi [\araa]
  {10.1146/annurev-astro-081915-023329}, \href
  {https://ui.adsabs.harvard.edu/abs/2017ARA&A..55..261K} {55, 261}

\bibitem[\protect\citeauthoryear{{Keer} \& {Jones}}{{Keer} \&
  {Jones}}{2015}]{keerJones2015}
{Keer} L.,  {Jones} D.~I.,  2015, \mn@doi [\mnras] {10.1093/mnras/stu2123},
  \href {https://ui.adsabs.harvard.edu/abs/2015MNRAS.446..865K} {446, 865}

\bibitem[\protect\citeauthoryear{{Keitel} et~al.,}{{Keitel}
  et~al.}{2019}]{keiteletal2019}
{Keitel} D.,  et~al., 2019, \mn@doi [\prd] {10.1103/PhysRevD.100.064058}, \href
  {https://ui.adsabs.harvard.edu/abs/2019PhRvD.100f4058K} {100, 064058}

\bibitem[\protect\citeauthoryear{{Khomenko} \& {Haskell}}{{Khomenko} \&
  {Haskell}}{2018}]{khomenkoHaskell2018}
{Khomenko} V.,  {Haskell} B.,  2018, \mn@doi [\pasa] {10.1017/pasa.2018.12},
  \href {https://ui.adsabs.harvard.edu/abs/2018PASA...35...20K} {35, e020}

\bibitem[\protect\citeauthoryear{{Kuiper} \& {Hermsen}}{{Kuiper} \&
  {Hermsen}}{2010}]{kuiperHermsen2010}
{Kuiper} L.,  {Hermsen} W.,  2010, The Astronomer's Telegram, \href
  {https://ui.adsabs.harvard.edu/abs/2010ATel.2446....1K} {2446, 1}

\bibitem[\protect\citeauthoryear{{Larson} \& {Link}}{{Larson} \&
  {Link}}{1999}]{larsonLink1999}
{Larson} M.~B.,  {Link} B.,  1999, \mn@doi [\apj] {10.1086/307532}, \href
  {https://ui.adsabs.harvard.edu/abs/1999ApJ...521..271L} {521, 271}

\bibitem[\protect\citeauthoryear{{Link}, {Epstein}  \& {Lattimer}}{{Link}
  et~al.}{1999}]{linkEpsteinLattimer1999}
{Link} B.,  {Epstein} R.~I.,   {Lattimer} J.~M.,  1999, \mn@doi [\prl]
  {10.1103/PhysRevLett.83.3362}, \href
  {https://ui.adsabs.harvard.edu/abs/1999PhRvL..83.3362L} {83, 3362}

\bibitem[\protect\citeauthoryear{{Lopez}, {Tiwari}, {Drago}, {Keitel},
  {Lazzaro}  \& {Prodi}}{{Lopez} et~al.}{2022}]{lopezetal2022}
{Lopez} D.,  {Tiwari} S.,  {Drago} M.,  {Keitel} D.,  {Lazzaro} C.,   {Prodi}
  G.~A.,  2022, \mn@doi [\prd] {10.1103/PhysRevD.106.103037}, \href
  {https://ui.adsabs.harvard.edu/abs/2022PhRvD.106j3037L} {106, 103037}

\bibitem[\protect\citeauthoryear{{Lorenz}, {Ravenhall}  \& {Pethick}}{{Lorenz}
  et~al.}{1993}]{lorenzRavenhallPethick1993}
{Lorenz} C.~P.,  {Ravenhall} D.~G.,   {Pethick} C.~J.,  1993, \mn@doi [\prl]
  {10.1103/PhysRevLett.70.379}, \href
  {https://ui.adsabs.harvard.edu/abs/1993PhRvL..70..379L} {70, 379}

\bibitem[\protect\citeauthoryear{{Lorimer} et~al.,}{{Lorimer}
  et~al.}{1995}]{lorimeretal1995}
{Lorimer} D.~R.,  et~al., 1995, \mn@doi [\apj] {10.1086/175230}, \href
  {https://ui.adsabs.harvard.edu/abs/1995ApJ...439..933L} {439, 933}

\bibitem[\protect\citeauthoryear{{Lower} et~al.,}{{Lower}
  et~al.}{2021}]{loweretal2021}
{Lower} M.~E.,  et~al., 2021, \mn@doi [\mnras] {10.1093/mnras/stab2678}, \href
  {https://ui.adsabs.harvard.edu/abs/2021MNRAS.508.3251L} {508, 3251}

\bibitem[\protect\citeauthoryear{{Lyne}, {Brinklow}, {Middleditch}, {Kulkarni}
  \& {Backer}}{{Lyne} et~al.}{1987}]{lyneetal1987}
{Lyne} A.~G.,  {Brinklow} A.,  {Middleditch} J.,  {Kulkarni} S.~R.,   {Backer}
  D.~C.,  1987, \mn@doi [\nat] {10.1038/328399a0}, \href
  {https://ui.adsabs.harvard.edu/abs/1987Natur.328..399L} {328, 399}

\bibitem[\protect\citeauthoryear{{Manchester}, {Hobbs}, {Teoh}  \&
  {Hobbs}}{{Manchester} et~al.}{2005}]{manchesteretal2005}
{Manchester} R.~N.,  {Hobbs} G.~B.,  {Teoh} A.,   {Hobbs} M.,  2005, \mn@doi
  [\aj] {10.1086/428488}, \href
  {https://ui.adsabs.harvard.edu/abs/2005AJ....129.1993M} {129, 1993}

\bibitem[\protect\citeauthoryear{{McKee} et~al.}{{McKee}
  et~al.}{2016}]{mckeeetal2016}
{McKee} J.~W.,  et~al., 2016, \mn@doi [\mnras] {10.1093/mnras/stw1442}, \href
  {https://ui.adsabs.harvard.edu/abs/2016MNRAS.461.2809M} {461, 2809}

\bibitem[\protect\citeauthoryear{{Modafferi}, {Moragues}  \&
  {Keitel}}{{Modafferi} et~al.}{2021}]{modafferietal2021}
{Modafferi} L.~M.,  {Moragues} J.,   {Keitel} D.,  2021, in Journal of Physics
  Conference Series. IOP, p. 012079 (\mn@eprint {arXiv} {2201.08785}),
  \mn@doi{10.1088/1742-6596/2156/1/012079}

\bibitem[\protect\citeauthoryear{{Moragues}, {Modafferi}, {Tenorio}  \&
  {Keitel}}{{Moragues} et~al.}{2023}]{moraguesetal2023}
{Moragues} J.,  {Modafferi} L.~M.,  {Tenorio} R.,   {Keitel} D.,  2023, \mn@doi
  [\mnras] {10.1093/mnras/stac3665}, \href
  {https://ui.adsabs.harvard.edu/abs/2023MNRAS.519.5161M} {519, 5161}

\bibitem[\protect\citeauthoryear{{Page} \& {Applegate}}{{Page} \&
  {Applegate}}{1992}]{pageApplegate1992}
{Page} D.,  {Applegate} J.~H.,  1992, \mn@doi [\apjl] {10.1086/186462}, \href
  {https://ui.adsabs.harvard.edu/abs/1992ApJ...394L..17P} {394, L17}

\bibitem[\protect\citeauthoryear{{Palfreyman}}{{Palfreyman}}{2024}]{palfreyman2024}
{Palfreyman} J.,  2024, The Astronomer's Telegram, 16615, 1

\bibitem[\protect\citeauthoryear{{Piekarewicz}, {Fattoyev}  \&
  {Horowitz}}{{Piekarewicz} et~al.}{2014}]{piekarewiczFattoyevHorowitz2014}
{Piekarewicz} J.,  {Fattoyev} F.~J.,   {Horowitz} C.~J.,  2014, \mn@doi [\prc]
  {10.1103/PhysRevC.90.015803}, \href
  {https://ui.adsabs.harvard.edu/abs/2014PhRvC..90a5803P} {90, 015803}

\bibitem[\protect\citeauthoryear{{Pines} \& {Alpar}}{{Pines} \&
  {Alpar}}{1985}]{pinesAlpar1985}
{Pines} D.,  {Alpar} M.~A.,  1985, \mn@doi [\nat] {10.1038/316027a0}, \href
  {https://ui.adsabs.harvard.edu/abs/1985Natur.316...27P} {316, 27}

\bibitem[\protect\citeauthoryear{{Pletsch} et~al.}{{Pletsch}
  et~al.}{2013}]{pletschetal2013}
{Pletsch} H.~J.,  et~al., 2013, \mn@doi [\apjl] {10.1088/2041-8205/779/1/L11},
  \href {https://ui.adsabs.harvard.edu/abs/2013ApJ...779L..11P} {779, L11}

\bibitem[\protect\citeauthoryear{{Pradhan}, {Pathak}  \&
  {Chatterjee}}{{Pradhan} et~al.}{2023}]{pradhanPathakChatterjee2023}
{Pradhan} B.~K.,  {Pathak} D.,   {Chatterjee} D.,  2023, \mn@doi [\apj]
  {10.3847/1538-4357/acef1f}, \href
  {https://ui.adsabs.harvard.edu/abs/2023ApJ...956...38P} {956, 38}

\bibitem[\protect\citeauthoryear{{Prix}, {Giampanis}  \& {Messenger}}{{Prix}
  et~al.}{2011}]{prixGiampanisMessenger2011}
{Prix} R.,  {Giampanis} S.,   {Messenger} C.,  2011, \mn@doi [\prd]
  {10.1103/PhysRevD.84.023007}, \href
  {https://ui.adsabs.harvard.edu/abs/2011PhRvD..84b3007P} {84, 023007}

\bibitem[\protect\citeauthoryear{{Radhakrishnan} \& {Cooke}}{{Radhakrishnan} \&
  {Cooke}}{1969}]{radhakrishnanCooke1969}
{Radhakrishnan} V.,  {Cooke} D.~J.,  1969, \aplett, \href
  {https://ui.adsabs.harvard.edu/abs/1969ApL.....3..225R} {3, 225}

\bibitem[\protect\citeauthoryear{{Ravenhall} \& {Pethick}}{{Ravenhall} \&
  {Pethick}}{1994}]{ravenhallPethick1994}
{Ravenhall} D.~G.,  {Pethick} C.~J.,  1994, \mn@doi [\apj] {10.1086/173935},
  \href {https://ui.adsabs.harvard.edu/abs/1994ApJ...424..846R} {424, 846}

\bibitem[\protect\citeauthoryear{{Reardon} et~al.}{{Reardon}
  et~al.}{2021}]{reardonetal2021}
{Reardon} D.~J.,  et~al., 2021, \mn@doi [\mnras] {10.1093/mnras/stab1990},
  \href {https://ui.adsabs.harvard.edu/abs/2021MNRAS.507.2137R} {507, 2137}

\bibitem[\protect\citeauthoryear{{Reardon} et~al.}{{Reardon}
  et~al.}{2023}]{ppta2023}
{Reardon} D.~J.,  et~al., 2023, \mn@doi [\apjl] {10.3847/2041-8213/acdd02},
  \href {https://ui.adsabs.harvard.edu/abs/2023ApJ...951L...6R} {951, L6}

\bibitem[\protect\citeauthoryear{{Ruderman}}{{Ruderman}}{1969}]{ruderman1969}
{Ruderman} M.,  1969, \mn@doi [\nat] {10.1038/223597b0}, \href
  {https://ui.adsabs.harvard.edu/abs/1969Natur.223..597R} {223, 597}

\bibitem[\protect\citeauthoryear{{Shaw} et~al.,}{{Shaw}
  et~al.}{2018}]{shawetal2018}
{Shaw} B.,  et~al., 2018, \mn@doi [\mnras] {10.1093/mnras/sty1294}, \href
  {https://ui.adsabs.harvard.edu/abs/2018MNRAS.478.3832S} {478, 3832}

\bibitem[\protect\citeauthoryear{{Shaw}, {Keith}, {Lyne}, {Mickaliger},
  {Stappers}, {Turner}  \& {Weltevrede}}{{Shaw} et~al.}{2021}]{shawetal2021}
{Shaw} B.,  {Keith} M.~J.,  {Lyne} A.~G.,  {Mickaliger} M.~B.,  {Stappers}
  B.~W.,  {Turner} J.~D.,   {Weltevrede} P.,  2021, \mn@doi [\mnras]
  {10.1093/mnrasl/slab038}, \href
  {https://ui.adsabs.harvard.edu/abs/2021MNRAS.505L...6S} {505, L6}

\bibitem[\protect\citeauthoryear{{Shibazaki} \& {Lamb}}{{Shibazaki} \&
  {Lamb}}{1989}]{shibazakiLamb1989}
{Shibazaki} N.,  {Lamb} F.~K.,  1989, \mn@doi [\apj] {10.1086/168062}, \href
  {https://ui.adsabs.harvard.edu/abs/1989ApJ...346..808S} {346, 808}

\bibitem[\protect\citeauthoryear{{Sidery}, {Passamonti}  \&
  {Andersson}}{{Sidery} et~al.}{2010}]{sideryPassamontiAndersson2010}
{Sidery} T.,  {Passamonti} A.,   {Andersson} N.,  2010, \mn@doi [\mnras]
  {10.1111/j.1365-2966.2010.16497.x}, \href
  {https://ui.adsabs.harvard.edu/abs/2010MNRAS.405.1061S} {405, 1061}

\bibitem[\protect\citeauthoryear{{Singh}}{{Singh}}{2017}]{singh2017}
{Singh} A.,  2017, \mn@doi [\prd] {10.1103/PhysRevD.95.024022}, \href
  {https://ui.adsabs.harvard.edu/abs/2017PhRvD..95b4022S} {95, 024022}

\bibitem[\protect\citeauthoryear{{Tenorio}, {Modafferi}, {Keitel}  \&
  {Sintes}}{{Tenorio} et~al.}{2022}]{tenorioetal2022}
{Tenorio} R.,  {Modafferi} L.~M.,  {Keitel} D.,   {Sintes} A.~M.,  2022,
  \mn@doi [\prd] {10.1103/PhysRevD.105.044029}, \href
  {https://ui.adsabs.harvard.edu/abs/2022PhRvD.105d4029T} {105, 044029}

\bibitem[\protect\citeauthoryear{{Walin}}{{Walin}}{1969}]{walin1969}
{Walin} G.,  1969, \mn@doi [Journal of Fluid Mechanics]
  {10.1017/S0022112069001662}, \href
  {https://ui.adsabs.harvard.edu/abs/1969JFM....36..289W} {36, 289}

\bibitem[\protect\citeauthoryear{{Wang} et~al.}{{Wang}
  et~al.}{2024}]{wangetal2024}
{Wang} H.,  et~al., 2024, The Astronomer's Telegram, 16619, 1

\bibitem[\protect\citeauthoryear{{Warszawski}, {Melatos}  \&
  {Berloff}}{{Warszawski} et~al.}{2012}]{warszawskiMelatosBerloff2012}
{Warszawski} L.,  {Melatos} A.,   {Berloff} N.~G.,  2012, \mn@doi [\prb]
  {10.1103/PhysRevB.85.104503}, \href
  {https://ui.adsabs.harvard.edu/abs/2012PhRvB..85j4503W} {85, 104503}

\bibitem[\protect\citeauthoryear{{Wilson} \& {Ho}}{{Wilson} \&
  {Ho}}{2024}]{wilsonHo2024}
{Wilson} O.~H.,  {Ho} W. C.~G.,  2024, \mn@doi [\prd]
  {10.1103/PhysRevD.109.083006}, \href
  {https://ui.adsabs.harvard.edu/abs/2024PhRvD.109h3006W} {109, 083006}

\bibitem[\protect\citeauthoryear{{Xu} et~al.}{{Xu} et~al.}{2023}]{cpta2023}
{Xu} H.,  et~al., 2023, \mn@doi [Research in Astronomy and Astrophysics]
  {10.1088/1674-4527/acdfa5}, \href
  {https://ui.adsabs.harvard.edu/abs/2023RAA....23g5024X} {23, 075024}

\bibitem[\protect\citeauthoryear{{Yim}}{{Yim}}{2022}]{yim2022}
{Yim} G.,  2022, PhD thesis, University of Southampton, UK

\bibitem[\protect\citeauthoryear{{Yim} \& {Jones}}{{Yim} \&
  {Jones}}{2020}]{yimJones2020}
{Yim} G.,  {Jones} D.~I.,  2020, \mn@doi [\mnras] {10.1093/mnras/staa2534},
  \href {https://ui.adsabs.harvard.edu/abs/2020MNRAS.498.3138Y} {498, 3138}

\bibitem[\protect\citeauthoryear{{Yim} \& {Jones}}{{Yim} \&
  {Jones}}{2023}]{yimJones2023}
{Yim} G.,  {Jones} D.~I.,  2023, \mn@doi [\mnras] {10.1093/mnras/stac3405},
  \href {https://ui.adsabs.harvard.edu/abs/2023MNRAS.518.4322Y} {518, 4322}

\bibitem[\protect\citeauthoryear{{Yim}, {Gao}, {Kang}, {Shao}  \& {Xu}}{{Yim}
  et~al.}{2024}]{yimetal2024}
{Yim} G.,  {Gao} Y.,  {Kang} Y.,  {Shao} L.,   {Xu} R.,  2024, \mn@doi [\mnras]
  {10.1093/mnras/stad3337}, \href
  {https://ui.adsabs.harvard.edu/abs/2024MNRAS.527.2379Y} {527, 2379}

\bibitem[\protect\citeauthoryear{{Yu} et~al.,}{{Yu} et~al.}{2013}]{yuetal2013}
{Yu} M.,  et~al., 2013, \mn@doi [\mnras] {10.1093/mnras/sts366}, \href
  {https://ui.adsabs.harvard.edu/abs/2013MNRAS.429..688Y} {429, 688}

\bibitem[\protect\citeauthoryear{{Zhou}, {G{\"u}gercino{\u{g}}lu}, {Yuan}, {Ge}
   \& {Yu}}{{Zhou} et~al.}{2022}]{zhouetal2022}
{Zhou} S.,  {G{\"u}gercino{\u{g}}lu} E.,  {Yuan} J.,  {Ge} M.,   {Yu} C.,
  2022, \mn@doi [Universe] {10.3390/universe8120641}, \href
  {https://ui.adsabs.harvard.edu/abs/2022Univ....8..641Z} {8, 641}

\bibitem[\protect\citeauthoryear{{Zubieta}, {Furlan}, {Palacio}, {Garcia},
  {Gancio}, {Lousto}, {Combi}  \& {PuMA Collaboration}}{{Zubieta}
  et~al.}{2024}]{zubietaetal2024}
{Zubieta} E.,  {Furlan} S.~B.~A.,  {Palacio} S.~d.,  {Garcia} F.,  {Gancio} G.,
   {Lousto} C.~O.,  {Combi} J.~A.,   {PuMA Collaboration} 2024, The
  Astronomer's Telegram, \href
  {https://ui.adsabs.harvard.edu/abs/2024ATel16608....1Z} {16608, 1}

\bibitem[\protect\citeauthoryear{{van Eysden} \& {Melatos}}{{van Eysden} \&
  {Melatos}}{2008}]{vanEysdenMelatos2008}
{van Eysden} C.~A.,  {Melatos} A.,  2008, \mn@doi [Classical and Quantum
  Gravity] {10.1088/0264-9381/25/22/225020}, \href
  {https://ui.adsabs.harvard.edu/abs/2008CQGra..25v5020V} {25, 225020}

\makeatother
\end{thebibliography}

%\begin{thebibliography}{99}
%\bibitem[\protect\citepauthoryear{Author}{2012}]{Author2012}
%\bibitem[\protect\citepauthoryear{Others}{2013}]{Others2013}
%\end{thebibliography}

%%%%%%%%%%%%%%%%%%%%%%%%%%%%%%%%%%%%%%%%%%%%%%%%%%

%%%%%%%%%%%%%%%%% APPENDICES %%%%%%%%%%%%%%%%%%%%%

\appendix

\section{Explicit form of A$_\text{2}$ and B$_\text{2}$}
%\section{Explicit form of $\mathbf{A_2}$ and $\mathbf{B_2}$}
\label{appendix_A_2_B_2}

In this appendix, we provide the explicit form of $A_2$ and $B_2$, first introduced in eq.\,(\ref{SNR_A_2_B_2}) of the main text. These equations come directly from Appendix~B of \cite{jaranowskiKrolakSchutz1998} but we state them here for convenience.
\begin{align}
    A_2(\delta, \psi, \iota, \lambda, \gamma) = F_2(\iota)e_1(\delta)\cos4\psi + G_2(\iota)e_2(\delta)~,
\end{align}
\begin{align}
    B_2(\alpha, \delta, \psi, \iota, \lambda, \gamma; T_\text{GW}) &= \frac{1}{\Omega_\oplus} \sum_{n=1}^{4} \sin\left(n \frac{\Omega_\oplus}{2} T_\text{GW}\right)\times \nonumber \\
    &~~~~\{C_{2,n}(\delta,\psi,\iota)\cos[n(\alpha - \phi_\oplus)] \nonumber \\
    &~~~~+ D_{2,n}(\delta,\psi,\iota)\sin[n(\alpha - \phi_\oplus)]\}~,
\end{align}
where $\Omega_\oplus$ is the rotational angular frequency of the Earth and $\phi_\oplus$ is the rotational phase of the Earth at $t=0$. In other words, the sum $\phi_\oplus + \Omega_\oplus t$ corresponds to the angle subtended between the meridian line that the detector lies on and the vernal equinox point (see Fig.\,\ref{figure_geometric_parameters}). There are also other functions that need defining:
\begin{align}
    C_{2,n}(\delta,\psi,\iota) &= F_2(\iota)[f_{1,n}(\delta)\cos4\psi + g_{1,n}(\delta)\sin4\psi] \nonumber \\
    &~~~~+ G_2(\iota)h_{1,n}(\delta)~, \\
    D_{2,n}(\delta,\psi,\iota) &= F_2(\iota)[f_{2,n}(\delta)\cos4\psi + g_{2,n}(\delta)\sin4\psi] \nonumber \\
    &~~~~+ G_2(\iota)h_{2,n}(\delta)~, \\
    F_2(\iota) &= \frac{1}{4}\sin^4\iota~, \\ 
    G_2(\iota) &=  \frac{1}{4} (1 + 6\cos^2\iota + \cos^4\iota)~,
\end{align}
where the functions $e_1$, $e_2$, $f_{k,n}$, $g_{k,n}$ and $h_{k,n}$ (for $k = 1, 2$ and $n = 1, 2, 3, 4$) are
{\allowdisplaybreaks
\begin{align}
    e_1(\delta) &= 4j_1 \cos^4\delta~,\nonumber \\
    e_2(\delta) &= 4j_2 - j_3 \cos2\delta + j_1 \cos^2 2\delta~,\nonumber \\
    f_{1,1}(\delta) &= - 4j_4 \cos^3\delta \sin\delta~,\nonumber \\
    f_{1,2}(\delta) &= j_5 \cos^2 \delta (3-\cos2\delta)~,\nonumber \\
    f_{1,3}(\delta) &= - j_6 (7 - \cos2\delta) \sin 2\delta~,\nonumber \\
    f_{1,4}(\delta) &= - j_7(35-28\cos2\delta + \cos4\delta)~,\nonumber \\
    f_{2,1}(\delta) &= - 28j_8 \cos^3\delta \sin\delta~,\nonumber \\
    f_{2,2}(\delta) &= -7j_9 \cos^2 \delta (3-\cos2\delta)~,\nonumber \\
    f_{2,3}(\delta) &= - j_{10} (7 - \cos2\delta) \sin 2\delta~,\nonumber \\
    f_{2,4}(\delta) &= - j_{11} (35-28\cos2\delta + \cos4\delta)~,\nonumber \\
    g_{1,1}(\delta) &= 28 j_{8} \cos^3 \delta~,\nonumber \\
    g_{1,2}(\delta) &= 28 j_{9} \cos^2 \delta \sin\delta~,\nonumber \\
    g_{1,3}(\delta) &= 2 j_{10} (5 - 3\cos2\delta)\cos\delta~,\nonumber \\
    g_{1,4}(\delta) &= 16 j_{11} (3 - \cos2\delta) \sin\delta~,\nonumber \\
    g_{2,1}(\delta) &= -4 j_{4} \cos^3 \delta~,\nonumber \\
    g_{2,2}(\delta) &= 4 j_{5} \cos^2 \delta \sin\delta~,\nonumber \\
    g_{2,3}(\delta) &= -2 j_{6} (5 - 3\cos2\delta)\cos\delta~,\nonumber \\
    g_{2,4}(\delta) &= -16 j_{7} (3 - \cos2\delta) \sin\delta~,\nonumber \\
    h_{1,1}(\delta) &= (j_{12} - j_{4} \cos2\delta)\sin2\delta~,\nonumber \\
    h_{1,2}(\delta) &= (j_{13} - j_{5} \cos2\delta)\cos^2\delta~,\nonumber \\
    h_{1,3}(\delta) &= 4 j_{6} \cos^3\delta \sin\delta~,\nonumber \\
    h_{1,4}(\delta) &= - 8 j_{7}\cos^4\delta \nonumber~,\\
    h_{2,1}(\delta) &= j_{8}(1 - 7 \cos2\delta)\sin2\delta~,\nonumber \\
    h_{2,2}(\delta) &= - j_{9}(5 - 7 \cos2\delta)\cos^2\delta~,\nonumber \\
    h_{2,3}(\delta) &= 4 j_{10} \cos^3\delta \sin\delta~,\nonumber \\
    h_{2,4}(\delta) &= - 8 j_{11}\cos^4\delta~,\nonumber
\end{align}}and all the $j_1$ to $j_{13}$ coefficients depend on the detector latitude $\lambda$ and orientation $\gamma$
{\allowdisplaybreaks
\begin{align}
    j_1(\lambda, \gamma) &= \frac{1}{256} (4 - 20\cos^2\lambda + 35 \sin^2 2\gamma\cos^4\lambda)~,\nonumber \\ 
    j_2(\lambda, \gamma) &= \frac{1}{1024} (68 - 20\cos^2\lambda - 13 \sin^2 2\gamma\cos^4\lambda)~,\nonumber \\ 
    j_3(\lambda, \gamma) &= \frac{1}{128} (28 - 44\cos^2\lambda + 5 \sin^2 2\gamma\cos^4\lambda)~,\nonumber \\ 
    j_4(\lambda, \gamma) &= \frac{1}{32} (2 - 7 \sin^2 2\gamma\cos^2\lambda) \sin 2\lambda~,\nonumber \\ 
    j_5(\lambda, \gamma) &= \frac{1}{32} (3 - 7\cos4\gamma - 7 \sin^2 2\gamma\cos^2\lambda) \cos^2\lambda~,\nonumber \\ 
    j_6(\lambda, \gamma) &= \frac{1}{96} (2\cos4\gamma + \sin^2 2\gamma\cos^2\lambda) \sin 2\lambda~,\nonumber \\ 
    j_7(\lambda, \gamma) &= \frac{1}{1024} (4\cos4\gamma\sin^2\lambda - \sin^2 2\gamma\cos^4\lambda)~,\nonumber \\ 
    j_8(\lambda, \gamma) &= \frac{1}{32} \sin4\gamma \cos^3\lambda~,\nonumber \\ 
    j_9(\lambda, \gamma) &= \frac{1}{32} \sin4\gamma \cos^2\lambda \sin\lambda~,\nonumber \\ 
    j_{10}(\lambda, \gamma) &= \frac{1}{192} \sin4\gamma (5 - 3\cos2\lambda)\cos\lambda~,\nonumber \\ 
    j_{11}(\lambda, \gamma) &= \frac{1}{1024} \sin4\gamma (3 - \cos2\lambda)\sin\lambda~,\nonumber \\ 
    j_{12}(\lambda, \gamma) &= \frac{1}{32} (14 - \sin^2 2\gamma \cos^2\lambda) \sin 2\lambda~,\nonumber \\ 
    j_{13}(\lambda, \gamma) &= \frac{1}{32} (9 - 5\cos4\gamma - 5 \sin^2 2\gamma \cos^2\lambda) \cos^2\lambda~.\nonumber 
\end{align}}

\section{Cleaning the different catalogues}
\label{appendix_cleaning}
In this appendix, we outline the changes made to the three publicly available databases used in the analysis. They are the ATNF Pulsar Catalogue \citep{manchesteretal2005}, the JBCA Glitch Catalogue \citep{espinozaetal2011, basuetal2022} and the ATNF Glitch Table \citep{manchesteretal2005}. Often, the changing of a pulsar's J-name was to ensure it matched with the other databases. The specific details of the cleaning process are listed below. 

ATNF Pulsar Catalogue (v2.1.1):
{\allowdisplaybreaks
\begin{itemize}
    \item Removed pulsars without frequency or distance measurements.
    \item Changed J-name of ``J0625+10'' to ``J0625+1015'' \citep{basuetal2022}.
    \item Changed J-name of ``J1636$-$2614'' to ``J1635$-$2614''. The former was used in \cite{burgayetal2019} but was updated in \cite{basuetal2022}.
    \item Changed J-name of ``J1844+00'' to ``J1844+0034'' \citep{espinozaetal2011}.
\end{itemize}}
JBCA Glitch Catalogue:
{\allowdisplaybreaks
\begin{itemize}
    \item The J-name for J1913+0446's second glitch was corrected to ``J1913+0446'' from ``J1914+0446''.
    \item Changed J-name of ``B0410+69'' to ``J0415+6954'' \citep{basuetal2022}.
    \item Changed J-name of ``B1254$-$10'' to ``J1257$-$1027'' \citep{delleretal2019}.
    \item Changed J-name of ``J1824$-$2452'' to ``J1824$-$2452A'' to be consistent with the ATNF Pulsar Database.
    \item Changed J-name of ``AX\_1838-0655'' to ``J1838$-$0655'' \citep{kuiperHermsen2010}.
    \item Changed J-name of ``J1844+00'' to ``J1844+0034'' \citep{espinozaetal2011}.
    \item ``J0417+35'' was not in standard form but we left it as it is \citep{basuetal2022}.
    \item Removed two glitches from J1341$-$6220 that occurred on MJD~58178 and MJD~58214 as both had no $\Delta \nu / \nu$ readings. This is because the two glitches occurred too close to one another for independent readings \citep{loweretal2021}.
    \item Removed an antiglitch from 1E\_2259+586 (J2301+5852) that occurred on MJD~54880 \citep{icdemBaykalInam2012}.
    \item Added a closing parenthesis for the error on the glitch size of J0631+1036 at MJD~54632.
    \item Removed five pulsars with no distance measurements affecting seven glitches: J1647$-$4552 (MJD~53999), J1838$-$0537 (MJD~55100), J2111+4606 (MJD~55700 and 57562), J1048$-$5937 (MJD~52386 and 54185; pulsar not in ATNF Pulsar Catalogue), M82-X2 (MJD~56685; pulsar not in ATNF Pulsar Catalogue).
\end{itemize}}
ATNF Glitch Table:
{\allowdisplaybreaks
\begin{itemize}
    \item Changed J-name of ``J0625+10'' to ``J0625+1015'' \citep{basuetal2022}.
    \item Changed J-name of ``B1254$-$10'' to ``J1257$-$1027'' \citep{delleretal2019}.
    \item Changed J-name of ``J1636$-$2614'' to ``J1635$-$2614''. The former was used in \cite{burgayetal2019} but was updated in \cite{basuetal2022}.
    \item Changed J-name of ``J1844+00'' to ``J1844+0034'' \citep{espinozaetal2011}.
    \item Removed two glitches from J1341$-$6220 that occurred on MJD~58178 and MJD~58214 as both had no $\Delta \nu / \nu$ readings. This is because the two glitches occurred too close to one another for independent readings \citep{loweretal2021}.
    \item Removed an antiglitch from 1E\_2259+586 (J2301+5852) that occurred on MJD~56035 \citep{archibaldetal2013}.
    \item Removed an antiglitch from J1522$-$5735 that occurred on MJD~55250 \citep{pletschetal2013}.
    \item Note that the MJD~54661 glitch for J1801$-$2451 was reported twice which was due to observations from two independent groups.
    \item Note that of the 621 glitches, 15 glitches have 2 recovery components, one glitch has 3 components and one glitch has 4 components.
    \item Removed four pulsars with no distance measurements affecting five glitches: J1422$-$6138 (MJD~55310 and 55450), J1822$-$1604 (MJD~56756), J1844$-$0346 (MJD~56135), J1906+0722 (MJD~55063).
\end{itemize}}

\section{Differences between the glitch catalogues}
\label{appendix_differences_between_glitch_catalogues}

In this appendix, we document the pulsars that are missing from the JBCA Glitch Catalogue but are found in the ATNF Glitch Table (and vice versa). We list the pulsars that are missing before the cleaning procedure (mentioned in Appendix~\ref{appendix_cleaning}) is applied.

The missing pulsars from the JBCA Glitch Catalogue that are found in the ATNF Glitch Table are:

J0908$-$4913,
J0954$-$5430,
J1015$-$5719,
J1050$-$5953,
J1141$-$6545,
J1422$-$6138*, 
J1550$-$5418,
J1602$-$5100,
J1636$-$2614\dag, 
J1645$-$0317,
J1703$-$4851,
J1706$-$4434,
J1722$-$3632,
J1822$-$1604*, 
J1844$-$0346*, 
J1852$-$0635,
J1906+0722*, 
J1910+1026,
J1915+1150,
J1939+2609,
J1947+1957,
and J1954+2529,

\noindent where the asterisks (*) show pulsars with no distance measurements and the dagger (\dag) represents J1636$-$2614 which does exist in the JBCA Glitch Catalogue but as J1635$-$2614 which we ultimately rename it to.

The missing pulsars from the ATNF Glitch Table that are found in the JBCA Glitch Catalogue are:

J0417+35,
J0726$-$2612,
J0738$-$4042,
J0855$-$3331,
J1048$-$5937,
J1635$-$2614\dag, 
J1647$-$4552*, 
J1730$-$3353,
J1809$-$0119,
J1821$-$1419,
J1832+0029,
J1838$-$0537*, 
J1838$-$0655,
J1844$-$0310,
J1907+0602,
J1948+2819,
J1949$-$2524,
J2022+2854,
J2111+4606*, 
and M82-X2,

\noindent where again, the asterisks represent pulsars with no distance measurement and the dagger represents J1635$-$2614 which is same as J1636$-$2614 in the ATNF Glitch Catalogue.

%%%%%%%%%%%%%%%%%%%%%%%%%%%%%%%%%%%%%%%%%%%%%%%%%%

% Don't change these lines
\bsp	% typesetting comment
\label{lastpage}
\end{document}